\documentclass[11pt,a4paper]{article} 
\usepackage{jheppub}

\usepackage[utf8]{inputenc} 
\usepackage{amsmath,tabu}
\usepackage{amsfonts,dsfont,mathrsfs} 
\usepackage{blkarray} 
\usepackage{subcaption} 
\usepackage{rotating} 
\usepackage{float}
\usepackage{tensor}
\usepackage{multirow}
\usepackage{adjustbox}
\allowdisplaybreaks

\usepackage[table]{xcolor}

\usepackage{makecell}
\setcellgapes{5pt}

\usepackage{graphicx}

\newcommand{\vl}{\mathbin{\rotatebox[origin=c]{-90}{$<$}}}
\newcommand{\ve}{\mathbin{\rotatebox[origin=c]{-90}{$=$}}}
\newcommand{\vleq}{\mathbin{\rotatebox[origin=c]{-90}{$\leq$}}}

\usepackage{pdflscape}
\usepackage{lscape}
\usepackage{rotating}

\usepackage{mathtools} 
\newcommand{\defeq}{\vcentcolon=}

\newcommand\restr[2]{{
  \left.\kern-\nulldelimiterspace 
  #1 
  \vphantom{\big|} 
  \right|_{#2} 
  }}
  
\def\symmlabel#1{$\phantom{.^2} #1 \phantom{.^2}$}

\usepackage{tikz} 

\usetikzlibrary{decorations.pathreplacing} 
\usetikzlibrary{decorations.pathmorphing} 
\usetikzlibrary{decorations.markings} 
\usetikzlibrary{arrows} 
\usetikzlibrary{shapes} 
\usetikzlibrary{matrix} 
\usetikzlibrary{positioning} 
\usepackage[english]{babel} 
\usepackage[autostyle]{csquotes}

\tikzset{cross/.style={cross out, draw=black}} 
\tikzset{circ/.style={circle,fill=white,draw=black}} 
\tikzset{hasse/.style={circle, fill,inner sep=2pt}} 
\tikzset{h/.style={circle, fill,inner sep=2pt}} 
\tikzset{ns/.style={circle, draw,inner sep=2pt}} 
\tikzset{dot/.style={circle,draw,fill=black}} 
\tikzset{gauge/.style={circle,draw}} 
\tikzset{flavor/.style={regular polygon,regular polygon sides=4, draw}} 
\tikzset{doublearrow/.style={ draw=black!75, color=black!75, thick, double distance=3pt, }} 
\tikzset{thirdline/.style={ draw=black!75, color=black!75, thick, }} 
\tikzset{middlearrow/.style={ decoration={markings, mark= at position 0.5 with {\arrow{#1}} , }, postaction={decorate} } } 
\tikzset{subquiver/.style={circle,draw,dashed,inner sep=0pt,minimum size=1cm}} 
\newcommand{\midarrow}{\tikz 
\draw[-triangle 90] (0,0) -- +(.1,0);} 
\newcommand{\midarrowrev}{\tikz 
\draw[-triangle 90] +(.1,0) -- (0,0);}
\newcommand{\midarrowup}{\tikz 
\draw[-triangle 90] +(0,0) -- (0,.1);}
\newcommand{\midarrowdown}{\tikz 
\draw[-triangle 90] +(0,.1) -- (0,0);}
\newcommand{\midarrowdiag}{\tikz 
\draw[-triangle 90] +(.1,.05) -- (0,0);}
\newcommand{\midarrowdiagtwo}{\tikz 
\draw[-triangle 90] +(0,0) -- (.05,.1);}

\newcommand{\abel}{\mathrm{abel}}
\newcommand{\Coul}{\mathcal C}
\newcommand{\tCoul}{\Tilde{\mathcal C}}
\newcommand{\hatCoul}{\hat{\mathcal C}}
\newcommand{\Higgs}{\mathcal H} 
\newcommand{\co}{\mathbb C} 
\newcommand{\Z}{\mathbb Z} 
\newcommand{\N}{\mathbb Z_{\geq 0}} 
\newcommand{\qu}{\mathbb H} 
\newcommand{\m}{\mathbf m} 
 
\newcommand{\rank}{\text {rank}} 
\def\tr#1{\mathrm{tr}#1}

\def\e#1{e_{\langle #1 \rangle}} 
\def\h#1{h_{#1}} 
\def\Ed#1{E^*_{\langle #1 \rangle}} 
\def\Hd#1{H^*_{#1}} 
\def\te#1{\Tilde{e}_{\langle #1 \rangle}}
\def\th#1{\Tilde{h}_{#1}} 
\def\tEd#1{\Tilde{E}^*_{\langle #1 \rangle}}  
\def\tHd#1{\Tilde{H}^*_{#1}} 
\def\hate#1{\hat{e}_{\langle #1 \rangle}}

\def\p#1{\varphi_{#1}} 
\def\m#1{M_{#1}} 
\def\up#1{u^+_{#1}} 
\def\um#1{u^-_{#1}}  
\def\upm#1{u^\pm_{#1}} 
\def\tp#1{\Tilde{\varphi}_{#1}} 
\def\tup#1{\Tilde{u}^+_{#1}} 
\def\tum#1{\Tilde{u}^-_{#1}} 
 
\def\hatp#1{\hat{\varphi}_{#1}} 
\def\hatup#1{\hat{u}^+_{#1}} 
\def\hatum#1{\hat{u}^-_{#1}}
\def\hatupm#1{\hat{u}^\pm_{#1}} 
\def\x#1{x_{#1}}

\def\hatx#1{\hat{x}_{#1}}
\def\sl#1{\mathfrak{sl}(#1,\mathbb{C})} 
\def\so#1{\mathfrak{so}(#1,\mathbb{C})} 
 
\def\usp#1{\mathfrak{usp}(#1,\mathbb{C})} 
\def\be{
\begin{equation}
	} 
	\def\ee{
\end{equation}
} 
\newcommand\pois[1]{\lbrace #1 \rbrace}
\def\mommap#1{N_{#1}} 
\def\PL{\mathrm{PL}}
\def\HS{\mathrm{HS}}
\def\HWG{\mathrm{HWG}}

\newcommand{\im}{\mathrm{i}}

\allowdisplaybreaks

\preprint{Imperial/TP/20/AH/05}

\title{Quiver origami: discrete gauging and folding} 
\author{Antoine Bourget, } 
\author{Amihay Hanany, } 
\author{Dominik Miketa} 
\affiliation{Theoretical Physics, The Blackett Laboratory\\
Imperial College London\\
SW7 2AZ United Kingdom}

\emailAdd{a.bourget@imperial,ac.uk, a.hanany@imperial.ac.uk, d.miketa16@imperial.ac.uk}

\abstract{We study two types of discrete operations on Coulomb branches of $3d$ $\mathcal{N}=4$ quiver gauge theories using both abelianisation and the monopole formula. We generalise previous work on discrete quotients of Coulomb branches and introduce novel \emph{wreathed} quiver theories. We further study \emph{quiver folding} which produces Coulomb branches of non-simply laced quivers.}


\begin{document}

\maketitle \flushbottom

\section{Introduction}

The purpose of this work is to clarify the relation between several concepts relating to $3d$ $\mathcal{N}=4$ Coulomb branches. It has been known since \cite{cremonesi_coulomb_2014} that the Coulomb branch monopole formula \cite{cremonesi_monopole_2014} can be extended to quivers in the form of non-simply laced framed Dynkin diagrams. However, while all the ingredients of a simply laced Dynkin diagram – gauge and flavor nodes, hypermultiplet links – are readily interpretable, it was unknown at the time what to make of the novel multiple link. Recently \cite{nakajima_coulomb_2019} argued that their Coulomb branches result from a discrete folding operation on Coulomb branches of simply laced quivers. We independently derive and illustrate the same claim through the method of abelianisation \cite{bullimore_coulomb_2017}. We also develop a second, related but distinct discrete operation which was previously studied in \cite{hanany_discrete_2018,hanany_discrete_2018-1}. Both aspects expand on our previous work in \cite{hanany_nilpotent_2019}.

\begin{figure}
	[t] \centering 
	\begin{tikzpicture}
		\node (Un) [gauge, label=left:{$U(n)$}]{}; 
		\node (Qp1) [subquiver, above left = of Un, label=left:{$Q'_1$}]{}; 
		\node (Qpk) [subquiver, below left = of Un, label=left:{$Q'_k$}]{}; 
		\node (Q1) [subquiver, above right = of Un, label=right:{$Q_1$}]{}; 
		\node (Qm) [subquiver, below right = of Un, label=right:{$Q_m$}]{}; 
		\draw (Qp1)--(Un); 
		\draw (Qpk)--(Un); 
		\draw (Un)--(Q1); 
		\draw (Un)--(Qm); 
		\draw[loosely dotted] ([shift={(Un)}]330:2) arc[radius=2, start angle=330, end angle=390]; 
		\draw[loosely dotted] ([shift={(Un)}]150:2) arc[radius=2, start angle=150, end angle=210];
		
		\node (sUn) [gauge, right = 8cm of Un, label=left:{$U(n)$}]{}; 
		\node (sQp1) [subquiver, above left = of sUn, label=left:{$Q'_1$}]{}; 
		\node (sQpk) [subquiver, below left = of sUn, label=left:{$Q'_k$}]{}; 
		\node (sQ) [subquiver, right = of sUn, label=right:{$Q_i\wr S_m$}]{}; 
		\draw (sQp1)--(sUn); 
		\draw (sQpk)--(sUn); 
		\draw (sUn)--(sQ); 
		\draw[loosely dotted] ([shift={(sUn)}]150:2) arc[radius=2, start angle=150, end angle=210];
		
		\node (hiddenL) [right = 2cm of Un]{}; 
		\node (hiddensR) [left = 2cm of sUn]{};
		
		\draw[thick, ->] (hiddenL) -- (hiddensR);
		
		\node (fUn) [gauge, below = 6cm of sUn, label=left:{$U(n)$}]{}; 
		\node (fQp1) [subquiver, above left = of fUn, label=left:{$Q'_1$}]{}; 
		\node (fQpk) [subquiver, below left = of fUn, label=left:{$Q'_k$}]{}; 
		\node (fQ) [subquiver, right = of fUn, label=right:{$Q_i$}]{}; 
		\draw (fQp1)--(fUn); 
		\draw (fQpk)--(fUn); 
		\draw[doublearrow] (fUn)-- node {\midarrow} (fQ); 
		\draw[loosely dotted] ([shift={(fUn)}]150:2) arc[radius=2, start angle=150, end angle=210];
		
		\node (hiddensB) [below = 1.5cm of sUn]{}; 
		\node (hiddenfA) [above = 1.5cm of fUn]{};
		
		\draw[thick, ->] (hiddensB) -- (hiddenfA);
		
		\node (hiddenBR) [below = 2cm of Un]{}; 
		\node (hiddenfAL) [left= 2cm of fUn]{};
		
		\draw[thick, ->] (hiddenBR) -- (hiddenfAL); 
	\end{tikzpicture}
	\caption[Wreathed and non-simply laced quivers]{(Top left) $k$ generic subquivers $Q'_1$ through $Q'_k$ and $m$ identical subquivers $Q_1$ through $Q_m$ are connected to a common central $U(n)$ node. (Top right) Wreathed quiver. (Bottom right) Non-simply laced quiver. The multiple link has valence $m$, here depicted for $m=2$.} 
\label{fig:discrete-operations} \end{figure}
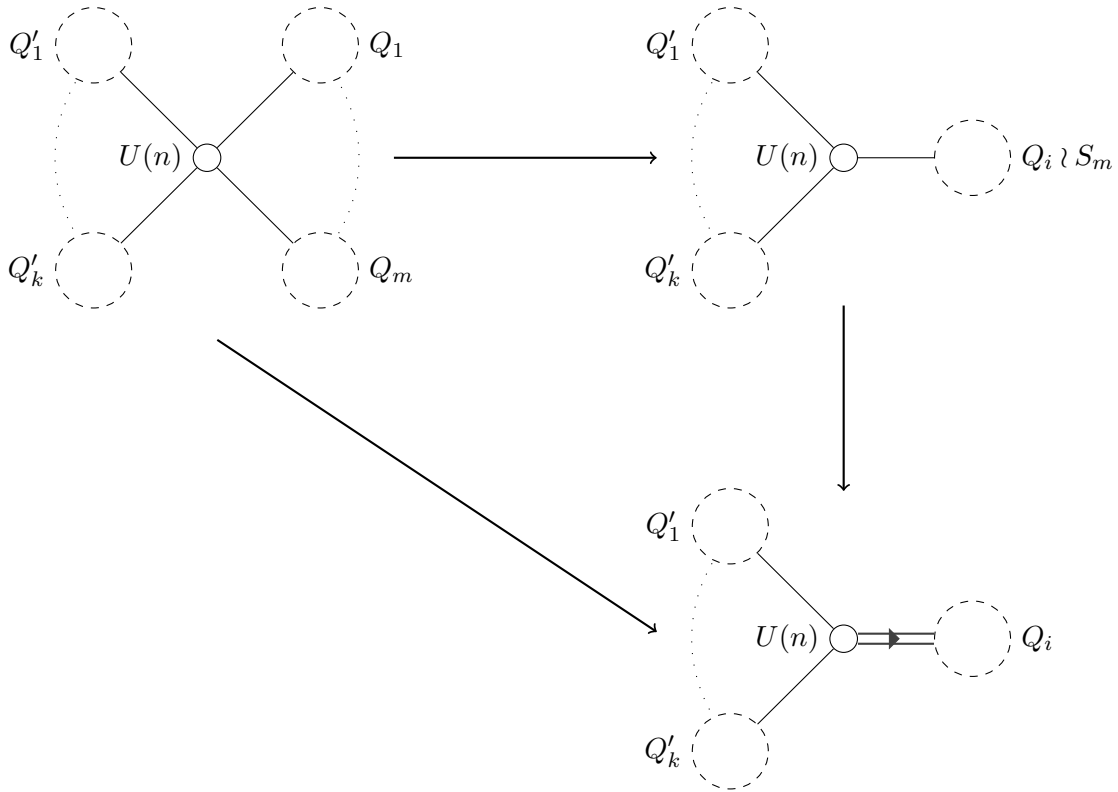

The main concepts, presented in Fig. \ref{fig:discrete-operations}, can be summarised as follows. 

Quivers with an automorphism possess a discrete symmetry relating gauge groups. By analogy with continuous gauge groups, it, or any of its subgroups, can be gauged\footnote{Discretely gauging string backgrounds is of course an old idea which has generated a lot of discussion, for example in \cite{marcus_tree-level_1982,dixon_strings_1985,dixon_strings_1986,dabholkar_lectures_1998,sen_duality_1996}, and our paper may be viewed as a new entry.}, and we demonstrate that this results in a theory whose Coulomb branch is a discrete quotient of the original, where the action by which we quotient is directly induced by the quiver automorphism (or subgroup thereof). This operation, which we call \emph{discrete gauging}, produces \emph{wreathed quivers}. Previous work \cite{hanany_discrete_2018,hanany_discrete_2018-1} generated similar results on the Coulomb branch by replacing $n$ $U(1)$ nodes by a $U(n)$ node with adjoint matter.

In contrast, \emph{quiver folding} relates Coulomb branches of pairs of simply laced and non-simply laced quiver gauge theories. To be clear, we show the action on the Coulomb branch and conjecture that one can view it as one effect of an action on the \emph{theory}\footnote{In the rest of this text we will elide the distinction between folding a quiver theory and folding its Coulomb branch, but wish to be clear that we present solid evidence only for the latter and at best circumstantial evidence for the former.}. However, we have been unsuccessful in our attempts to write down the path integral or compute the Higgs branch of folded theories. Coulomb branches of balanced $A_{2n-1}$–type quivers, ie. framed linear quivers satisfying the balance condition\footnote{A node is balanced when the contributions of gauge and matter to the RG flow of the gauge coupling exactly cancel out \emph{assuming the quiver is understood as a $4d$ theory}. Assuming simply laced unitary quivers without loops, this amounts to the condition that twice the node's rank equals the sum of all surrounding (gauge or matter) nodes' ranks.} and exhibiting $\sl{2n}$ symmetry on the Coulomb branch, can be ``folded" into Coulomb branches of balanced $C_n$–type quivers with $\usp{2n}$ symmetry. Balanced $D_n$–type quiver Coulomb branches, ie. Coulomb branches of balanced framed quivers shaped like $D_n$ Dynkin diagrams, can be ``folded" into Coulomb branches of balanced $B_{n-1}$–type quivers. $G_2$–type quiver Coulomb branches can be similarly obtained from $D_4$–type quivers while $F_4$–type quivers are folded $E_6$–type quivers. The folded spaces are fixed points under the group action induced by the quiver automorphism and we show that they are symplectic leaves of spaces obtained by discretely gauging their respective original Coulomb branches. In some cases distinct subgroups of the quiver automorphism can give identical sets of fixed points (eg. $S_3$ and $\Z_3$ of the $D_4$ affine quiver) and their folded spaces coincide; as a result, there are ``fewer" folded than wreathed quivers.

\begin{figure}[t]
	\centering 
	\includegraphics[width=
	\textwidth]{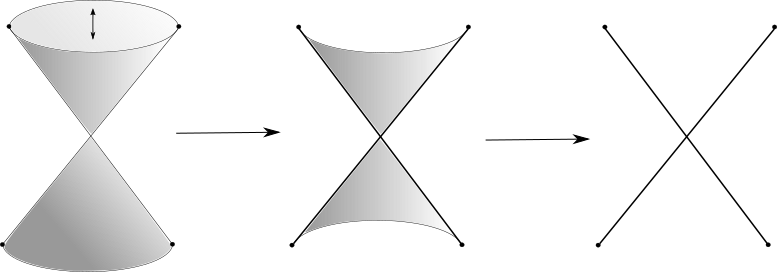} 
	\caption[Geometric interpretation of discrete gauging and folding]{(left) Initial Coulomb branch with highlighted $\mathbb{Z}_2$ symmetry. (middle) Coulomb branch of the discretely gauged quiver depicted as an orbifold of the original space. Note that bold edges form a singular subspace under the $\mathbb{Z}_2$ symmetry. (right) Coulomb branch of the folded quiver, the subspace fixed under the $\mathbb{Z}_2$ symmetry.} 
\label{fig:geometric_interpretation} \end{figure}

Actions of both discrete gauging and folding on the Coulomb branch are readily interpreted through a geometric lens, see Figure \ref{fig:geometric_interpretation}. We claim that, since discrete gauging is implemented by restricting the chiral ring to invariants of a symmetry group action $\Gamma$, the resulting space is an orbifold of the initial Coulomb branch under $\Gamma$ – and since the Poisson structure respects this group action, the orbifold inherits a natural symplectic structure. If the original space is a nilpotent orbit of some algebra then the orbifold is sometimes, but not always, a nilpotent orbit of the relevant folded algebra, but it is in any case symmetric under the folded algebra's action.

Folding, on the other hand, reduces the Coulomb branch to the fixed subspace under the same group action $\Gamma$. We show that it has a Poisson structure and, since the fixed subspace is (a singular) part of the corresponding orbifold, the Hasse diagram \cite{bourget_higgs_2020} of the folded space is a subdiagram of the orbifold's Hasse diagram. In all known cases a nilpotent orbit folds to another nilpotent orbit (of the folded algebra). 
This situation is reminiscent of a general phenomenon identified in \cite{achar_geometric_2013}, in which orbits in the small affine Grassmannian for an algebraic group $G$ (the subvariety of the affine Grassmannian corresponding to the so-called small coweights of $G$; see \cite{futureAG} for a friendly introduction addressed to physicists) possess a $\mathbb{Z}_2$ global involution -- here these orbits would be depicted as the left portion of Figure \ref{fig:geometric_interpretation}. Some of these orbits can be mapped to so-called \emph{Reeder pieces} which are the union of two nilpotent orbits of $G$, one which can be identified with a $\mathbb{Z}_2$ quotient of the affine Grassmannian slice, and the other as the $\mathbb{Z}_2$ fixed points -- respectively the middle and right parts of Figure \ref{fig:geometric_interpretation}. Coulomb branches of framed unitary ADE quivers were identified with slices in the associated affine Grassmannian \cite{braverman_coulomb_2019}, following the construction \cite{nakajima_towards_2016,braverman_towards_2019}, and the $\mathbb{Z}_2$ involution of \cite{achar_geometric_2013} is realized on the quiver as leg permutations like in Figure \ref{fig:discrete-operations}. As a consequence, several of the examples discussed below follow from the geometric point of view from these previous works; the present paper sheds a new light on this topic by providing quivers for each of the three spaces, and giving formulas to compute the Hilbert series and HWGs of their closures.

Figure \ref{tabTerminology} features a third discrete action called \emph{crossing}. Flavorless simply laced quiver theories possess a certain freedom of reparametrisation: the gauge group $G$ factorises as $G/U(1) \times U(1)$, with the $U(1)$ factor contributing a (geometrically uninteresting) factor of $\qu$ to the Coulomb branch, which is discarded by convention. Crucially, while the choice of $U(1)$ is somewhat constrained, the allowed options are in practice equivalent and one can in particular choose to ungauge any given $U(1)$ node without affecting the Coulomb branch. The situation is modified for non-simply laced quivers, where ungaugings on opposite sides of the directed multiple link give rise to pairs of Coulomb branches where one is the discrete quotient of the other. We list this case for the sake of completeness, but do not study it further in this paper. The reader could instead consult the recent treatment in \cite{hanany_ungauging_2020}.

\begin{figure}[t] 
\centering 
	\hspace*{-.5cm}\begin{tabular}
    	{| >{\centering\arraybackslash} m{2.2cm} | >{\centering\arraybackslash} m{4.2cm} | >{\centering\arraybackslash} m{2.1cm} | >{\centering\arraybackslash} m{2.1cm} | >{\centering\arraybackslash} m{3cm} |} 
		\hline  Operation on gauge theory & Quiver description & Resulting quiver & String background action & Operation on Coulomb branch \\ \hline
		\hline Discrete gauging & 
		\begin{tikzpicture}
			\node (g1) at (0,0) [gauge, label=below:{$a$}]{}; 
			\node (g2) at (1,.5) [gauge, label=right:{$b$}]{}; 
			\node (g3) at (1,-.5) [gauge, label=right:{$b$}]{}; 
			\draw (g2)--(g1)--(g3); 
			\node at (1.85,0) {$\rightarrow$};
			\node (g4) at (2.5,0) [gauge, label=below:{\symmlabel{a}}]{}; 
			\node (g5) at (3.5,0) [gauge, label=below:{\symmlabel{b\wr S_2}}]{}; 
			\draw (g4)--(g5); 
		\end{tikzpicture} &
		Wreathed quiver &
		Orbifold &
		Discrete quotient (dimension preserving)
		\\
		\hline Folding & 
		\begin{tikzpicture}
			\node (g1) at (0,0) [gauge, label=below:{$a$}]{}; 
			\node (g2) at (1,.5) [gauge, label=right:{$b$}]{}; 
			\node (g3) at (1,-.5) [gauge, label=right:{$b$}]{}; 
			\draw (g2)--(g1)--(g3); 
			\node at (1.85,0) {$\rightarrow$};
			\node (g4) at (2.5,0) [gauge, label=below:{\symmlabel{a}}]{}; 
			\node (g5) at (3.5,0) [gauge, label=below:{\symmlabel{b}}]{}; 
			\draw[doublearrow] (g4)-- node {\midarrow} (g5); 
		\end{tikzpicture} &
		Non-simply laced quiver &
		Orientifold &
		Fixed points \hspace{1cm} (not dimension preserving)
		\\
		\hline Crossing & 
		\begin{tikzpicture}
			\node (g1) at (0,1) [gauge, label=above:{$1$}]{}; 
			\node (g2) at (0,0) [gauge, label=below:{$a$}]{}; 
			\node (g3) at (1,0) [gauge, label=below:{$b$}]{}; 
			\node (g4) at (1,1) [flavor, label=above:{$1$}]{}; 
			\draw (g1)--(g2);
			\draw[doublearrow] (g2)-- node {\midarrow} (g3); 
			\draw (g3)--(g4);
			\node at (1.85,0) {$\rightarrow$};
			\node (g5) at (2.5,1) [flavor, label=above:{$1$}]{}; 
			\node (g6) at (2.5,0) [gauge, label=below:{$a$}]{}; 
			\node (g7) at (3.5,0) [gauge, label=below:{$b$}]{}; 
			\node (g8) at (3.5,1) [gauge, label=above:{$1$}]{}; 
			\draw (g5)--(g6);
			\draw[doublearrow] (g6)-- node {\midarrow} (g7); 
			\draw (g7)--(g8);
		\end{tikzpicture} &
		Non-simply laced quiver &
		? &
		Discrete quotient \hspace{1cm} (dimension preserving)
		\\
		\hline 
	\end{tabular}
	\caption{Discrete actions on the quiver} 
\label{tabTerminology} \end{figure}

\subsection*{Kostant-Brylinski reductions}

In \cite{brylinski_nilpotent_1994} the authors identified that discrete quotients of certain minimal nilpotent orbits were equivalent to (generically non-minimal) nilpotent orbits of other algebras; their results are summarised in Figure \ref{tabKostantBrylinski}\footnote{\cite{kobak_classical_1996} provide more examples of discrete and non-discrete quotients in nilpotent orbits.}. The same pattern is observed in discrete gauging and we claim that our construction is a physical realisation of their cases 1,2,3,4 and 9. We empirically confirmed this conjecture using both Hilbert series and abelianisation methods as in \cite{hanany_nilpotent_2019} up to low but non-trivial rank.  The lines painted in green (cases 2, 3, 4 and 9) correspond to wreathed simply laced quivers. Case 1, painted in red, stands apart because of the non-simply laced initial quiver; although the moduli space can be described algebraically using abelianised variables, the explicit implementation of the monopole formula for non-simply laced wreathed quivers is postponed for future investigations. 

A recent work \cite{hanany_ungauging_2020} showed that cases 5, 6 and 7 (yellow in Figure \ref{tabKostantBrylinski}) occur in Coulomb branches of non-simply laced quivers. The $\mathbb{Z}_n$ quotient corresponds to gauging a $U(1)$ node on the ``long" end of an edge of multiplicity $n$ and ungauging another $U(1)$ node on the ``short" end. The quiver realisations of all eight known cases are collected in Figure \ref{tabKostantBrylinski}.

Case number 8 still presents a challenge, and we are not aware of any quiver realisation of the corresponding $\mathbb{Z}_2^2$ quotient. However the HWGs are under control, and are discussed briefly at the end of Section \ref{sectionHWG}.

\begin{landscape}
\begin{figure}
	[p] \centering 
	\begin{tabular}
	    {| >{\centering\arraybackslash} m{0.3cm} | >{\centering\arraybackslash} m{1cm} |  >{\centering\arraybackslash} m{0.8cm} | >{\centering\arraybackslash} m{2.5cm} |  >{\centering\arraybackslash} m{2.3cm} |  >{\centering\arraybackslash} m{1.2cm} | >{\centering\arraybackslash} m{5.2cm} |  >{\centering\arraybackslash} m{5.2cm} |}
		\hline  & $\mathfrak{g}$ & $\Tilde{\mathfrak{g}}$ & $V$ & $M$ & $\mathrm{dim}_{\mathbb{H}} M$ & Quiver for $\mathfrak{g}$ & Quiver for $\Tilde{\mathfrak{g}}$  \\   \hline 
		\hline 
\rowcolor{red!10}		1 & $B_3$  & $G_2$ & $\mathbb{C}^7$ & $\mathcal{O}_{[3,2^2]}^{G_2}$ & 4 & 
		\vspace{0.06cm}
		\scalebox{.88}{\begin{tikzpicture}
		    \node (g1) at (-1,-.5) [gauge, label=left:{1}]{};
			\node (f2) at (-1,.5) [flavor, label=left:{1}]{};
			\node (g2) at (0,0) [gauge, label=below:{2}]{};
			\node (g3) at (1,0) [gauge, label=below:{1}]{};
			\draw (g1)--(g2)--(f2); 
			\draw[doublearrow] (g2)-- node {\midarrow} (g3);
		\end{tikzpicture}}
		& 
		\vspace{0.06cm}
		\scalebox{.88}{\begin{tikzpicture}
			\node (f2) at (2,0) [flavor, label=below:{1}]{};
			\node (g1) at (0,0.5) [gauge, label=left:{1}]{}; 
			\node (g2) at (1,0) [gauge, label=below:{2}]{}; 
			\node (g3) at (0,-0.5) [gauge, label=left:{1}]{};
			\draw (g1)--(g2)--(f2);
			\draw[doublearrow] (g2)-- node {\midarrowdiag} (g3);
			\draw[<->] (0,0.25) -- (0,-0.25);
		\end{tikzpicture}}  \\ \hline
\rowcolor{green!10}		2  & $D_{n+1}$& $B_n$ & $\mathbb{C}^{2n+1}$ & $\mathbb{Z}_2 \cdot \mathcal{O}_{[3,1^{2n-2}]}^{B_n}$ & $2n-1$ & 	\vspace{0.06cm}\scalebox{.88}{\begin{tikzpicture}
			\node (g1) at (-1,.5) [flavor, label=left:{$1$}]{}; 
			\node (g2) at (-1,-.5) [gauge, label=left:{$1$}]{}; 
			\node (g3) at (0,0) [gauge, label=below:{\symmlabel{2}}]{}; 
			\node (g4) at (1,0) {$\cdots$}; 
			\node (g5) at (2,0) [gauge, label=below:{\symmlabel{2}}]{}; 
			\node (g6) at (3,.5) [gauge, label=right:{$1$}]{}; 
			\node (g7) at (3,-.5) [gauge, label=right:{$1$}]{}; 
			\draw (g2)--(g3)--(g4)--(g5)--(g6); 
			\draw (g1)--(g3); 
			\draw (g5)--(g7); 
		\end{tikzpicture}} & \vspace{0.08cm}\scalebox{.88}{\begin{tikzpicture}
			\node (g1) at (-1,.5) [flavor, label=left:{$1$}]{}; 
			\node (g2) at (-1,-.5) [gauge, label=left:{$1$}]{}; 
			\node (g3) at (0,0) [gauge, label=below:{\symmlabel{2}}]{}; 
			\node (g4) at (1,0) {$\cdots$}; 
			\node (g5) at (2,0) [gauge, label=below:{\symmlabel{2}}]{}; 
			\node (g6) at (3,0) [gauge, label=below:{\symmlabel{[1]\wr S_2}}]{}; 
			\draw (g2)--(g3)--(g4)--(g5)--(g6); 
			\draw (g1)--(g3); 
		\end{tikzpicture}} \\ \hline
\rowcolor{green!10}			3 & $A_{2n-1}$ & $C_n$  & $(\Lambda^2 \mathbb{C}^{2n})/\mathbb{C}$ & $\mathbb{Z}_2 \cdot \mathcal{O}_{[2^2,1^{2n-4}]}^{C_n}$ & $2n-1$ & \vspace{0.16cm}\scalebox{.8}{\begin{tikzpicture}
			\node (g1) at (-1,0) [gauge, label=below:{$1$}]{}; 
			\node (g2) at (0,.5) [gauge, label=above:{$1$}]{}; 
			\node (g3) at (1,.5) {$\cdots$}; 
			\node (g4) at (2,.5) [gauge, label=above:{$1$}]{}; 
			\node (g5) at (3,.5) [flavor, label=above:{$1$}]{}; 
			\node (g12) at (0,-.5) [gauge, label=below:{$1$}]{}; 
			\node (g13) at (1,-.5) {$\cdots$}; 
			\node (g14) at (2,-.5) [gauge, label=below:{$1$}]{}; 
			\node (g15) at (3,-.5) [flavor, label=below:{$1$}]{}; 
			\draw (g1)--(g2)--(g3)--(g4)--(g5); 
			\draw (g1)--(g12)--(g13)--(g14)--(g15); 
		\end{tikzpicture}} & \vspace{0.26cm}\scalebox{.8}{\begin{tikzpicture}
			\node (g1) at (0,0) [gauge, label=below:{\symmlabel{1}}]{}; 
			\node (g2) at (1,0) [gauge, label=below:{\symmlabel{[1}}]{}; 
			\node (g3) at (2,0) {$\cdots$}; 
			\node (g4) at (3,0) [gauge, label=below:{\symmlabel{1}}]{}; 
			\node (g5) at (4,0) [flavor, label=below:{\symmlabel{\quad 1]\wr S_2}}]{}; 
			\draw (g1)--(g2)--(g3)--(g4)--(g5); 
		\end{tikzpicture}} \\ \hline
\rowcolor{green!10}			4& $E_6$ & $F_4$  & $\mathbb{C}^{26}$ & $\mathbb{Z}_2 \cdot\mathcal{O}_{[3,2^8,1^7]}^{F_4}$ & 11 & \vspace{0.12cm}\scalebox{.8}{\begin{tikzpicture}
			\node (g1) at (0,0) [flavor, label=below:{$1$}]{}; 
			\node (g2) at (1,0) [gauge, label=below:{$2$}]{}; 
			\node (g3) at (2,0) [gauge, label=below:{$3$}]{}; 
			\node (g4) at (3,.5) [gauge, label=below:{$2$}]{}; 
			\node (g5) at (3,-.5) [gauge, label=below:{$2$}]{}; 
			\node (g6) at (4,.5) [gauge, label=below:{$1$}]{}; 
			\node (g7) at (4,-.5) [gauge, label=below:{$1$}]{}; 
			\draw (g1)--(g2)--(g3)--(g4)--(g6);
			\draw (g3)--(g5)--(g7);
		\end{tikzpicture}} &  \vspace{0.4cm}\scalebox{.8}{\begin{tikzpicture}
			\node (g1) at (0,0) [flavor, label=below:{\symmlabel{1}}]{}; 
			\node (g2) at (1,0) [gauge, label=below:{\symmlabel{2}}]{}; 
			\node (g3) at (2,0) [gauge, label=below:{\symmlabel{3}}]{}; 
			\node (g5) at (3,0) [gauge, label=below:{\symmlabel{[2}}]{}; 
			\node (g7) at (4,0) [gauge, label=below:{\symmlabel{\quad 1] \wr S_2}}]{}; 
			\draw (g1)--(g2)--(g3)--(g5)--(g7);
		\end{tikzpicture}}  \\ \hline
\rowcolor{yellow!10}		5 & $G_2$  & $A_2$& $\mathbb{C}^{3} \oplus \Lambda^2 \mathbb{C}^{3}$ & $\mathbb{Z}_3 \cdot\mathcal{O}_{[3]}^{A_2}$ & 3&  \vspace{0.26cm}\scalebox{.8}{\begin{tikzpicture}
			\node (g1) at (-1,0) [flavor, label=below:{$1$}]{}; 
			\node (g3) at (0,0) [gauge, label=below:{$2$}]{}; 
			\node (g4) at (1,0) [gauge, label=below:{$1$}]{}; 
			\draw (g1)--(g3)--(g4);
			\draw[doublearrow] (g3)-- node {\midarrow} (g4); 
			\draw[thirdline] (g3)--(g4); 
		\end{tikzpicture}} &  \vspace{0.26cm}\scalebox{.8}{\begin{tikzpicture}
			\node (g1) at (-1,0) [gauge, label=below:{$1$}]{}; 
			\node (g3) at (0,0) [gauge, label=below:{$2$}]{}; 
			\node (g4) at (1,0) [flavor, label=below:{$3$}]{}; 
			\draw (g1)--(g3)--(g4);
		\end{tikzpicture}}  \\ \hline
\rowcolor{yellow!10}		6 & $B_{n}$ & $D_n$ & $\mathbb{C}^{2n}$ & $\mathbb{Z}_2 \cdot \mathcal{O}_{[3,1^{2n-3}]}^{D_n}$ & $2n-2$ & \vspace{0.06cm}\scalebox{.8}{\begin{tikzpicture}
			\node (g1) at (0,-.5) [gauge, label=left:{$1$}]{}; 
			\node (g2) at (1,0) [gauge, label=below:{$2$}]{}; 
			\node (g22) at (0,.5) [flavor, label=left:{$1$}]{}; 
			\node (g3) at (2,0) [gauge, label=below:{$2$}]{}; 
			\node (g4) at (3,0) {$\cdots$}; 
			\node (g5) at (4,0) [gauge, label=below:{$2$}]{}; 
			\node (g6) at (5,0) [gauge, label=below:{$1$}]{}; 
			\draw (g1)--(g2)--(g3)--(g4)--(g5);
			\draw (g22)--(g2);
			\draw[doublearrow] (g5)-- node {\midarrow} (g6); 
		\end{tikzpicture}} &  \vspace{0.06cm}\scalebox{.8}{\begin{tikzpicture}
			\node (g1) at (0,-.5) [gauge, label=left:{$1$}]{}; 
			\node (g2) at (1,0) [gauge, label=below:{$2$}]{}; 
			\node (g22) at (0,.5) [gauge, label=left:{$1$}]{}; 
			\node (g3) at (2,0) [gauge, label=below:{$2$}]{}; 
			\node (g4) at (3,0) {$\cdots$}; 
			\node (g5) at (4,0) [gauge, label=below:{$2$}]{}; 
			\node (g6) at (5,0) [flavor, label=below:{$2$}]{}; 
			\draw (g1)--(g2)--(g3)--(g4)--(g5)--(g6);
			\draw (g22)--(g2);
		\end{tikzpicture}}  \\ \hline
\rowcolor{yellow!10}			7 & $F_4$  & $B_4$& $\mathbb{C}^{16}$ & $\mathbb{Z}_2 \cdot \mathcal{O}_{[2^4,1]}^{B_4}$ & 8 & \vspace{0.26cm}\scalebox{.8}{\begin{tikzpicture}
			\node (g1) at (0,0) [flavor, label=below:{$1$}]{}; 
			\node (g2) at (1,0) [gauge, label=below:{$2$}]{}; 
			\node (g3) at (2,0) [gauge, label=below:{$3$}]{}; 
			\node (g4) at (3,0) [gauge, label=below:{$2$}]{}; 
			\node (g5) at (4,0) [gauge, label=below:{$1$}]{}; 
			\draw (g1)--(g2)--(g3);
			\draw (g4)--(g5);
			\draw[doublearrow] (g3)-- node {\midarrow} (g4); 
		\end{tikzpicture}} & \vspace{0.26cm}\scalebox{.8}{\begin{tikzpicture}
			\node (g1) at (0,0) [gauge, label=below:{$1$}]{}; 
			\node (g2) at (1,0) [gauge, label=below:{$2$}]{}; 
			\node (g3) at (2,0) [gauge, label=below:{$3$}]{}; 
			\node (g4) at (3,0) [gauge, label=below:{$2$}]{}; 
			\node (g5) at (4,0) [flavor, label=below:{$1$}]{}; 
			\draw (g1)--(g2)--(g3);
			\draw (g4)--(g5);
			\draw[doublearrow] (g3)-- node {\midarrow} (g4); 
		\end{tikzpicture}}  \\ \hline
\rowcolor{black!10}		8& $F_4$  & $D_4$ & $\mathbb{C}^{8} \oplus \mathbb{C}^{8} \oplus \mathbb{C}^{8}$ & $\mathbb{Z}_2^2 \cdot \mathcal{O}_{[3,2^2,1]}^{D_4}$ & 8  & \vspace{0.26cm}\scalebox{.8}{\begin{tikzpicture}
			\node (g1) at (0,0) [flavor, label=below:{$1$}]{}; 
			\node (g2) at (1,0) [gauge, label=below:{$2$}]{}; 
			\node (g3) at (2,0) [gauge, label=below:{$3$}]{}; 
			\node (g4) at (3,0) [gauge, label=below:{$2$}]{}; 
			\node (g5) at (4,0) [gauge, label=below:{$1$}]{}; 
			\draw (g1)--(g2)--(g3);
			\draw (g4)--(g5);
			\draw[doublearrow] (g3)-- node {\midarrow} (g4); 
		\end{tikzpicture}} &   \\ \hline
\rowcolor{green!10}		9& $D_4$  & $G_2$ & $\mathbb{C}^{7} \oplus \mathbb{C}^{7}$ & $S_3 \cdot \mathcal{O}_{[3^2,1]}^{G_2}$ & 5 & \vspace{0.06cm}\scalebox{.8}{\begin{tikzpicture}
			\node (g1) at (-1,0) [flavor, label=below:{$1$}]{}; 
			\node (g2) at (1,0) [gauge, label=right:{$1$}]{}; 
			\node (g3) at (0,0) [gauge, label=below:{$2$}]{}; 
			\node (g4) at (1,.5) [gauge, label=right:{$1$}]{}; 
			\node (g5) at (1,-.5) [gauge, label=right:{$1$}]{}; 
			\draw (g1)--(g3)--(g4);
			\draw (g2)--(g3)--(g5);
		\end{tikzpicture}} & \vspace{0.4cm}\scalebox{.8}{\begin{tikzpicture}
			\node (g1) at (-1,0) [flavor, label=below:{\symmlabel{1}}]{}; 
			\node (g3) at (0,0) [gauge, label=below:{\symmlabel{2}}]{}; 
			\node (g4) at (1,0) [gauge, label=below:{\symmlabel{[1]\wr S_3}}]{}; 
			\draw (g1)--(g3)--(g4);
		\end{tikzpicture}} \\  
		\hline 
	\end{tabular}
	\caption[Augmented Kostant-Brylinski table]{Augmented Table 1 of \cite{brylinski_nilpotent_1994}. $\mathfrak{g}$ ($\Tilde{\mathfrak{g}}$) is the symmetry algebra of the original (reduced) space. $V\simeq\mathfrak{g}/\Tilde{\mathfrak{g}}$. $M$ is the reduced space with discrete group prefactor $\Gamma$ if the original is its $\Gamma$-cover. The initial (reduced) space is the Coulomb branch of the first (second) quiver, respectively. See Section \ref{sec:g2-4dim} for more details on the first row and an explanation of the arrow connecting two $U(1)$ nodes.} 
\label{tabKostantBrylinski} \end{figure}
\end{landscape}

\section{Preliminaries}

\subsection{Folding of Dynkin diagrams} \label{sec:algebra}

\begin{figure}
	[p] 
	\centering
	\begin{tabular}
	    {| >{\centering\arraybackslash} m{0.8cm} | >{\centering\arraybackslash} m{0.8cm} |  >{\centering\arraybackslash} m{3.7cm} | >{\centering\arraybackslash} m{3.5cm} |  >{\centering\arraybackslash} m{4cm} |} \hline 
	    $\mathfrak{g}$ & $\Tilde{\mathfrak{g}}$ & Dynkin diagram of $\mathfrak{g}$ & Dynkin diagram of $\Tilde{\mathfrak{g}}$ & projection \\ \hline 
		\hline $A_{2n-1}$ & $C_n$ &
		\begin{tikzpicture}
			\node (gn) at (0,0) [gauge, label=right:{$n$}]{}; 
			\node (gn-1) at (-.5,1) [gauge, label=left:{$n-1$}]{}; 
			\node (gn+1) at (.5,1) [gauge, label=right:{$n+1$}]{}; 
			\node (g1) at (-.5,3) [gauge, label=left:{$1$}]{}; 
			\node (g2) at (-.5,2) [gauge, label=left:{$2$}]{}; 
			
			\node (g2n-2) at (.5,2) [gauge, label=right:{$2n-2$}]{}; 
			\node (g2n-1) at (.5,3) [gauge, label=right:{$2n-1$}]{}; 
			\draw (g1)--(g2); 
			\draw[dashed] (g2)--(gn-1);
			\draw (g1)--(g2);
			\draw[dashed] (gn+1)--(g2n-2); 
			\draw (g2n-2)--(g2n-1); 
			\draw (gn-1)--(gn)--(gn+1);
		\end{tikzpicture}\hspace{0.2cm} &
		\begin{tikzpicture}
			\node (g1) [gauge, label=right:{$1$}]{}; 
			\node (g2) [gauge, below of=g1, label=right:{$2$}]{}; 
			\node (gn-1) [gauge, below of=g2, label=right:{$n-1$}]{}; 
			\node (gn) [gauge, below of=gn-1, label=right:{$n$}]{}; 
			\draw (g1)--(g2); 
			\draw[dashed] (g2)--(gn-1); 
			\draw[doublearrow] (gn-1)-- node {\midarrowup} (gn); 
		\end{tikzpicture} &
        $S_2: \begin{cases} 1,2n-1\mapsto 1 \\ \vdots \\ n\pm 1\mapsto n-1 \\ n\mapsto n \end{cases}$\\ \hline
        $D_{n+1}$ & $B_n$ & 
        \begin{tikzpicture}
			\node (g1) [gauge, label=right:{1}]{}; 
			\node (g2) [gauge, below of=g1, label=right:{2}]{}; 
			\node (gn-1) [gauge, below of=g2, label=right:{$n-1$}]{}; 
			\node (gn) [gauge, below left of=gn-1, label=below:{$n$}]{}; 
			\node (gn+1) [gauge, below right of=gn-1, label=below:{$n+1$}]{}; 
			\draw (g1)--(g2); 
			\draw[dashed] (g2)--(gn-1); 
			\draw (gn-1)--(gn); 
			\draw (gn-1)--(gn+1); 
		\end{tikzpicture} &
        \begin{tikzpicture}
			\node (g1) [gauge, label=right:{$1$}]{}; 
			\node (g2) [gauge, below of=g1, label=right:{$2$}]{}; 
			\node (gn-1) [gauge, below of=g2, label=right:{$n-1$}]{}; 
			\node (gn) [gauge, below of=gn-1, label=right:{$n$}]{}; 
			\draw (g1)--(g2); 
			\draw[dashed] (g2)--(gn-1); 
			\draw[doublearrow] (gn-1)-- node {\midarrowdown} (gn); 
		\end{tikzpicture} &
        $S_2: \begin{cases} 1 \mapsto 1 \\ \vdots \\ n-1 \mapsto n-1 \\ n,n+1 \mapsto n \end{cases}$ \\ \hline
		$E_6$ & $F_4$ & 
		\begin{tikzpicture}
			\node (g1) at (-.5,2) [gauge, label=left:{1}]{}; 
			\node (g2) at (-.5,1) [gauge, label=left:{2}]{}; 
			\node (g3) at (0,0) [gauge, label=right:{3}]{}; 
			\node (g4) at (.5,1) [gauge, label=right:{4}]{}; 
			\node (g5) at (.5,2) [gauge, label=right:{5}]{}; 
			\node (g6) at (0,-1) [gauge, label=right:{6}]{}; 
			\draw (g1)--(g2)--(g3)--(g4)--(g5); 
			\draw (g3)--(g6); 
		\end{tikzpicture} &
		\begin{tikzpicture}
			\node (g1) [gauge, label=right:{1}]{}; 
			\node (g2) [gauge, below of=g1, label=right:{2}]{}; 
			\node (g3) [gauge, below of=g2, label=right:{3}]{}; 
			\node (g4) [gauge, below of=g3, label=right:{4}]{}; 
			\draw (g1)--(g2); 
			\draw[doublearrow] (g2)-- node {\midarrowup} (g3); 
			\draw (g3)--(g4);
		\end{tikzpicture} & 
		$S_2: \begin{cases} 1,5 \mapsto 1 \\ 2,4 \mapsto 2 \\ 3 \mapsto 3 \\ 6 \mapsto 4 \end{cases}$ \\ \hline
		$D_4$ & $G_2$ &
		\begin{tikzpicture}
			\node (g1) [gauge, label=right:{1}]{}; 
			\node (g2) [gauge, below of=g1, label=right:{2}]{}; 
			\node (g3) [gauge, below left of=g2, label=below:{$3$}]{}; 
			\node (g4) [gauge, below right of=g2, label=below:{$4$}]{}; 
			\draw (g1)--(g2);  
			\draw (g2)--(g3); 
			\draw (g2)--(g4); 
		\end{tikzpicture} &
        \begin{tikzpicture}
			\node (g1) [gauge, label=right:{$1$}]{}; 
			\node (g2) [gauge, below of=g1, label=right:{$2$}]{};
			\draw (g1)--(g2);
			\draw[doublearrow] (g1)-- node {\midarrowup} (g2); 
			\draw[thirdline] (g1)--(g2); 
		\end{tikzpicture}
		& $S_3: \begin{cases} 1,3,4 \mapsto 1 \\ 2 \mapsto 2 \end{cases}$ \\ \hline
		$B_3$ & $G_2$ &
		\begin{tikzpicture}
			\node (g1) at (-.5,1) [gauge, label=right:{$1$}]{}; 
			\node (g2) at (0,0) [gauge, label=right:{$2$}]{};
			\node (g3) at (.5,1) [gauge, label=right:{$3$}]{}; 
			\draw (g1)--(g2); 
			\draw[doublearrow] (g2)-- node {\midarrowdiagtwo} (g3); 
		\end{tikzpicture} &
		\begin{tikzpicture}
			\node (g1) [gauge, label=right:{$1$}]{}; 
			\node (g2) [gauge, below of=g1, label=right:{$2$}]{};
			\draw (g1)--(g2);
			\draw[doublearrow] (g1)-- node {\midarrowup} (g2); 
			\draw[thirdline] (g1)--(g2); 
		\end{tikzpicture} & $S_2: \begin{cases} 1,3 \mapsto 1 \\ 2 \mapsto 2 \end{cases}$ \\
		\hline 
	\end{tabular}
	\caption[Foldable simple Lie algebras]{Foldable simple Lie algebras. Note that numbers label nodes and do not indicate gauge groups as these are not quiver theories. The $S_2$ last row is a special case treated in several places in the main text.} 
\label{table:foldable_algebras} \end{figure}

Some pairs of simple Lie algebras can be related by an operation called \textit{folding} \cite{bump_lie_2013}, which acts on an algebra's Dynkin diagram and its internal structure. In a prototypical example, the $D_4$ algebra folds into $B_3$; in other words, rotations in eight dimensions are restricted to seven. Moreover, we show the $D_4$ Chevalley-Serre basis folds to its $B_3$ counterpart.

For simplicity\footnote{Results for semi-simple Lie algebras follow the same pattern.}, let $\mathfrak{g}$ be a complex simple Lie algebra associated to a Dynkin diagram of type $A_{2n-1}$, $D_{n+1}$ or $E_6$. $\mathfrak{g}$ has a (up to some choices of sign) canonical Chevalley-Serre basis obeying
\begin{align}\label{eq:CW1}
	\lbrack H_a,H_b\rbrack &= 0\\
	\lbrack H_a, E_{\pm i}\rbrack &= \pm\kappa_{ia}E_{\pm i} \label{eq:cartansimple}\\
	\label{eq:CW3}\lbrack E_{i},E_{-i}\rbrack &= H_i\\
	\label{eq:CW4}\lbrack E_{\pm i},\cdot\rbrack^{1-\kappa_{ji}}E_{\pm j} &= 0. 
\end{align}
where $H_a$ span the Cartan subalgebra, $E_{\pm i}$ are step operators and the indices $1 \leq i, a \leq \rank{(\mathfrak{g})}$ range over nodes of the diagram as in Figure \ref{table:foldable_algebras}. $\kappa$ is the Cartan matrix. Throughout this work we will often use a matrix realisation of the Chevalley-Serre basis, in which case we follow the construction of \cite{hanany_nilpotent_2019}.

Dynkin diagrams can be folded if there is a graph automorphism such that no node is linked to its own image under the automorphism. In particular, the diagrams for $A_{2n-1}$, $D_n$ or $E_6$ satisfy this constraint as they possess $S_2$ graph automorphisms, while the special case $D_4$ is invariant under $S_3$. In a unique case, $B_3$ folds to $G_2$ despite lacking an obvious graph automorphism (see Figure 30.14 in \cite{bump_lie_2013}). The associated algebra $\mathfrak{g}$ is then folded to $\Tilde{\mathfrak{g}}$ by the following recipe.

First let us denote the set of automorphisms by $\Gamma$, which is in practice either $S_2$ or $S_3$, and its elements by $\pi \in \Gamma$. We write 
\be \pi(i)=j \ee
to express that under the automorphism $\pi$ the $i$-th node is mapped to the $j$-th node. The fact that $\pi$ is a Dynkin diagram automorphism translates into the following invariance of the Cartan matrix under the action of $\pi$: $\kappa_{\pi(i) \pi(a)} = \kappa_{ia}$.  

We define the folding function $f$ taking as input nodes of the unfolded Dynkin diagram and mapping them to appropriate nodes in the folded diagram. Consequently, $f\circ \pi = f$. As an example, take $A_{2n-1}$ which folds to $C_n$ and think of $f$ as acting on indices $i$ of the original linear diagram. $f$ acts as $f(1) = f(2n-1) = 1$, $f(2) = f(2n-2)=2$ and so on, but $f(n)=n$. 

The folding procedure is now easily stated:
\begin{align}
	\Tilde{H}_a &= \sum_{b:f(b)=a} H_b\\
	\Tilde{E}_{\pm i} &= \sum_{j:f(j)=i} E_{\pm j} 
\end{align}
This defines the Chevalley-Serre basis for the folded algebra $\Tilde{\mathfrak{g}}$. In the case of $A_{2n-1}$, the folded algebra is indeed $C_n$. 

Special care must be taken when folding non-simple roots. Sometimes a sign change is required to preserve the algebra homomorphism $\mathfrak{g}\rightarrow f(\mathfrak{g})$. Consider the case of $A_3 \rightarrow C_2$. $A_3$ includes two elements $E_{12} = - \lbrack E_1, E_2 \rbrack$ and $E_{23} = - \lbrack E_2, E_3 \rbrack$. According to the definition of folding just given, $\Tilde{E}_1 = E_1+E_3$ and $\Tilde{E}_2 = E_2$. Then it follows that 
$$\Tilde{E}_{12} = -\lbrack \Tilde{E}_1, \Tilde{E}_2 \rbrack
= - \lbrack E_1 + E_3, E_2 \rbrack = -\left( \lbrack E_1,E_2 \rbrack + \lbrack E_3, E_2 \rbrack \right) = E_{12} - E_{23}.$$
In this specific case it is clear that the sign flips because the third node, which comes after the second, is mapped to the first, which comes before the second. Likewise it is clear that such a scenario will never occur in the case $D_{n+1}\rightarrow B_n$ and only comes into play for $A_{2n-1}\rightarrow C_n$, $B_3\rightarrow G_2$, $D_4\rightarrow G_2$ and $E_6\rightarrow F_4$. 

The interested reader can easily check (\ref{eq:CW1})-(\ref{eq:CW3}) and (\ref{eq:CW4}) with a bit more effort. To illustrate the typical calculation, we will confirm (\ref{eq:cartansimple}) for $A_5$ folding to $C_3$. The Cartan matrix of $C_3$ is 
\begin{equation}
	\kappa = \left(
	\begin{matrix}
		2 &-1 &0 \\
		-1 & 2 & -1 \\
		0 & -2 & 2 
	\end{matrix}
	\right) 
\end{equation}
and 
\begin{align}
	\lbrack \Tilde{H}_2, \Tilde{E}_3 \rbrack &= \lbrack H_2 + H_4, E_3 \rbrack = -E_3 - E_3 = -2\Tilde{E}_3 = \kappa_{32} \Tilde{E}_3\\
	\lbrack \Tilde{H}_3, \Tilde{E}_2 \rbrack &= \lbrack H_3, E_2 + E_4 \rbrack = -E_2 - E_4 = -\Tilde{E}_2 = \kappa_{23} \Tilde{E}_2 
\end{align}

Folded Lie algebras sometimes preserve additional tensors. In the case of $C_n$ there exists a tensor $J$ such that for every $X$ in $C_n$
\be
X^T J + J X = 0. \label{eq:cn-condition}
\ee
We can also reverse this statement: every $X$ in $A_{2n-1}$ which satisfies (\ref{eq:cn-condition}) is in $C_n$. 

In our convention $J$ assumes the following form:
\be
J = 
\left(\begin{matrix} 
0 & 0 & 0 & \dots & 0 & 0 & 1 \\
0 & 0 & 0 & \dots & 0 & -1 & 0 \\
0 & 0 & 0 & \dots & 1 & 0 & 0 \\
\vdots & \vdots & \vdots & \vdots & \vdots & \vdots & \vdots \\
0 & 0 & -1 & \dots & 0 & 0 & 0\\
0 & 1 & 0 & \dots & 0 & 0 & 0\\
-1 & 0 & 0 & \dots & 0 & 0 & 0
\end{matrix}\right)
\ee

The other case of this type is $G_2$, which is the subalgebra of $SO(7)$ preserving the following rank 3 antisymmetric tensor $\phi$:
$$
\sum_{a'} \phi_{a'bc} X_{a'a} + \sum_{b'} \phi_{ab'c} X_{b'b} + \sum_{c'} \phi_{abc'} X_{c'c} = 0
$$
for all $X\in G_2$. Given our choice of Chevalley-Serre basis the tensor is defined as
$$
\phi_{127}=-\phi_{136}=-\phi_{145}=\phi_{235}=-\phi_{246}=-\phi_{347}=-\phi_{567}=1
$$
with the remaining values either fixed by antisymmetry or equal to 0.

The dual Chevalley-Serre basis of linear forms $\lbrace X^*_i \rbrace$ is defined to obey $X^*_i (X_j) = \delta_{ij}$ for all $X_i$ in the Chevalley-Serre basis. In practice we realise $X^*_i$ as square matrices of the same dimension as $X_i$ and represent the evaluation as the linear extension of
\be
X^*_i (X_j) = \langle X_i^*, X_j\rangle = \langle X_j, X^*_i\rangle = \tr\left( X^*_i X_j \right).
\ee

The dual Chevalley-Serre bases of ``parent" and folded algebras are related: 
\begin{align}
	\Tilde{H}^*_a &= \frac{1}{\#_a} \sum_{b:f(b)=a} H^*_b\\
	\Tilde{E}^*_{\pm i} &= \frac{1}{\#_i} \sum_{j:f(j)=i} E^*_{\pm j} 
\end{align}
where $\#_i$ denotes the \textit{multiplicity} of node $i$ defined as \be \#_i = \left|\lbrace j:f(j)=i \rbrace\right| \ee

For example: 
\begin{equation}
	\langle \Tilde{H}_a, \Tilde{H}^*_b \rangle = \frac{1}{\#_b} \sum_{\substack{c:f(c)=a \\
	d:f(d)=b}} \langle H_c, H^*_d \rangle = \frac{1}{\#_b} \sum_{\substack{c:f(c)=a \\
	d:f(d)=b}} \delta_{cd} = \frac{1}{\#_b} \#_b \delta_{ab} = \delta_{ab} 
\end{equation}
where the second-to-last equality follows from the fact that $c=d$ can only occur if both fold to the same node, ie. $a=b$, and that this can happen for $\#_b$ joint choices of (identical) $c$ and $d$.

We close this section with a brief discussion of the aforementioned case of $B_3$ folding to $G_2$ despite a lack of graph automorphisms. This is easily elucidated with a quick detour through $D_4$: 
\begin{align}
	\Tilde{H}^B_3 &= H^D_3 + H^D_4 \\
	\Tilde{H}^G_2 &= H^D_1 + H^D_3 + H^D_4 = \Tilde{H}^B_1 + \Tilde{H}^B_3 
\end{align}
where we decorate each Cartan generator with a subscript denoting its algebra. As illustrated - and the pattern holds up for remaining $G_2$ basis elements - $G_2$ can be expressed as a folding of $B_3$ in the same way that $B_3$ is a folding of $D_4$.

\subsection{The monopole formula} \label{sec:monopole-formula}

The monopole formula is an important computational tool in the study of Coulomb branch chiral rings which was introduced and subsequently generalised in \cite{cremonesi_monopole_2014,cremonesi_coulomb_2014}. The formula calculates the chiral ring Hilbert series by counting monopole operators labelled by their conformal dimension and, optionally, action under $U(1)$ global symmetries arising from unitary gauge nodes.

Consider a simply laced quiver. The underlying graph is made of a set of vertices $V$ and a set of (unoriented) edges $E \subset S^2(V)$. To each vertex $v \in V$ is associated a gauge group $U(n_v)$, and to each edge $e \in E$ is associated a hypermultiplet in the bifundamental of $U(n_{v}) \times U(n_{v'})$ where $e=(v,v')$. Finally, we have a set of flavor vertices $F \neq \emptyset$ with global symmetries $SU(n_f)$ for $f \in F$, and a set of edges $E' \subset V \times F$. An edge $e'=(v,f)$ stands for $n_f$ hypermultiplets in the fundamental representation of $U(n_v)$. The total gauge group is 
\begin{equation}
    G = \prod\limits_{v \in V} U(n_v) \, 
\end{equation}
and it has rank 
\begin{equation}
    r = \sum\limits_{v \in V} n_v \, . 
\end{equation}
The Weyl group is 
\begin{equation}
\label{Weylgroup}
    W = \prod\limits_{v \in V} S_{n_v} \, . 
\end{equation}

A magnetic charge is an element $m \in \mathbb{Z}^r$. For $H$ a subgroup of $S_r$ and $m$ a magnetic charge, we define the stabiliser  
\begin{equation}
    H (m) = \{ g \in H | g\cdot m=m\} \, . 
\end{equation}
The conformal dimension $\Delta (m)$ is defined by 
\begin{equation}
    2 \Delta (m) = \sum\limits_{(v,v') \in E} \sum\limits_{i=1}^{n_v} \sum\limits_{i'=1}^{n_{v'}} |m_{v,i} - m_{v',i'}|  + \sum\limits_{(v,f) \in E'} \sum\limits_{i=1}^{n_v} n_f |m_{v,i}| - \sum\limits_{v \in V} \sum\limits_{i=1}^{n_v} \sum\limits_{j=1}^{n_{v}} |m_{v,i} - m_{v,j}| \, . \label{eq:conf-dim-simply-laced}
\end{equation}
Then the (unrefined) Hilbert series for the Coulomb branch of the quiver is given by the monopole formula, which can be written as 
\begin{equation}
\label{monopole}
    \HS(t) = \HS_W(t) = \frac{1}{|W|} \sum\limits_{m \in \mathbb{Z}^r} \sum\limits_{\gamma \in W(m)} \frac{t^{2 \Delta (m)}}{\det \left( 1-t^2 \gamma\right)} \, ,  
\end{equation}
where $W$ is the Weyl group (\ref{Weylgroup}). The form of this formula slightly differs from the one presented in \cite{cremonesi_monopole_2014}; in particular note the absence of the Casimir symmetry factors and the summation on the whole magnetic lattice. One can show that despite these superficial differences the formulas are equivalent. The form (\ref{monopole}) is better suited for generalisation to wreathed quivers in Section \ref{sectionDG}. 

This formula can be further refined by labelling each monopole insertion with its charge under the topological symmetry $q_v$. We only need introduce $|V|$ extra fugacities $z_v$:
\be
\label{eq:monopole-formula-refined}
    \HS_\text{ref}(t, z_v) = \HS_{\text{ref},W}(t, z_v) = \frac{1}{|W|} \sum\limits_{m \in \mathbb{Z}^r} \sum\limits_{\gamma \in W(m)} \frac{\left(\prod_v z_v^{q_v(m)}\right) t^{2 \Delta (m)}}{\det \left( 1-t^2 \gamma\right)} \, .  
\ee

Reference \cite{cremonesi_coulomb_2014} further improved the formula with the addition of non-simply laced quivers to the world of quiver gauge theories. While they were not explicitly constructed (say, as Lagrangian theories), it was relatively straightforward to modify the monopole formula such that, when computed for non-simply laced quivers, the results made sense and followed the pattern of their simply-laced cousins. In particular, it is well known that balanced quivers' Coulomb branch symmetry enhances according to the Dynkin diagram which the quiver resembles. For example, balanced linear quivers exhibit $A_n$ symmetry. Non-simply laced balanced quivers were found to have $B_n$, $C_n$, $F_4$ or $G_2$ symmetry. 

The only difference introduced by non-simply laced quivers to the monopole formula is a modification of (\ref{eq:conf-dim-simply-laced}) to
\be
\begin{split}
2 \Delta (m) =& \sum\limits_{(v,v') \in E} \sum\limits_{i=1}^{n_v} \sum\limits_{i'=1}^{n_{v'}} |\kappa_{v,v'}m_{v,i} - \kappa_{v',v}m_{v',i'}|  + \sum\limits_{(v,f) \in E'} \sum\limits_{i=1}^{n_v} n_f |m_{v,i}| 
\\&- \sum\limits_{v \in V} \sum\limits_{i=1}^{n_v} \sum\limits_{j=1}^{n_{v}} |m_{v,i} - m_{v,j}|
\end{split}
\ee
where $\kappa$ is defined as follows:
\begin{itemize}
    \item $\kappa_{vv}=2$
    \item $\kappa_{vv'}=\kappa_{v'v}=-n$ if $v$ and $v'$ are connected by $n$ undirected edges
    \item $\kappa_{vv'}=-n,\kappa_{v'v}=-1$ if $v$ and $v'$ are connected by an $n$-valent directed edge from $v$ to $v'$
\end{itemize}
The similarity to Cartan matrices is, of course, not coincidental and reappears in the abelianised formalism.

\subsubsection*{Reading relations off the Hilbert series}

We briefly describe the method by which we extract chiral ring relations from the Coulomb branch Hilbert series. Assume that the Hilbert series is refined with fugacities $z_i$ counting charge under a Cartan subalgebra of the Coulomb branch symmetry algebra $\mathfrak{g}$. The Hilbert series expands as
\be
\HS(t,z_i) = \sum_{s\in\N} p_s(z_i) t^s
\ee
where $p_s(z_i)$ are characters of $\mathfrak{g}$. 

We first state the general strategy for a nilpotent orbit, whose coordinate ring is generated by a single (co)adjoint representation with $R$-symmetry spin 1. The quaternionic dimension of each Coulomb branch is easily calculated by summing up gauge ranks, which is unaffected by discrete gauging. Knowing the dimension and global symmetry, we can look up the space in \cite{hanany_quiver_2017}\footnote{This paper differs from the present paper in the simple root convention for $G_2$: for the present paper the (co)adjoint representation goes by $\lbrack 01 \rbrack$ whereas in \cite{hanany_quiver_2017} the two labels are swapped.}.

We could then expand the highest weight generating function, comparing (polynomial) coefficients of $t^{2n}$ to the character representation of the $n$-th symmetric product $\mathrm{Sym}^n \mathrm{adj}(\mathfrak{g})$ and find missing representations suggesting the existence of relations. Or we can perform the same computation in a more elegant fashion using the plethystic logarithm:
\be
\PL(\HS(t,z_i)) = \sum_{k=1}^\infty \frac{\mu(k)}{k} \log\left(\HS\left(t^k, z_i^k\right)\right)= \sum_{s=1}^\infty g_s(z_i) t^s - \sum_{s=1}^\infty r_s(z_i) t^s
\ee
where $\mu(k)$ is the M\"{o}bius function and $g_s(z_i)$ and $r_s(z_i)$ are characters of $\mathfrak{g}$. If the space is a complete intersection, the list of $g_s$ and $r_s$ is finite and they correspond respectively to the generators and relations of the Coulomb branch. The minimal set of relations is typically present in the first few orders of $t$. For example, the (closure of the) minimal nilpotent orbit of any simple algebra $\mathfrak{g}$ (whose coordinate ring is generated by one coadjoint generator \cite{namikawa_characterization_2018}) is described by a set of Joseph relations \cite{joseph_minimal_1976,gaiotto_argyres-seiberg_2010} of its coordinate ring. They are always necessarily quadratic in the coadjoint generator. In more general cases we go to slightly higher order, $t^6$ or $t^8$. Then, whenever feasible, we verify that that the full set of relations are identified: we calculate the Hilbert series of a ring defined by $\mathrm{dim}\ \mathfrak{g}$ generators subject to the relations in question and compare it to tabulated expressions.

This procedure is only slightly modified in the few isolated cases in this paper where the Coulomb branch is not a nilpotent orbit. The chiral ring is then generated by more generators, which are in these particular cases also coadjoint. Their contribution will be visible in the PL.

\subsection{Abelianisation}

Coulomb branch chiral rings of simply-laced unitary quiver theories can be explicitly constructed following the prescriptions of \cite{bullimore_coulomb_2017,dedushenko_coulomb_2018,dedushenko_coulomb_2019}. One first defines the abelianised ring, which is then reduced by the action of the gauge symmetry's Weyl group. 

Let $i$ index the vertices and hence gauge group factors of a quiver gauge theory. Each gauge node $G_i$ contributes several basic variables to the ring: $\up{i,a}$, $\um{i,a}$ and $\p{i,a}$, where $1\leq a \leq \mathrm{rank}\ G_i$. We will sometimes blur the distinction between the three types of variables by dropping all identifying information except for the node and gauge indices, leaving only $\x{i,a}$. The variables satisfy \textit{abelianised relations}
\be \up{i,a} \um{i,a} = -\frac{\prod_{w\in \mathcal{R}}\langle{w,\vec\varphi}\rangle^{\left| w_{i,a} \right|}}{\prod_{\alpha\in\Phi}\langle\alpha,\vec{\varphi}\rangle^{\left| \alpha_{i,a} \right|}}\label{eq:abelrel} \ee
where $\vec{\varphi} = \left( \p{1,1}, \dots, \p{n, \mathrm{rank}\ G_n, \m{1,1}, \dots \m{n,N_f^n}} \right) $, $N_f^i$ is the number of fundamental flavors on the $i$-th node, and both the roots $\alpha$ and weights $w$ are expressed as weights in the weight basis of the theory's gauge group $G=\prod_{i=1}^n G_i$. 

For example $U(2)$ with 4 fundamental flavors comes with the following matter and gauge representations:
\be
\mathcal{R} = \lbrace (1,0;-1,0,0,0), \dots (1, 0;0,0,0,-1), (0,1;-1,0,0,0), \dots (0,1;0,0,0,-1) \rbrace
\ee
with the first two charges belonging to $U(2)$ and the last four belonging to $SU(4)$ and
\be
\Phi = \lbrace (1, -1), (-1, 1) \rbrace
\ee
where the charges are associated to $U(2)$.

The Coulomb branch is a symplectic space and its chiral ring carries a Poisson bracket, which descends from a bracket defined on the abelianised ring:
\begin{align}
	\pois{\p{i,a},\upm{i,a}} &= \pm\upm{i,a} \label{eq:poiss1}\\
	\pois{\up{i,a},\um{i,a}} &= \frac{\partial}{\partial \p{i,a}} \frac{\prod_{w\in \mathcal{R}}\langle{w,\vec\varphi}\rangle^{\left| w_{i,a} \right|}}{\prod_{\alpha\in\Phi}\langle\alpha,\vec{\varphi}\rangle^{\left| \alpha_{i,a} \right|}}\\
    \pois{\upm{i,a},\upm{j,b}} &= \pm \kappa_{ij}\frac{\upm{i,a}\upm{j,b}}{\p{i,a}-\p{j,b}} \label{eq:poiss3}
\end{align}
where $\kappa_{ij}$ is defined as in Section \ref{sec:monopole-formula}.

Note that the last Poisson bracket generates new elements of the abelianised ring which cannot be expressed by adding and multiplying the basic variables. The abelianised ring is, at least in all the cases we study, freely generated as a Poisson algebra with the variables $\upm{i,a}$ and $\p{i,a}$.

To recover the physical chiral ring $\co[\Coul]$ we reduce the abelianised chiral ring $\co[\Coul_\mathrm{\abel}]$ to its Weyl-invariant subring $\co[\Coul_\mathrm{\abel}]^{\mathcal{W}_G} = \co[\Coul]$. The chiral ring is believed to be finitely generated in all cases, and is known to have this property in every case we consider. 

Finally, we point out that our notion of the abelianised ring is a departure from that of \cite{bullimore_coulomb_2017}, where it is introduced in (4.9) (with minor notational differences) as  
\begin{equation*}
\co[\mathcal{M}_C^{\abel}] = \left( \co[ \lbrace \upm{A} \rbrace, \lbrace \p{a} \rbrace, \lbrace (M_j^W)^{-1} \rbrace_{j \in \mathrm{roots}} ] / (\text{abelianised relations}) \right)^{\mathcal{W}_G}.
\end{equation*}
In essence, this ``abelianised" ring is the Weyl-invariant part of a ring generated by all abelianised monopoles, scalar operators and all inverse masses (ie. $1/(\p{a}-\p{b})$) modulo relations between abelianised monopoles. The authors follow with clarification that this is emphatically \emph{not} the Coulomb branch chiral ring, since a) the inverse masses are not defined everywhere on it (ie. when $\p{a}=\p{b}$), and even if one restricts to the ``discriminant locus" of the Coulomb branch (that is, the points where $\p{a}\neq\p{b}$), b) this ring contains \emph{extra} elements. The precise relation between $\co[\mathcal{M}_C^{\abel}]$ and the Coulomb branch chiral ring is, in general, unknown.

We focus on an altogether different type of ``abelianised" ring, which we denote $\co [\Coul_{\abel}]$. It is generated, as a Poisson algebra with the standard abelianised Poisson bracket, by all the abelianised monopole variables $\upm{v, i}$ and scalars $\p{v,i}$, modulo standard abelianised relations. We then posit that the true Coulomb branch chiral ring is the Weyl-invariant subring\footnote{\cite{bullimore_coulomb_2017} suggest in their Section 6.3 that the Coulomb branch chiral ring can be generated as a Poisson algebra by a certain small set of Weyl-invariant operators. Our two approaches are ultimately equivalent, but we impose Weyl-invariance as an extra condition on the resulting Poisson algebra, while they demand it of the generators themselves.} $\co [\Coul_{\abel}]^{\mathcal{W}_G}$.

To see that the two ``abelianised" rings are inequivalent, consider the theory $U(2)$ with 4 flavors. Then the element $1/(\p{1}-\p{2})^2$ is included in $\co[\mathcal{M}_C^{\abel}]$, but not in $\co [\Coul_{\abel}]^{\mathcal{W}_G}$. The inverse masses never appear in $\co [\Coul_{\abel}]$, and even the abelianised relation
\begin{equation}
    \up{1} \um{1} = - \frac{\prod_{1\leq i \leq 4} (\p{1}-\m{i})}{(\p{1}-\p{2})^2}
\end{equation}
can be interpreted without introducing inverse masses. In our view, this relation simply expresses the fact that 
\begin{equation}
    \up{1} \um{1} (\p{1}-\p{2})^2 = - \prod_{1\leq i \leq 4} (\p{1}-\m{i}),
\end{equation}
which is satisfied on every generic point of the Coulomb branch. The relation of course becomes problematic on points where $\p{1}=\p{2}$, which is a sign that we are dealing with the wrong variables and relations. The correct variables are components of two coadjoint tensors, which themselves satisfy several tensorial relations. Abelianised relations are crucial in deriving and testing the form of these coadjoint tensors, but ultimately serve as a ladder to be thrown away once the tensors are obtained.

Since $\co [\Coul_{\abel}]$ actually serves as an active computational precursor to the Coulomb branch chiral ring, we reserve it the name \emph{abelianised ring} despite any confusion it may cause when compared to a similarly denoted ring in \cite{bullimore_coulomb_2017}.

\subsection{Construction of Coulomb branch multiplets}

Abelianised variables $\p{i,a}$ have weight 2 under the $R$-symmetry\footnote{The $R$-symmetry is assumed to be the $SU(2)$ factor acting non-trivially on the Coulomb branch. An operator's weight is twice its conformal dimension.}, while the Poisson bracket scales with weight $-2$. Weights of $\upm{i,a}$ can be read off from (\ref{eq:abelrel}). The Coulomb branch chiral ring of any good or ugly theory is graded by $R$-symmetry weights as $\co [\Coul] = \sum_{i\in \N} \co [\Coul]_i$ where $\co [\Coul]_i$ is the vector space of all Coulomb branch chiral ring operators with $R$-symmetry weight $i$.

Any Coulomb branch operator $\mathcal{O}$ with well-defined $R$-symmetry weight $j$ defines a map $\pois{\mathcal{O},\cdot}:\co [\Coul]_i\rightarrow \co [\Coul]_{i+j-2}$ and therefore operators in $\co [\Coul]_2$ form a closed Poisson algebra. This algebra is precisely the symmetry algebra $\mathfrak{g}$ of the Coulomb branch and all operators in $\co [\Coul]_i$ necessarily assemble into tensors of the $\co [\Coul]_2$ algebra $\mathfrak{g}$. In this paper we focus almost exclusively on good (in fact, balanced) theories whose Coulomb branch chiral rings are generated by operators in $\co [\Coul]_2$ and whose symmetry algebra $\mathfrak{g}$ is simple. Consequently, $\co [\Coul]_2$ operators assemble into a single (coadjoint) representation of $\mathfrak{g}$ – the moment map of the symmetry – which has a matrix realisation for all cases in this article. We will also consider one case in which the ring is generated by $\co [\Coul]_2$ operators transforming in the coadjoint representation along with another set of $\co [\Coul]_4$ operators, also in the coadjoint representation.

In \cite{hanany_nilpotent_2019} two of this paper's authors developed a prescription for coadjoint chiral ring generators following \cite{bullimore_coulomb_2017}. Let $X_k \in \mathfrak{g}$ form a basis of $\mathfrak{g}$ satisfying $[ X_k, X_l ] =\sum_m c_{klm} X_m $. There is a basis of $\co [\Coul]_2$ formed by $\mathcal{O}_k $ such that $\pois{\mathcal{O}_k,\mathcal{O}_l}=\sum_m c_{klm} \mathcal{O}_m $. If $X^*_k$ are dual to $X_k$, ie. $\langle X^*_k, X_l \rangle = \delta_{kl}$, the moment map $N$\footnote{We reserve the usual symbol for moment maps, $\mu$, for highest weight fugacities.} is explicitly constructed as 

\be N = \sum_k \mathcal{O}_k X^*_k. \label{eq:moment_map} \ee

This definition guarantees that $\langle N, \cdot \rangle$ acts as a Lie algebra homomorphism: \be \pois{\langle N, X_k \rangle,\langle N, X_l \rangle} = \langle N, [X_k, X_l] \rangle \ee

The moment map $N$ satisfies certain matrix relations, which can be inferred from the Hilbert series of the Coulomb branch; this combined approach is called the \emph{synthetic method} in \cite{hanany_nilpotent_2019}. We construct moment maps for all examples in this paper and explicitly check the expected relations. By finding total agreement between predictions of Hilbert series and our explicit construction of folded quivers, we will argue for specific interpretations of the multiple link from \cite{cremonesi_coulomb_2014} and an operation called \emph{discrete gauging} (similar to but distinct from that studied in \cite{hanany_discrete_2018,hanany_discrete_2018-1}), all at the level of individual and explicitly constructed operators.

\section{Discrete gauging}
\label{sectionDG}

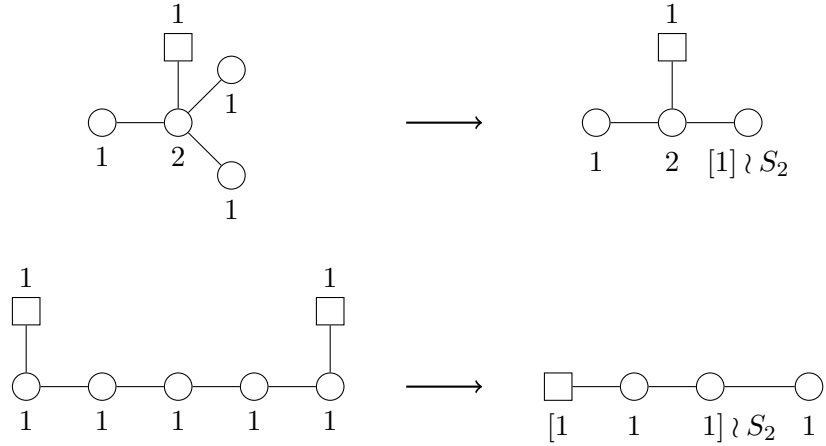
\begin{figure}
	[t] 
\centering 
	\begin{tikzpicture}
		\node (dg1) at (-1,0) [gauge, label=below:{1}]{}; 
		\node (dg2) at (0,0) [gauge, label=below:{2}]{}; 
		\node (dg3) at (.7,.7) [gauge, label=below:{1}]{};
		\node (dg4) at (.7,-.7) [gauge, label=below:{1}]{}; 
		\node (df2) at (0,1) [flavor, label=above:{1}]{}; 
		\draw (dg1)--(dg2); 
		\draw (dg2)--(dg3); 
		\draw (dg2)--(dg4); 
		\draw (dg2)--(df2); 
		
		\node (bg1) at (5.5,0) [gauge, label=below:{\symmlabel{1}}]{}; 
		\node (bg2) at (6.5,0) [gauge, label=below:{\symmlabel{2}}]{}; 
		\node (bg3) at (7.5,0) [gauge, label=below:{\symmlabel{[1]\wr S_2}}]{}; 
		\node (bf2) at (6.5,1) [flavor, label=above:{1}]{}; 
		\draw (bg1)--(bg2)--(bg3); 
		\draw (bg2)--(bf2); 
		
		\draw[->,thick] (3,0) -- (4,0);
		
		\node (ag1) at (-2,-3.5) [gauge, label=below:{1}]{}; 
		\node (ag2) at (-1,-3.5) [gauge, label=below:{1}]{}; 
		\node (ag3) at (0,-3.5) [gauge, label=below:{1}]{}; 
		\node (ag4) at (1,-3.5) [gauge, label=below:{1}]{}; 
		\node (ag5) at (2,-3.5)  [gauge, label=below:{1}]{}; 
		\node (af1) at (-2,-2.5) [flavor, label=above:{1}]{}; 
		\node (af5) at (2,-2.5) [flavor, label=above:{1}]{}; 
		\draw (ag1)--(ag2)--(ag3)--(ag4)--(ag5); 
		\draw (ag1)--(af1); 
		\draw (ag5)--(af5); 
		
		\node (cg1) at (6,-3.5) [gauge, label=below:{\symmlabel{1}}]{}; 
		\node (cg2) at (7,-3.5) [gauge, label=below:{\symmlabel{\qquad 1]\wr S_2}}]{}; 
		\node (cg3) at (8.3,-3.5) [gauge, label=below:{\symmlabel{1}}]{}; 
		\node (cf1) at (5,-3.5) [flavor, label=below:{\symmlabel{[1}}]{}; 
		\draw (cf1)--(cg1)--(cg2)--(cg3); 
		
		\draw[->, thick] (3,-3.5) -- (4,-3.5);
	\end{tikzpicture}
	\caption[Wreathing examples]{Quivers on the left wreathe into quivers on the right.} \label{fig:wreathing-examples}
\end{figure}

Our first example of a discrete quiver operation, discrete gauging, orbifolds the Coulomb branch by a subgroup of the quiver's automorphisms. Another operation, which also acts on the Coulomb branch as an orbifold, was previously studied in \cite{hanany_discrete_2018,hanany_discrete_2018-1}\footnote{In contrast to our treatment, these works did not claim to discretely gauge the theory, but restricted their claims only to effects on the Coulomb branch.}. Ours differs in several respects: it preserves the dimension of the Higgs branch as well as the Coulomb branch, allows for consistent and successive discrete gauging of nodes into ``larger" nodes and generalises beyond acting on a collection of $n$ $U(1)$ nodes (which form a $U(n)$ node with adjoint matter in \cite{hanany_discrete_2018,hanany_discrete_2018-1}) to acting on $n$ copies of arbitrary gauge groups or ``legs" of the quiver.

It is possible to discretely gauge any quiver of the type depicted in the top left corner of Fig. \ref{fig:discrete-operations}, ie. one with $m$ identical legs\footnote{A leg can have arbitrary shape and in particular need not be linear.} $Q_i$ (and potentially other legs $Q'_j$) connected to a single common node which we call the \emph{pivot}\footnote{It may be possible to discretely gauge quivers without a pivot node but we do not have a successful case to present.}. One discretely gauges the $m$ identical legs by extending the overall gauge group with the symmetric group $S_m$, or a subgroup thereof, which permutes the gauge factors associated with each leg. We say that we have \emph{gauged the quiver's automorphism}. For example, three legs composed solely of $U(1)$ nodes will arrange into a $U(1)\wr S_3$ node, while two legs with $U(2)\times U(1)$ gauge nodes will combine to give $\left(U(2)\times U(1)\right)\wr S_2$, with $S_2$ simultaneously exchanging $U(2)$ and $U(1)$ factors.

Our strategy in this section consists of the following steps. We first demonstrate the existence of a well-defined orbifolding operation on the Coulomb branch, giving results consistent with existing literature. Then we suggest that the operation acts on the quiver as a whole in a way that can be deduced from the Coulomb branch action and that the results of this operation should be viewed as quivers in their own right, even if they often cannot be written down using existing notation; we introduce the concept of \emph{wreathed quivers} to get over this difficulty; see Figure \ref{fig:wreathing-examples} for two examples. We support this claim by generalising the monopole formula to this family of quivers and computing an example, as well as calculating a few wreathed quiver Higgs branches. We also conjecture that a well known Higgs branch operation is the 3d mirror to this operation on the Coulomb branch.

\subsection{Wreath product}

We pause for a moment to introduce the notion of the \emph{wreath product} $G \wr \Gamma$ of a group $G$ by a permutation group $\Gamma \subseteq S_n$ (the integer $n$ is understood in the notation $G \wr \Gamma$, which we could denote $G \wr_n \Gamma$ if there is a risk of confusion) \cite{kerber_applied_2013}. As a set, we define 
\begin{equation}
\label{definitionWreath}
  G \wr_n \Gamma \equiv   G \wr \Gamma = \left( \prod\limits_{i=1}^n G_i \right) \times \Gamma \, , 
\end{equation}
where the $\times$ denotes the Cartesian product of sets, \emph{not} the direct product of groups. There are $n$ copies $G_1$, ..., $G_n$ of the group $G$. An element of $(g,\sigma) \in G \wr \Gamma$ is an ordered list of $n$ elements $g_i$ of $G$ together with a permutation $\sigma \in \Gamma$. The group multiplication  law is given, for $(g,\sigma) \in G \wr \Gamma$ and $(g',\sigma') \in G \wr \Gamma$, by 
\begin{equation}
    (g,\sigma) \cdot (g',\sigma') = (g \sigma (g') , \sigma \sigma ') \, , \textrm{  with  } (g\sigma (g'))_i = g_i g'_{\sigma^{-1} (i)} \, . 
\end{equation}
Intuitively, $G \wr \Gamma$ is the direct product of $n$ copies of $G$, which can in addition be permuted by $\Gamma$. 

In this paper we consider wreath products where $G$ is a unitary group $U(r)$, or more generally a direct product of finitely many unitary groups $U(r_1) \times \cdots \times U(r_k)$. In this case, in particular in the quivers, we extend the usual shorthand notation in which $U(r)$ is replaced by the rank $r$, and we write $r \wr \Gamma$ for $U(r) \wr \Gamma$, and more generally $[r_1 \quad \cdots \quad r_k] \wr \Gamma$ for $(U(r_1) \times \cdots \times U(r_k)) \wr \Gamma$.

\subsection{Action on the Coulomb branch}

We will first study this procedure through the lens of Coulomb branch abelianisation. The goal is to show that the Coulomb branch can be reduced to an orbifold by an automorphism of the quiver. 

Since each node contributes several variables to the abelianised chiral ring, there is an induced $S_m$ action permuting them. For any $\pi \in S_m$, we have

\begin{align}
	\pi(\x{i,a})&\defeq\x{\pi(i),a}
\end{align}
Action on more complicated (polynomial or rational) functions of these variables is defined by action on indices of the full expression. For example \be \pi(\up{i,a}\up{j,b}) = \up{\pi(i),a} \up{\pi(j), b}. \ee
Note that mass parameters should be treated as \emph{numbers} (parameters) rather than ring elements (VEVs); therefore $\pi$ does \emph{not} act on them, ie. \be \pi(\m{i,a})=\m{i,a}. \ee In fact, this constraint forces \be \pi(\m{i,a})=\m{i,a}=\m{\pi(i),a}. \ee To see this consider the $A_5$ theory which gauges to the bottom right quiver in Fig. \ref{fig:wreathing-examples}:
\begin{align}
    \pi(\up{1}\um{1}) &= \pi\left( - (\p{1}-\p{2})(\p{1}-\m{1}) \right) = -(\p{5}-\p{4})(\p{5}-\m{1})\\
    \up{5}\um{5} &= - (\p{5}-\p{4})(\p{5}-\m{5})
\end{align}
Since $\pi(\up{1}\um{1}) = \up{5}\um{5}$, the two mass parameters must be equal to preserve symmetry under $\pi$. This is a sensible constraint: if $\m{1}\neq \m{5}$ then the mass deformation  breaks the quiver's $S_2$ symmetry.

We should check that the form of the Poisson brackets (\ref{eq:poiss1})-(\ref{eq:poiss3}) is compatible with this action in the sense that $\pois{\pi(x), \pi(y)} = \pi(\pois{x,y})$.
\begin{align}
	\pois{\pi(\p{i,a}),\pi(\upm{i,a}}) &= \pois{\p{\pi(i),a},\upm{\pi(i),a}} =\pm\upm{\pi(i),a} = \pi\left(\pois{\p{i,a}, \upm{i,a}}\right) \\
	\pois{\pi(\up{i,a}),\pi(\um{i,a})} &= \pois{\up{\pi(i),a},\um{\pi(i),a}} = \frac{\partial}{\partial \p{\pi(i),a}} \frac{\prod_{w\in \mathcal{R}}\langle{w,\pi(\vec\varphi)}\rangle^{\left| w_{i,a} \right|}}{\prod_{\alpha\in\Phi}\langle\alpha,\pi(\vec{\varphi})\rangle^{2\left| \alpha_{i,a} \right|}} = \pi\left(\pois{\up{i,a},\um{i,a}}\right) \\
	\pois{\pi(\upm{i,a}),\pi(\upm{j,b})} &= \pois{\upm{\pi(i),a},\upm{\pi(j),b}} = \pm \kappa_{ij}\frac{\upm{\pi(i),a}\upm{\pi(j),b}}{\p{\pi(i),a}-\p{\pi(j),b}} = \pi\left(\pois{\upm{i,a},\upm{j,b}}\right) 
\end{align}

The first line is clearly compatible with the action. The second line also succeeds with a simple relabelling: $w_{\pi(i)} \leftrightarrow w_i$ and $\alpha_{\pi(i)}\leftrightarrow\alpha_i$. The third line is similarly preserved because $\kappa_{\pi(i)\pi(j)} = \kappa_{ij}$ is a consequence of the automorphism. In fact, it is noteworthy that the third line forces the action of $\pi$ to preserve connectedness while the second line enforces identical gauge and matter content on each leg $Q_i$.

To implement the quotient on the Coulomb branch chiral ring, it is enough to declare that only $S_m$–invariant operators are physical. This is easily done through the use of a projector: \be P(\cdot) = \frac{1}{m!} \sum_{\pi \in S_m} \pi(\cdot).\label{eq:projector} \ee Every operator of the form $P(\mathcal{O})$ is physical.

The effect on the Coulomb branch is then transparent. If $\Coul$ and $\tCoul$ are Coulomb branches of, respectively, the original quiver and discretely gauged quivers, the two spaces are related by
\be
\tCoul = \Coul / S_m
\ee
ie. the discretely gauged Coulomb branch is a $S_m$ orbifold of the original space. This construction leads to new Coulomb branches which were previously unknown, provided that they are orbifolds of known Coulomb branches.

Note that nothing prevents generalisation from $S_m$ to arbitrary subgroups $\Gamma$ of $S_m$, for instance the alternating group $A_m$ or cyclic group $\Z_m$. We investigate one such example in Section \ref{sec:g2-z3}.

The projector acts remarkably simply on moment maps of type $AD$ quivers as studied in \cite{hanany_nilpotent_2019}. In fact, if $\mommap{A_{2n-1}}$ is the moment map for a type $A_{2n-1}$ quiver then

\be P(\mommap{A_{2n-1}})=\mommap{C_n} \ee is a $C_n$ moment map and $P$ acts component-wise. Similarly, \be P(\mommap{D_{n+1}})=\mommap{B_n}. \ee

To see this action on an example, and to illustrate why its action on moment maps is so simple, consider the top left quiver in Fig. \ref{fig:folding-examples}. Select the Chevalley-Serre basis of $D_4$ and its operator counterpart, ie. the basis of operators in $\co [ \Coul ]_2$ which replicates (\ref{eq:CW1})-(\ref{eq:CW4}) with Poisson brackets. We will denote the algebra elements and their duals with capital letters, reserving lower case letters for operators with appropriate commutation relations. In this notation, the moment map is 
\begin{equation}
    \mommap{D_4} = \sum_{1\leq i \leq 4} \h{i} \Hd{i} + \sum_{\alpha \in \Phi^+} \left(e_{\alpha} E^*_{\alpha} + e_{-\alpha} E^*_{-\alpha}\right) \, . 
\end{equation}
This sum above contains 28 terms, and to exhibit the action of $P$ we show what happens to 5 of them: 
\be
\begin{split}
    \mommap{D_4} = &\h{1} \Hd{1} + \h{2} \Hd{2} +  \h{3} \Hd{3} + \h{4} \Hd{4} \\
    &+ \e{1,0,0,0} \Ed{1,0,0,0} + \e{0,1,0,0} \Ed{0,1,0,0} + \e{0,0,1,0} \Ed{0,0,1,0} + \e{0,0,0,1} \Ed{0,0,0,1} \\ 
    &+ \e{1,1,0,0} \Ed{1,1,0,0} + \e{0,1,1,0} \Ed{0,1,1,0} + \e{0,1,0,1} \Ed{0,1,0,1} \\
    &+ \e{1,1,1,0} \Ed{1,1,1,0} + \e{1,1,0,1} \Ed{1,1,0,1} + \e{0,1,1,1} \Ed{0,1,1,1} \\
    &+ \e{1,1,1,1} \Ed{1,1,1,1} + \e{1,2,1,1} \Ed{1,2,1,1}
\end{split}
\ee
where the dots stand for the other 23 terms. 
The projector acts on operators: 
\begin{align*}
    P(\mommap{D_4}) =& \sum_{1\leq i \leq 4} P(\h{i}) \Hd{i} + \sum_{\alpha \in \Phi^+} \left(P(e_{\alpha}) E^*_{\alpha} + P(e_{-\alpha}) E^*_{-\alpha}\right)\\
    =& P(\h{1}) \Hd{1} + P(\h{2}) \Hd{2} +  P(\h{3}) \Hd{3} + P(\h{4}) \Hd{4} \\
    &+ P(\e{1,0,0,0}) \Ed{1,0,0,0} + P(\e{0,1,0,0}) \Ed{0,1,0,0} \\
    &+ P(\e{0,0,1,0}) \Ed{0,0,1,0} + P(\e{0,0,0,1}) \Ed{0,0,0,1} \\ 
    &+ P(\e{1,1,0,0}) \Ed{1,1,0,0} + P(\e{0,1,1,0}) \Ed{0,1,1,0} + P(\e{0,1,0,1}) \Ed{0,1,0,1} \\
    &+ P(\e{1,1,1,0}) \Ed{1,1,1,0} + P(\e{1,1,0,1}) \Ed{1,1,0,1} + P(\e{0,1,1,1}) \Ed{0,1,1,1} \\
    &+ P(\e{1,1,1,1}) \Ed{1,1,1,1} + P(\e{1,2,1,1}) \Ed{1,2,1,1} \\
    =& \h{1} \Hd{1} + \h{2} \Hd{2} + \frac{\h{3}+\h{4}}{2} \Hd{3} + \frac{(\h{3}+\h{4})}{2} \Hd{4} \\
    &+ \e{1,0,0,0} \Ed{1,0,0,0} + \e{0,1,0,0} \Ed{0,1,0,0} \\
    &+ \frac{\e{0,0,1,0}+\e{0,0,0,1}}{2} \Ed{0,0,1,0} + \frac{\e{0,0,1,0}+\e{0,0,0,1}}{2} \Ed{0,0,0,1} \\
    &+ \e{1,1,0,0} \Ed{1,1,0,0} + \frac{\e{0,1,1,0}+\e{0,1,0,1}}{2} \Ed{0,1,1,0} + \frac{\e{0,1,1,0}+\e{0,1,0,1}}{2} \Ed{0,1,0,1} \\
    &+ \frac{\e{1,1,1,0}+\e{1,1,0,1}}{2} \Ed{1,1,1,0} + \frac{\e{1,1,1,0}+\e{1,1,0,1}}{2} \Ed{1,1,0,1} + \e{0,1,1,1} \Ed{0,1,1,1} \\
    &+ \e{1,1,1,1} \Ed{1,1,1,1} + \e{1,2,1,1} \Ed{1,2,1,1} \\
    =& \h{1} \Hd{1} + \h{2} \Hd{2} + (\h{3}+\h{4}) \frac{\Hd{3}+\Hd{4}}{2} \\
    &+ \e{1,0,0,0} \Ed{1,0,0,0} + \e{0,1,0,0} \Ed{0,1,0,0} \\
    &+ (\e{0,0,1,0}+\e{0,0,0,1}) \frac{\Ed{0,0,1,0}+\Ed{0,0,0,1}}{2} \\
    &+ \e{1,1,0,0} \Ed{1,1,0,0} + (\e{0,1,1,0}+\e{0,1,0,1}) \frac{\Ed{0,1,1,0} + \Ed{0,1,0,1}}{2} \\
    &+ (\e{1,1,1,0}+\e{1,1,0,1}) \frac{\Ed{1,1,1,0} + \Ed{1,1,0,1}}{2} + \e{0,1,1,1} \Ed{0,1,1,1} \\
    &+ \e{1,1,1,1} \Ed{1,1,1,1} + \e{1,2,1,1} \Ed{1,2,1,1} \\
    =& \th{1} \tHd{1} + \th{2} \tHd{2} + \th{3} \tHd{3} \\
    &+ \te{1,0,0} \tEd{1,0,0} + \te{0,1,0} \tEd{0,1,0} + \te{0,0,1} \tEd{0,0,1} \\
    &+ \te{1,1,0} \tEd{1,1,0} + \te{0,1,1} \tEd{0,1,1}\\
    &+ \te{1,1,1} \tEd{1,1,1} + \te{0,1,2} \tEd{0,1,2} \\
    &+ \te{1,1,2} \tEd{1,1,2} + \te{1,2,2} \tEd{1,2,2} \\
    =& \mommap{B_3} \tag{\stepcounter{equation}\theequation}
\end{align*}
where we defined 
\begin{align}\label{eq:B3-1}
	\th{3} &= \h{3} + \h{4} = 2(\p{3}+\p{4}) - 2 (\p{2,1}+\p{2,2})\\
	\label{eq:B3-2}\te{0,0,1} &= \e{0,0,1,0} + \e{0,0,0,1} = \up{3}+\up{4}\\
	\label{eq:B3-3}\te{0,1,2} &= \e{0,1,1,1} = \sum_{a=1,2}\frac{\up{2,a}\up{3}\up{4}}{(\p{2,a}-\p{3})(\p{2,a}-\p{4})} 
\end{align}
and the remaining operators follow the same pattern $\te{a,b,2c} = \e{a,b,c,c}$ or $\te{a,b,c} = \e{a,b,c,0} + \e{a,b,0,c}$ if $c\neq 0$ and $\te{a,b,0} = \e{a,b,0,0}$ otherwise. Notice especially how the prefactor from the operator becomes the inverse multiplicity required in the definition of the new dual basis.

Just as $\mommap{D_4}$ satisfies  certain matrix relations which identify  the space it parametrises as the (closure of the) minimal nilpotent orbit of $D_4$, so does $\mommap{B_3}$ obey several relations appropriate for a $B_3$ nilpotent orbit. The space should be an orbifold of the minimal orbit of $D_4$, so it should in particular have the same quaternionic dimension, namely 5. That is precisely the dimension of the next-to-minimal orbit of $B_3$ with the HWG \cite{hanany_quiver_2016}
\be
\HWG(t,\mu_i)=\frac{1}{(1-\mu_2 t^2)(1-\mu_1^2 t^4)}
\ee
This space is parametrised by a matrix $M$ satisfying the relations (computed using standard plethystic techniques)\footnote{A similar set of relations appears in \cite{cheng_coulomb_2017}, albeit for the next-to-minimal orbit of $D_4$. The methods employed therein can be extended to the present case: given a general nilpotent orbit, one can construct the quiver for which it is the Higgs branch and look for matrix relations implied by the $F$-terms.}
\begin{align}
    t^4 \lbrack 000 \rbrack&: \tr{N^2} = 0 \nonumber \\
    t^4 \lbrack 002 \rbrack&: N \wedge N = 0 \label{eq:relationsB3} \\ 
    t^6 \lbrack 010 \rbrack&: N^3 = 0 \nonumber
\end{align}
We describe a relation by its $R$-symmetry weight appearing in the exponent of $t$ and the global symmetry representation in which it transforms. This often, but not always, specifies the tensorial form of the relation, which we provide on the other side of the colon. The notation $N\wedge N$ should be understood as the contraction $\sum_{lmno}\epsilon_{ijklmno}N_{lm}N_{no}$ with the rank 7 antisymmetric invariant tensor of $\mathfrak{so}(7)$.

One can check that the moment map $\mommap{B_3}$ satisfies the identities in (\ref{eq:relationsB3}) modulo abelianised relations. To show that there exist no \textit{other} independent relations, or generators for that matter, one can calculate the Hilbert series of the ring as described below. This computation shows that indeed (\ref{eq:relationsB3}) form the complete set of relations for the next-to-minimal orbit of $\mathfrak{so}(7)$. Note that this is an instance of Case 2 of the Kostant-Brylinski Figure \ref{tabKostantBrylinski}.

\subsection{Wreathed quivers}

The previous section establishes that some Coulomb branches can be orbifolded by a quiver automorphism. We will now argue that the orbifold can also be recovered as the Coulomb branch of the original quiver after gauging its automorphism. It is natural to ask if the resulting theory is also a quiver theory which could be studied without reference to the original, ungauged theory. This is indeed possible, albeit at the cost of generalising the notion of a quiver theory to \emph{wreathed quivers}.

Traditionally a quiver theory is described by a quiver diagram in which nodes represent gauge or flavor groups and links represent appropriately charged matter. Wreathed quiver theories add wreathed legs denoted by $(\cdot)\wr S_n$ with an associated \emph{wreathing group} $S_n$. See Fig. \ref{fig:wreathing-examples} for two prototypical examples. The top right quiver has a single wreathed node while the bottom right quiver is an example of a quiver with a longer wreathed leg.

\subsubsection*{Abelianisation of wreathed quivers}

The Coulomb branch of a wreathed quiver can be studied through abelianisation with relatively minor changes, but it is cumbersome to write them down in full generality. We  find much greater clarity in (entirely equivalent) abelianised calculations performed on discretely gauged non-wreathed quivers. In practice, this amounts to keeping the indices, Poisson and abelianised chiral structure from the non-wreathed quiver while imposing invariance under the projector \ref{eq:projector}. For illustrative purposes, and to draw a link to \cite{hanany_discrete_2018,hanany_discrete_2018-1}, we present two particularly simple examples depicted in Fig. \ref{fig:wreathing-examples}.

There are very few new elements in the wreathed quiver theory depicted in the top right quiver of Fig. \ref{fig:wreathing-examples}. The third node brings six variables $\upm{3,a}$ and $\p{3,a}$, $a\in \lbrace 1,2 \rbrace$, much like a $U(2)$ node would. The wreathing group acts similarly to a Weyl group in that it permutes the index $a$ and all physical operators are invariant under it. 

Abelianised relations on the middle node read 
\be \up{2,a} \um{2,a} = - \frac{(\p{2,a}-\m{2})(\p{2,a}-\p{1})(\p{2,a}-\p{3,1})(\p{2,a}-\p{3,2})}{(\p{2,1}-\p{2,2})^2} \ee and the relations on the third node are essentially unchanged: \be \up{3,a} \um{3,a} = - (\p{3,a}-\p{2,1})(\p{3,a}-\p{2,2}). \ee 
Interestingly, the latter can be read in two ways: either as the relation of a $U(1)\wr S_2$ node, or as \be \up{3,a} \um{3,a} = - \frac{(\p{3,a}-\p{2,1})(\p{3,a}-\p{2,2})}{(\p{3,1}-\p{3,2})^2} (\p{3,1}-\p{3,2})^2, \ee which is appropriate for a $U(2)$ node with adjoint matter. This explains why in \cite{hanany_discrete_2018,hanany_discrete_2018-1} a ``bouquet" of $n$ $U(1)$ nodes combined into $U(n)$ with adjoint matter: at the level of the Coulomb branch, there is no difference between $U(1)\wr S_n$ and $U(n)$ with adjoint matter.

The case of the bottom right quiver in Fig. \ref{fig:wreathing-examples} is slightly more subtle. The first and second gauge nodes, which are inside the scope of a two-fold wreathing, each come with six variables $\upm{3,a}$ and $\p{3,a}$, $a\in \lbrace 1,2 \rbrace$. However, the pattern of abelianised relations, which can be determined by consistency with the discrete gauging of the bottom left quiver in Fig. \ref{fig:wreathing-examples}, is as follows: 
\begin{align}
	\up{1,a}\um{1,a} &= -(\p{1,a}-\m{1})(\p{1,a}-\p{2,a})\\
	\up{2,a}\um{2,a} &= -(\p{2,a}-\p{1,a})(\p{2,a}-\p{3})\\
	\up{3}\um{3} &= -(\p{3}-\p{2,1})(\p{3}-\p{2,2}) 
\end{align}

Note in particular that the index $a$ ``stretches" across several nodes (but not the mass variable, which is shared by all legs). The wreathing group $S_2$ again acts on this index, and invariance under it is a necessary prerequisite for operator physicality. The Coulomb branch has $C_3$ symmetry and the moment map parametrises the next-to-minimal nilpotent orbit of this algebra. Its components include:
\begin{align}
    \e{\pm 1,0,0}&= \upm{1,1}+\upm{1,2}\\
    \e{0,\pm 1,0}&= \upm{2,1}+\upm{2,2}\\
    \e{0,0,\pm 1}&= \upm{3}\\
    \e{\pm 1, \pm 1, 0} &= \frac{\upm{1,1}\upm{2,1}}{\p{1,1}-\p{2,1}} + \frac{\upm{1,2}\upm{2,2}}{\p{1,2}-\p{2,2}} \\
    \e{0, \pm 1, \pm 1} &= \frac{\upm{2,1}\upm{3}}{\p{2,1}-\p{3}} + \frac{\upm{2,2}\upm{3}}{\p{2,2}-\p{3}}\\
    \e{\pm 1, \pm 1, \pm 1} &= \frac{\upm{1,1}\upm{2,1}\upm{3,1}}{(\p{1,1}-\p{2,1})(\p{2,1}-\p{3})} \nonumber\\
    &\quad + \frac{\upm{1,2}\upm{2,2}\upm{3,1}}{(\p{1,2}-\p{2,2})(\p{2,2}-\p{3})} \\
    \e{0, \pm 2, \pm 1} &= \frac{\upm{2,1}\upm{2,2}\upm{3}}{(\p{2,1}-\p{3})(\p{2,2}-\p{3})}\\
    \e{\pm 1, \pm 2, \pm 1} &= \frac{\upm{1,1}\upm{2,1}\upm{2,2}\upm{3,1}}{(\p{1,1}-\p{2,1})(\p{2,1}-\p{3})(\p{2,2}-\p{3})} \nonumber\\
    &\quad + \frac{\upm{1,2}\upm{2,1}\upm{2,2}\upm{3,1}}{(\p{1,2}-\p{2,1})(\p{2,1}-\p{3})(\p{2,2}-\p{3})} \\
    \e{\pm 2, \pm 2, \pm 1} &= \frac{\upm{1,1}\upm{1,2}\upm{2,1}\upm{2,2}\upm{3,1}}{(\p{1,1}-\p{2,1})(\p{1,2}-\p{2,2})(\p{2,1}-\p{3})(\p{2,2}-\p{3})} \\
    \h{1} &= 2(\p{1,1}+\p{1,2})-(\p{2,1}+\p{2,2})\\
    \h{2} &= -(\p{1,1}+\p{3,1})+2(\p{2,1}+\p{2,2})-2\p{3}\\
    \h{3} &= -(\p{2,1}+\p{2,2})+2\p{3}
\end{align}

\subsection{Monopole formula for wreathed quivers}

Consider now a wreathed quiver. To compute the monopole formula, we need to replace the Weyl group in (\ref{monopole}) with an appropriate discrete group, necessarily a subgroup of $S_r$ which contains the Weyl group $W$ of the gauge group $G$ and leaves $\Delta(m)$ invariant. Generically, several such discrete groups exist. Our choice, which we dub $W_\Gamma$, is the wreath product of the Weyl group $W$ by the wreathing group $\Gamma$:  
\begin{equation}
\label{inclGamma}
   W_{\Gamma} = W \wr \Gamma \, , \qquad \qquad  W \subseteq W_{\Gamma} \subseteq S_r \, . 
\end{equation}
Formula (\ref{monopole}) generalises readily for such a group, setting 
\begin{equation}
\label{monopoleSym}
\boxed{
     \HS_{\Gamma}(t) = \frac{1}{|W_{\Gamma}|} \sum\limits_{m \in \mathbb{Z}^r} \sum\limits_{\gamma \in W_{\Gamma}(m)} \frac{t^{2 \Delta (m)}}{\det \left( 1-t^2 \gamma \right)} } \, .  
\end{equation}
This is the monopole formula for the wreathed quiver. 

\subsubsection*{A comment on computational complexity}

The monopole formula in the form (\ref{monopoleSym}) is very time-consuming to evaluate numerically in a series expansion in $t$. For such a task, it is preferable to preprocess it somewhat, using the high level of symmetry that it presents. In particular, if the group $\Gamma$ can be written as a product of two groups $W_{\Gamma} = W_1 \times W_2$, then one can split the summation into two sums. 

This procedure involves finding a subset of $\mathbb{Z}^r$ which contains exactly one element of each orbit of $W_{\Gamma}$. In the context of Weyl groups, or more generally Coxeter groups, this is called a fundamental chamber. For instance, if $W_{\Gamma}=W$ as in (\ref{Weylgroup}), then this group can be used to order the magnetic charges in increasing order for each node. Then one uses the identity 
\begin{equation}
\label{casimir}
    P_{U(n)} (t^2 ; m_1 , \dots , m_N) := P_{S_n} (t ; m_1 , \dots , m_n) = \frac{1}{n!} \sum\limits_{\gamma \in S_n (m)} \frac{1}{\det \left( 1-t^2 \gamma \right)}
\end{equation}
for the Casimir factors as defined in the appendix of \cite{cremonesi_monopole_2014}. This is done in the usual way of presenting the monopole formula. 

For wreathed quivers, $W_{\Gamma}$ does not decompose in general as a direct product of symmetry groups. One can introduce symmetry factors exactly as in (\ref{casimir}), via 
\begin{equation}
\label{casimirGamma}
    P_{W_{\Gamma}} (t^2 ; m) = \frac{1}{|W_{\Gamma}|} \sum\limits_{\gamma \in W_{\Gamma} (m)} \frac{1}{\det \left( 1-t^2 \gamma \right)} \, . 
\end{equation}
The formula (\ref{monopoleSym}) can then be rewritten 
\begin{equation}
\label{monopoleSymRephrased}
    \HS_\Gamma (t) = \sum\limits_{m \in \mathrm{Weyl}(G \wr \Gamma)} P_{W_{\Gamma}} (t;m) t^{2 \Delta(m)} \, ,  
\end{equation}
where $\mathrm{Weyl}(G \wr \Gamma)$ is a principal Weyl chamber for the group $G \wr \Gamma$.  
We now illustrate the procedure on three examples and most explicitly on the third. 

\subsubsection*{Example 1 : subgroups of $S_3$}

Consider the quiver corresponding to the affine $D_4$ Dynkin diagram (see the first column of Figure \ref{table:g2-reductions}). One of the four rank-one nodes is a flavor node, so we can define the graph by the vertices 
\begin{equation}
\label{notationD4}
    V=\{ a,b,c,d\} \qquad F = \{e\} 
\end{equation}
where $a$ denotes the central node, and 
\begin{equation}
    \qquad E = \{  (a,b) , (a,c) , (a,d) \} \qquad E' = \{(a,e)\} \, . 
\end{equation}
The corresponding ranks are $n_a=2$, $n_b=n_c=n_d=n_e=1$. The total gauge group is $G = U(2) \times U(1)^3$ with rank $r=5$. The Weyl group is $W = S_2$. The magnetic charges are elements $m = (m_{a,1} , m_{a,2} , m_b,m_c,m_d) \in \mathbb{Z}^5$ and the conformal dimension is given by
\begin{equation}
    2 \Delta (m) = \sum\limits_{i=1}^2 (|m_{a,i} - m_b| + |m_{a,i} - m_c| + |m_{a,i} - m_d| + |m_{a,i}| ) - 2 |m_{a,1} - m_{a,2}| \,  . 
\end{equation}
The group $S_5$ includes 156 subgroups which can be gathered into 19 conjugacy classes. These 19 classes are partially ordered and form a Hasse diagram. For $\Delta$ to be invariant, we have to select those groups $\Gamma$ which are subgroups of $S_2 \times S_3$ (this is also known as the dihedral group $D_{12}$), and moreover to satisfy (\ref{inclGamma}) the groups  $\Gamma$ have to contain $S_2$ as a subgroup. Out of the 19 classes of subgroups, 9 are subgroups of $D_{12}$, and out of these 9, 6 contain a $S_2$ as a subgroup. However there are two equivalent but non-conjugate $S_2$ subgroups of $D_{12}$, and we have to pick one of them. We are then left with 4 classes of subgroups, which can be identified with the 4 classes of subgroups of $S_3$ ($S_3$, $\mathbb{Z}_3$, $\mathbb{Z}_2$ and $1$). Clearly, in this simple example this analysis is slightly superfluous and the result could have been guessed. We end up with 4 inequivalent groups $\Gamma$, and we can readily evaluate the expression (\ref{monopoleSym}): 
\begin{eqnarray}
\HS_{\mathbb{Z}_2} &=& \frac{(1 + t^2) (1 + 17 t^2 + 48 t^4 + 17 t^6 + t^8)}{(1 - t^2)^{10}} \\
\HS_{\mathbb{Z}_2 \times \mathbb{Z}_2} &=& \frac{(1 + t^2) (1 + 10 t^2 + 20 t^4 + 10 t^6 + t^8)}{(1 - t^2)^{10}} \\
\HS_{\mathbb{Z}_2 \times \mathbb{Z}_3} &=& \frac{(1 + t^2) (1 + 3 t^2 + 20 t^4 + 3 t^6 + t^8)}{(1 - t^2)^{10}} \\
\HS_{\mathbb{Z}_2 \times S_3} &=& \frac{(1 + t^2) (1 + 3 t^2 + 6 t^4 + 3 t^6 + t^8)}{(1 - t^2)^{10}}  
\end{eqnarray}
which identify the spaces as the (closure of the) minimal nilpotent orbit of $SO(8)$, next to minimal of $SO(7)$, double cover of the subregular orbit of $G_2$ \cite{achar_geometric_2013}, and the subregular orbit of $G_2$.

\subsubsection*{Example 2 : subgroups of $S_4$}

\begin{figure}[t]
    \centering
    \begin{tabular}{c|c|c}
        Name & Generators & Cardinality \\ \hline 
        Trivial & - & 1  \\
        $S_2$ & $(12)$  & 2 \\
        Double transposition &$(12)(34)$ & 2\\
        $\mathbb{Z}_4$ & $(1234)$ & 4 \\
        Normal Klein & $(12)(34)$, $(13)(24)$ & 4 \\ 
        Non-normal Klein &$(12)$, $(34)$ & 4 \\
        $\mathrm{Dih}_4$ & $(1234)$ ,$(13)$   & 8 \\ 
        $\mathbb{Z}_3$ & $(123)$, &3 \\ 
        $S_3$ & $(12)$ , $(13)$  & 6 \\ 
        $A_4$ & $(123)$ , $(124)$ & 12 \\ 
        $S_4$ & $(12)$ , $(13)$ , $(14)$ & 24 \\ 
    \end{tabular}
    \caption{Subgroups of $S_4$ }
    \label{tableS4Subgroups}
\end{figure}

We now consider the same quiver as in the previous example, namely the affine $D_4$ quiver, but we use the fact that the gauge group of the theory is really 
\begin{equation}
\label{gaugegroupS4}
    \frac{U(1)^4 \times U(2)}{U(1)}
\end{equation}
where the $U(1)$ acts diagonally. This form makes the $S_4$ symmetry of the quiver explicit, and this $S_4$ contains as a subgroup the $S_3$ which is studied in the previous example. Following the approach of this section, one can define a wreathed quiver for each conjugacy class of subgroups of $S_4$. Part of the results presented here already appear in unpublished summer work by Siyul Lee \cite{siyul}, where the cycle index technique was used. The group $S_4$ admits 30 subgroups that can be organized into 11 conjugacy classes, as listed in Figure \ref{tableS4Subgroups}, where we give a name to each class of subgroups. 

For each subgroup, one can compute the Hilbert series (\ref{monopoleSym}). The results are gathered in Figure \ref{figHasseS4}, where they are arranged in the shape of the Hasse diagram of the subgroups of $S_4$. We give some details about the computation in appendix \ref{appendixS4}. We also give the first orders of the series expansions of these Hilbert series, along with their plethystic logarithms, in Figure \ref{tablehsS4}. The coefficient of the $t^2$ term gives the dimension of the isometry group of the Coulomb branch. 

\begin{figure}[t]
    \centering
    \hspace*{-1cm}\begin{tabular}{c|c|c}
        Subgroup & Perturbative Hilbert series & PLog \\ \hline 
        Trivial & $1+28 t^2+300 t^4+1925 t^6+8918 t^8+...$ & $28 t^2-106 t^4+833 t^6-8400 t^8+...$  \\
        $S_2$ & $1+21 t^2+195 t^4+1155 t^6+5096 t^8+...$ & $21 t^2-36 t^4+140 t^6-784 t^8+...$ \\
        Double transposition & $1+16 t^2+160 t^4+985 t^6+4522 t^8+...$ & $16 t^2+24 t^4-215 t^6+522 t^8+...$\\
        $\mathbb{Z}_4$ & $1+9 t^2+83 t^4+497 t^6+2270 t^8+...$ & $9 t^2+38 t^4-10 t^6-586 t^8+...$ \\
        Normal Klein & $1+10 t^2+90 t^4+515 t^6+2324 t^8+...$ & $10 t^2+35 t^4-55 t^6-396 t^8+...$ \\ 
        Non-normal Klein & $1+15 t^2+125 t^4+685 t^6+2898 t^8+...$ & $15 t^2+5 t^4-70 t^6+273 t^8+...$ \\
        $\mathrm{Dih}_4$ & $1+9 t^2+69 t^4+356 t^6+1485 t^8+...$ & $9 t^2+24 t^4-25 t^6-165 t^8+...$ \\ 
        $\mathbb{Z}_3$ & $1+14 t^2+118 t^4+693 t^6+3094 t^8+...$ & $14 t^2+13 t^4-49 t^6-56 t^8+...$ \\ 
        $S_3$ & $1+14 t^2+104 t^4+539 t^6+2184 t^8+...$ & $14 t^2-t^4-7 t^6+7 t^8+...$ \\ 
        $A_4$ & $1+8 t^2+48 t^4+223 t^6+896 t^8+...$ & $8 t^2+12 t^4+7 t^6+0t^8+...$ \\ 
        $S_4$ & $1+8 t^2+48 t^4+210 t^6+771 t^8+...$ & $8 t^2+12 t^4-6 t^6-21 t^8+...$ \\ 
    \end{tabular}
    \caption{Wreathed quivers obtained from the affine $D_4$ quiver by acting on the legs by all subgroups of $S_4$. }
    \label{tablehsS4}
\end{figure}

\begin{landscape}
\begin{figure}
    \centering
\vspace*{-2.5cm}\hspace*{-3cm}\begin{tikzpicture}
			\node[draw] (S4) at (0,-2) {\begin{tabular}{c}
			     24 \hfill $S_4$ \hfill $\mathfrak{su}(3)$ \\$\frac{1+3 t^2+13 t^4+25 t^6+46 t^8+48 t^{10}+\dots+t^{20}}{\left(1-t^2\right)^{10} \left(1+t^2\right)^5} $	\end{tabular} }; 
			\node[draw] (A4) at (-1,1) {\begin{tabular}{c}
			     12 \hfill $A_4$  \hfill $\mathfrak{su}(3)$ \\$\frac{1+3 t^2+13 t^4+38 t^6+106 t^8+126 t^{10}+\dots+t^{20}}{\left(1-t^2\right)^{10} \left(1+t^2\right)^5} $	\end{tabular} }; 
			\node[draw] (D4) at (-9,1) {\begin{tabular}{c}
			     8 \hfill $\mathrm{Dih}_4$ \hfill $\mathfrak{u}(3)$ \\$\frac{1+4 t^2+29 t^4+71 t^6+150 t^8+162 t^{10}+\dots+t^{20}}{\left(1-t^2\right)^{10} \left(1+t^2\right)^5} $	\end{tabular} }; 
			\node[draw] (S3) at (6,2.5) {\begin{tabular}{c}
			     6 \hfill $S_3$ \hfill $G_2$ \\$\frac{\left(1+t^2\right) \left(1+3 t^2+6 t^4+3 t^6+t^8\right)}{\left(1-t^2\right)^{10}} $	\end{tabular} }; 
			\node[draw] (Z4) at (-15,7) {\begin{tabular}{c}
			    4 \hfill  $\mathbb{Z}_4$ \hfill $\mathfrak{u}(3)$ \\$\frac{1+4 t^2+43 t^4+142 t^6+300 t^8+364 t^{10}+\dots+t^{20}}{\left(1-t^2\right)^{10} \left(1+t^2\right)^5} $	\end{tabular} }; 
			\node[draw] (NK) at (-9,5) {\begin{tabular}{c}
			     4 \hfill Normal Klein  \hfill $\mathfrak{sp}(2)$  \\$\frac{1+5 t^2+45 t^4+130 t^6+314 t^8+354 t^{10}+\dots+t^{20}}{\left(1-t^2\right)^{10} \left(1+t^2\right)^5} $	\end{tabular} }; 
			\node[draw] (NNK) at (-3,7) {\begin{tabular}{c}
			    4 \hfill  Non-normal Klein \hfill $\mathfrak{su}(4)$  \\$ \frac{1+10 t^2+55 t^4+150 t^6+288 t^8+336 t^{10}+\dots+t^{20}}{\left(1-t^2\right)^{10} \left(1+t^2\right)^5}$	\end{tabular} }; 
			\node[draw] (Z3) at (6,8.5) {\begin{tabular}{c}
			    3 \hfill $\mathbb{Z}_3$ \hfill $G_2$ \\$\frac{1+4 t^2+23 t^4+23 t^6+4 t^8+t^{10}}{\left(1-t^2\right)^{10}} $	\end{tabular} }; 
			\node[draw] (DT) at (-9,10) {\begin{tabular}{c}
			     2 \hfill Double Transposition \hfill $\mathfrak{u}(4)$ \\
			    $ \frac{\left(1+3 t^2+6 t^4+3 t^6+t^8\right) \left(1+8 t^2+55 t^4+64 t^6+55 t^8+8 t^{10}+t^{12}\right)}{\left(1-t^2\right)^{10} \left(1+t^2\right)^5}$
			\end{tabular} }; 
			\node[draw] (S2) at (-1,10) {\begin{tabular}{c}
			    2 \hfill $S_2$ \hfill $\mathfrak{so}(7)$ \\$\frac{\left(1+t^2\right) \left(1+10 t^2+20 t^4+10 t^6+t^8\right)}{\left(1-t^2\right)^{10}}$	\end{tabular} }; 
			\node[draw] (S1) at (0,13) {\begin{tabular}{c}
			    1 \hfill Trivial \hfill $\mathfrak{so}(8)$ \\$\frac{\left(1+t^2\right) \left(1+17 t^2+48 t^4+17 t^6+t^8\right)}{\left(1-t^2\right)^{10}}$	\end{tabular} }; 
			\draw (S4)--(D4)--(Z4)--(DT)--(NK)--(D4)--(NNK)--(DT)--(S1)--(S2)--(S3)--(S4);
			\draw (NK)--(A4)--(S4);
			\draw (A4)--(Z3)--(S1);
			\draw (S2)--(NNK);
			\draw (S3)--(Z3);
\end{tikzpicture}
    \caption[Hasse diagram of $S_4$ subgroups and corresponding wreaths]{Hasse diagram of the 11 conjugacy classes of subgroups of $S_4$ with the Hilbert series for the Coulomb branch of the corresponding wreathed affine $D_4$ quiver. Dots in the numerators can be filled in using the fact that each polynomial is palindromic. In each box, the number on the left is the order of the group $\Gamma$, and the algebra on the right is the global symmetry of the Coulomb branch. The whole diagram possesses a symmetry exchanging $S_4 \leftrightarrow \textrm{Trivial}$, $S_3 \leftrightarrow \mathbb{Z}_4$, $\mathrm{Dih}_4$ and Double Transposition, Normal and Non-normal Klein, and fixes $\mathbb{Z}_4$. This is not obvious in the depiction because of planarity constraints. }
    \label{figHasseS4}
\end{figure}
\end{landscape}

\subsubsection*{Example 3 : wreath product of non Abelian groups}

We now consider the quiver 
\begin{equation}
\label{quiverD5}
    \begin{tikzpicture}
			\node (g1) at (0,0) [gauge, label=below:{1}, label=above:{\textcolor{red}{$a$}}]{}; 
			\node (g2) at (1,0) [gauge, label=below:{2}, label=above:{\textcolor{red}{$b$}}]{}; 
			\node (g3) at (2,0) [gauge, label=below:{3}, label=above:{\textcolor{red}{$c$}}]{}; 
			\node (g4) at (3,1) [gauge, label=below:{2}, label=above:{\textcolor{red}{$d$}}]{}; 
			\node (g5) at (4,1) [flavor, label=below:{1}]{}; 
			\node (g6) at (3,-1) [gauge, label=below:{2}, label=above:{\textcolor{red}{$e$}}]{}; 
			\node (g7) at (4,-1) [flavor, label=below:{1}]{}; 
			\draw (g1)--(g2)--(g3)--(g4)--(g5);
			\draw (g3)--(g6)--(g7);
		\end{tikzpicture}
\end{equation}
whose Coulomb branch is the closure of the nilpotent orbit of $\mathfrak{so}(10)$ associated with the partition $[2^4,1^2]$. The letters in red give our assignment of magnetic charge for the various gauge groups. The rank is $r=10$ and the Weyl group is $W=S_1 \times S_2 \times S_3 \times S_2 \times S_2$. In order to preserve $\Delta$, we need a symmetry of the quiver, which is given by permutation of the two legs containing the nodes $d$ and $e$. So there are only two allowed groups $W_{\Gamma}$, namely $W_{\Gamma}=W$ and an extension $W_{\Gamma}$ of $W$ of index 2. Let's focus on this second group. 

The factors $S_1 \times S_2 \times S_3$ in $W$ are unaffected by the permutation, so we omit them in the matrix discussion that follows. The commutant of this part in $S_{10}$ is $S_4$, which acts by permuting the four magnetic fugacities $(d_1,d_2,e_1,e_2)$. The group $W_{\Gamma}$ is then the product $W_{\Gamma} = S_1 \times S_2 \times S_3 \times \Gamma '$ where $S_2 \times S_2 \subset \Gamma ' \subset S_4$. We can describe $\Gamma '$ explicitly as generated by the following two permutation matrices: 
\begin{equation}
\left(
\begin{array}{cccc}
 0 & 1 & 0 & 0 \\
 1 & 0 & 0 & 0 \\
 0 & 0 & 1 & 0 \\
 0 & 0 & 0 & 1 \\
\end{array}
\right) \, , \qquad 
\left(
\begin{array}{cccc}
 0 & 0 & 1 & 0 \\
 0 & 0 & 0 & 1 \\
 1 & 0 & 0 & 0 \\
 0 & 1 & 0 & 0 \\
\end{array}
\right)  \, . 
\end{equation}
This group is isomorphic to the dihedral group of order 8, that we denote by $\mathrm{Dih}_4$. With this explicit description, it is now possible to evaluate (\ref{monopoleSym}), and one finds the Hilbert series for the Coulomb branch of the wreathed quiver, 
\begin{equation}
    \HS_{S_1 \times S_2 \times S_3 \times \mathrm{Dih}_4} = \frac{1 + 14 t^2 + 106 t^4 + 454 t^6 + 788 t^8 + 454 t^{10} + 106 t^{12} + 
 14 t^{14} + t^{16}}{(1 - t^2)^{20} (1 + t^2)^{-2}} \, . 
\end{equation}
The corresponding HWG and other data concerning this space are gathered in the middle column of Figure \ref{tab:d5nnmin-b4}. 

However, the sum involved in the computation is difficult to evaluate in practice, and it is useful to use the symmetries to avoid unnecessary repetitions, as explained above. In the present case, the sum in (\ref{monopoleSym}) for $W_{\Gamma} = S_1 \times S_2 \times S_3 \times \mathrm{Dih}_4$ becomes (\ref{monopoleSymRephrased}) where the sum over the Weyl chamber is given by: 
\begin{equation}
\label{changeSum}
 \frac{1}{|W_{\Gamma}|} \sum\limits_{m \in \mathbb{Z}^{10}}  \longrightarrow \sum\limits_{a} \sum\limits_{b_1 \leq b_2} \sum\limits_{c_1 \leq c_2 \leq c_3} \left[  \sum\limits_{\mbox{\tiny\ensuremath{\begin{array}{ccc}
        d_1 & \leq & d_2 \\
           &   & \vl   \\
          e_1 & \leq & e_2 \\
    \end{array}}}} + \sum\limits_{\mbox{\tiny\ensuremath{\begin{array}{ccc}
        d_1 & \leq & d_2 \\
         \vleq  &   & \ve   \\
          e_1 & \leq & e_2 \\
    \end{array}}}} \right]  
\end{equation}
The notation here should be clear, with the charges $m=(a,b_1,b_2,c_1,c_2,c_3,d_1,d_2,e_1,e_2)\in \mathbb{Z}^{10}$ denoted with the letters as in (\ref{quiverD5}). The first three sums in the right hand side of (\ref{changeSum}) exploit in the standard way the symmetric groups, which allow to order the charges. The same principle is used to get the summation range over indices $(d_1,d_2,e_1,e_2)$. Inside the sum, one of course introduces symmetry factors (\ref{casimirGamma}). Let's now explain the summation range for the last four indices in (\ref{changeSum}). 

$\Gamma '$ is the dihedral group $\mathrm{Dih}_4$, of order 8, or the group of symmetries of the square 
\begin{equation}
        \begin{tikzpicture}
			\node at (0,0) [label=left:{\textcolor{red}{$e_1$}}]{}; 
			\node at (0,2) [label=left:{\textcolor{red}{$d_1$}}]{}; 
			\node at (2,0) [label=right:{\textcolor{red}{$d_2$}}]{}; 
			\node at (2,2) [label=right:{\textcolor{red}{$e_2$}}]{}; 
			\draw (0,0)--(0,2)--(2,2)--(2,0)--(0,0);
		\end{tikzpicture}
\end{equation}
The elements of $\Gamma '$ are listed in Figure \ref{tableDih4}, with some of their properties. Without entering into the details of the theory of Coxeter groups, let us note that the Weyl chambers in $\mathbb{R}^4$ are delimited by subspaces fixed by the order 2 elements in the group. Formally, the Weyl group of a simple Lie algebra, the principal Weyl chamber is defined as the set of charges $m$ which satisfy the inequalities
\begin{equation}
\label{fundChamber1}
    \alpha \cdot m \geq 0
\end{equation}
for every simple root $\alpha$. 
However, in the present case the order 2 elements don't necessarily fix a hyperplane in $\mathbb{R}^4$ (the $-1$ eigenspace can have dimension $> 1$). The condition (\ref{fundChamber1}) then has to be replaced by a more general condition, which we now explain on our example. We leave the study of the general case, and the connection with Coxeter group theory, for future work.

\begin{figure}[t]
    \centering
    \begin{tabular}{|c|c|c|c|}
     \hline 
        Permutation & Order  & $-1$ eigenspace & Inequality \\  \hline \hline
        Identity & 1  &  & \\
        $d_1 \leftrightarrow d_2$ & 2  &  $\left( \begin{array}{cccc}
            -1 & 1 & 0 & 0
        \end{array} \right)$ & $d_1 \leq d_2$ \\
        $e_1 \leftrightarrow e_2$ & 2  & $\left( \begin{array}{cccc}
            0 & 0 & -1 & 1
        \end{array} \right)$ & $e_1 \leq e_2$ \\
        $d_1 \leftrightarrow d_2$, $e_1 \leftrightarrow e_2$ & 2  & $\left( \begin{array}{cccc}
            -1 & 1 & 0 & 0 \\ 0 & 0 & -1 & 1
        \end{array} \right)$ & \begin{tabular}{c}
             $e_1 < e_2$ or   \\
            $e_1 = e_2$ and $d_1 \leq d_2$
        \end{tabular} \\
        $d_1 \leftrightarrow e_1$, $d_2 \leftrightarrow e_2$ & 2 & $\left( \begin{array}{cccc}
            -1 & 0 & 1 & 0 \\ 0 & -1 & 0 & 1
        \end{array} \right)$  &  \begin{tabular}{c}
             $d_2 < e_2$ or   \\
            $d_2 = e_2$ and $d_1 \leq e_1$
        \end{tabular} \\
        $d_1 \leftrightarrow e_2$, $d_2 \leftrightarrow e_1$   & 2 &  $\left( \begin{array}{cccc}
             0 & -1 & 1 & 0 \\ -1 & 0 & 0 & 1 
        \end{array} \right)$  &  \begin{tabular}{c}
             $d_1 < e_2$ or   \\
            $d_1 = e_2$ and $d_2 \leq e_1$
        \end{tabular} \\
        $d_1 \rightarrow e_1 \rightarrow d_2 \rightarrow e_2 \rightarrow d_1$ & 4 & & \\
        $d_1 \rightarrow e_2 \rightarrow d_2 \rightarrow e_1 \rightarrow d_1$ & 4 &  & \\ \hline 
    \end{tabular}
    \caption[Elements of $\mathrm{Dih}_4$]{Elements of the group $\mathrm{Dih}_4$. In the first column, they are presented as permutations, acting on $(d_1,d_2,e_1,e_2)$. The second column is the order of the element, the third gives a basis of the $-1$ eigenspace in the $(d_1,d_2,e_1,e_2)$ representation. The last column displays the condition imposed by (\ref{fundChamber2}).  }
    \label{tableDih4}
\end{figure}

The elements of order 2 in $W_{\Gamma}$ can be read from Figure \ref{tableDih4}. For every element $\alpha$ of order 2 in $W_{\Gamma}$, seen as a group of endomorphisms of its representation space, we pick a basis $(\delta^{\alpha}_i)$ of the kernel of this endomorphism in a consistent way (with $i=1 , \dots , \mathrm{dim \, ker} (\alpha)$). This is done in the third column of Figure \ref{tableDih4}. The principal Weyl chamber is then defined by 
\begin{equation}
\label{fundChamber2}
    \delta^{\alpha} \cdot m \geq 0 \, , 
\end{equation}
which generalizes (\ref{fundChamber1}). The subtlety in (\ref{fundChamber2}) comes from the cases where $\mathrm{dim \, ker} (\alpha) > 1$. When this is the case, $\delta^{\alpha} \cdot m$ is an element of $\mathbb{R}^{\mathrm{dim \, ker} (\alpha)}$ and we need to say what we mean by $\geq$. A simple choice, which we adopt here, is to pick the lexicographic order
\begin{equation}
    (x,y) \leq (x',y') \Leftrightarrow y<y' \textrm{ or } (y=y' \textrm{ and } x \leq x') \, . 
\end{equation}
Doing this for every order 2 element in Figure \ref{tableDih4}, we get the conditions listed in the last column of that figure. Combining all these conditions together, we obtain the summation range in (\ref{changeSum}).

\subsection{HWG for wreathed quivers}
\label{sectionHWG}

We now explain how to perform the orbifold at the level of the HWG. The starting point is the HWG of the Coulomb branch $\mathcal{C}$ of a quiver, which can be wreathed by a finite permutation group $\Gamma$. The goal is to compute the HWG for $\mathcal{C}/\Gamma$.

In the following, we give the general prescription, and at the same time we illustrate with three examples $\Gamma = \Z_2$, $\Z_3$ and $S_3$ to keep the discussion concrete. We first recall that the group $\Gamma$ has a well defined character table, which is a square matrix whose columns are labelled by conjugacy classes of elements of $\Gamma$, and whose rows are labelled by irreducible representations of $\Gamma$. For our three examples, these character tables are 
\begin{equation}
\label{charTable}
	\begin{array}{c|cc}
	\Gamma = \mathbb{Z}_2	& 1 & -1  \\ \hline
		\textrm{Cardinality}  & 1 & 1 \\ 
		\hline 
		\mathbf{1} & 1 & 1 \\
		\epsilon & 1& -1\\
	\end{array}
\end{equation}
\begin{equation}
\label{charTable2}
	\begin{array}{c|ccc}
	\Gamma = \mathbb{Z}_3	& 1 & \omega & \omega^2 \\  \hline
		\textrm{Cardinality}  & 1 & 1 & 1\\ 
		\hline 
		\mathbf{1} & 1 & 1 & 1\\
		\mathbf{f} & 1& \omega & \omega^2\\
		\overline{\mathbf{f}} & 1 &\omega^2 &\omega 
	\end{array}
	\qquad 
	\begin{array}{c|ccc}
\Gamma = S_3		& \mathrm{Id} & \textrm{3-cycles} & \textrm{2-cycles} \\  \hline
		\textrm{Cardinality}  & 1 & 2 & 3\\ 
		\hline \mathbf{1} & 1 & 1& 1\\
		\varepsilon & 1& 1&-1 \\
		\mathbf{2} &2 &-1 &0 
	\end{array}
\end{equation}
These character tables contain in each entry the trace of the matrices of the conjugacy class in the corresponding representation. One way to refine this information is to give, instead of the trace, the list (unordered, and with repetitions allowed) of the eigenvalues of these matrices. We will need these eigenvalues in equation (\ref{eqhwgquotient}). On our three examples, we get 
\begin{equation}
\label{charTableEigen2a}
	\begin{array}{c|cc}
	\Gamma = \mathbb{Z}_2	& 1 & -1  \\  \hline
		\textrm{Cardinality}  & 1 & 1 \\ 
		\hline 
		\mathbf{1} &\{1\} &\{1\}\\
		\epsilon &\{1\}& \{-1\}\\
	\end{array}
\end{equation}
\begin{equation}
\label{charTableEigen2b}
	\begin{array}{c|ccc}
	\Gamma = \mathbb{Z}_3	& 1 & \omega & \omega^2 \\  \hline
		\textrm{Cardinality}  & 1 & 1 & 1\\ 
		\hline 
		\mathbf{1} & \{1\} & \{1\} & \{1\}\\
		\mathbf{f} & \{1\}& \{\omega\} & \{\omega^2\} \\
		\overline{\mathbf{f}} & \{1\} & \{\omega^2\} & \{\omega\} 
	\end{array}
	\qquad 
	\begin{array}{c|ccc}
	\Gamma = S_3	& \mathrm{Id} & \textrm{3-cycles} & \textrm{2-cycles} \\  \hline
		\textrm{Cardinality} & 1 & 2 & 3\\ 
		\hline \mathbf{1} & \{1\} &\{1\}& \{1\}\\
		\varepsilon & \{1\}& \{1\}&\{-1\} \\
		\mathbf{2} &\{1,1\} &\{\omega , \omega^2\} & \{1,-1\} 
	\end{array}
\end{equation}
Of course in each case the sum of the eigenvalues listed in (\ref{charTableEigen2a}), (\ref{charTableEigen2b}) gives the characters (\ref{charTable}), (\ref{charTable2}). Let us call $C_j$ the conjugacy classes ($j=1,\dots,n$, with $n$ the number of conjugacy classes, and $C_1$ is the class of the identity element), $c_j = |C_j|$ their cardinalities, $\rho_i$ the irreducible representations ($i=1,\dots,n$, and $\rho_1$ is the trivial representation), and $d_i$ their dimensions. Finally we denote by $\Lambda_{i,j}$ the list of eigenvalues for $C_j$ in $\rho_i$. For a representation $R$ which is not irreducible, we similarly denote by $\Lambda_{R,j}$ the list of eigenvalues of the class $C_j$ in the representation $R$. The elements of $\Lambda_{R,j}$ are denoted $\lambda_{R,j}^k$ for $k=1,\dots, \mathrm{dim} \, R $. This list is easily obtained from the decomposition of $R$ in irreducible representations. Note that $\lambda_{R,1}^k=1$ for all $k$. 

We now show how to compute the HWG for an orbifold Coulomb branch based on an initial Coulomb branch that admits a \emph{finite} HWG. We say that $\mathrm{\HWG} ( \mathcal{C} )$ is finite is there exist two lists of monomials, that we denote $(M_1 , \dots , M_K)$ and $(M'_1 , \dots , M'_{K'})$, in the highest weight fugacities $(\mu_l)$ and the variable $t$, so that the HWG is
\begin{equation}
  \mathrm{\HWG} ( \mathcal{C} ) =    \frac{\prod\limits_{k'=1}^{K'} (1-M'_{k'})}{\prod\limits_{k=1}^{K} (1-M_{k})} \, . 
\end{equation}
We assume that $\mathrm{\HWG} ( \mathcal{C} )$ can be written in that way; this is a non-trivial assumption, as it is known that many Coulomb branches don't satisfy it. 

The fact that $\Gamma$ is a symmetry group translates into the fact that to the above expression are associated two representations $R$ and $R'$ of $\Gamma$, not necessarily irreducible, of respective dimensions $K$ and $K'$, such that the numerator and the denominator of the above expression transform according to these representations. Then $\mathrm{\HWG} ( \mathcal{C} )$ can be written 
\begin{equation}
  \mathrm{\HWG} ( \mathcal{C} ) =    \frac{\prod\limits_{k'=1}^{K'} (1- \lambda_{R',1}^{k'} M'_{k'})}{\prod\limits_{k=1}^{K} (1- \lambda_{R,1}^k M_{k})} \, . 
\end{equation}
From this expression it is then straightforward to deduce the conjectured HWG for the orbifold 
\begin{equation}
\label{eqhwgquotient}
    \HWG ( \mathcal{C} / \Gamma   ) = \frac{1}{|\Gamma|}  \sum\limits_{j=1}^n c_j \times   \frac{\prod\limits_{k'=1}^{K'} (1- \lambda_{R',j}^{k'} M'_{k'})}{\prod\limits_{k=1}^{K} (1- \lambda_{R,j}^k M_{k})} \, . 
\end{equation}

We illustrate how this formula works in practice on the example of the $D_4$ affine quiver. All HWGs and quivers are gathered in Figure \ref{table:g2-reductions}. Consider for instance the HWGs written in terms of $G_2$ fugacities. The closure of the minimal nilpotent orbit of $D_4$ has HWG equal to $\mathrm{PE} \left[ 2\mu_1 t^2 + \mu_2 t^2 + \mu_2 t^4 \right]$. The identification of the irreducible representations is as follows: 
\begin{eqnarray}
\mathbb{Z}_2 & \quad : \quad & \mathrm{\HWG} ( \mathcal{C}   ) =  \mathrm{PE} \left[ \mathbf{1} \mu_1 t^2 + \mathbf{\varepsilon} \mu_1 t^2 +  \mathbf{1}  \mu_2 t^2 + \mathbf{\varepsilon}  \mu_2 t^4 \right] \\ 
\mathbb{Z}_3 & \quad : \quad & \mathrm{\HWG} ( \mathcal{C}   ) =    \mathrm{PE} \left[ \mathbf{f} \mu_1 t^2 + \overline{\mathbf{f}} \mu_1 t^2 +  \mathbf{1}  \mu_2 t^2 + \mathbf{1}  \mu_2 t^4 \right] \\ 
S_3 & \quad : \quad & \mathrm{\HWG} ( \mathcal{C}   ) =   \mathrm{PE} \left[ \mathbf{2} \mu_1 t^2  +  \mathbf{1}  \mu_2 t^2 + \mathbf{\varepsilon}  \mu_2 t^4 \right] 
\end{eqnarray}
We then use equation (\ref{eqhwgquotient}) to obtain
\begin{eqnarray}
\mathrm{\HWG} ( \mathcal{C} / \mathbb{Z}_2  ) &=& \frac{1}{2} \left( \frac{1}{(1-\mu_1 t^2)^2(1-\mu_2 t^2)(1- \mu_2 t^4)} \right.  \nonumber \\ & & \qquad  \left.  +\frac{1}{(1-\mu_1 t^2)(1+\mu_1 t^2)(1-\mu_2 t^2)(1+ \mu_2 t^4)} \right)  \nonumber \\
&=& \frac{1- \mu_1^2\mu_2^2 t^{12}}{(1-\mu_1 t^2)(1-\mu_2 t^2)(1-\mu_1^2 t^4)(1- \mu_1\mu_2 t^6)(1- \mu_2^2 t^8)} \, . 
\end{eqnarray}
\begin{eqnarray}
\mathrm{\HWG} ( \mathcal{C} / \mathbb{Z}_3  ) &=& \frac{1}{3}  \sum\limits_{i=0}^2  \frac{1}{(1-\omega^i \mu_1 t^2)(1- \omega^{-i}\mu_1 t^2)(1-\mu_2 t^2)(1- \mu_2 t^4)}  \nonumber  \\ 
&=& \frac{(1- \mu_1^6 t^{12})}{(1-\mu_2 t^2)(1- \mu_2 t^4)(1- \mu_1^2 t^4)(1- \mu_1^3 t^6)^2} \, . 
\end{eqnarray}
\begin{eqnarray}
\mathrm{\HWG} ( \mathcal{C} / S_3  ) &=& \frac{1}{6} \left( \frac{1}{(1-\mu_1 t^2)^2(1-\mu_2 t^2)(1- \mu_2 t^4)} \right.  \nonumber \\ & & \qquad  \left.   + \frac{2}{(1- \omega \mu_1 t^2)(1-\omega^2 \mu_1 t^2)(1-\mu_2 t^2)(1- \mu_2 t^4)}  \right.   \nonumber\\ & & \qquad  \left. + \frac{3}{(1-\mu_1 t^2)(1 +  \mu_1 t^2)(1-\mu_2 t^2)(1 + \mu_2 t^4)} \right)   \nonumber \\ 
&=& \frac{1-\mu_1^6 \mu_2^2 t^{20}}{(1-\mu_2 t^2)(1-\mu_1^2 t^4)(1-\mu_1^3 t^6 )(1-\mu_2^2 t^8 )(1-\mu_1^3 \mu_2 t^{10} )} \, . 
\end{eqnarray}
This reproduces the results in \cite{hanany_ungauging_2020}.

\paragraph{Eighth case of Table \ref{tabKostantBrylinski}}
We can apply similar methods to the eighth line of Table \ref{tabKostantBrylinski}. The HWG for the minimal nilpotent orbit of $F_4$, written in terms of $D_4$ fugacities, is $\mathrm{PE}[(\mu_1 + \mu_2 + \mu_3 + \mu_4 )t^2]$. The weights $\mu_1$, $\mu_3$ and $\mu_4$ correspond to the external nodes of the Dynkin diagram. In order to perform the $\mathbb{Z}_2^2$ quotient, we charge them under the three distinct $\mathbb{Z}_2$ subgroups and apply formula (\ref{eqhwgquotient}). This way one gets the HWG 
\begin{equation}
    \frac{1}{4}\sum\limits_{\epsilon_1 = \pm 1 } \sum\limits_{\epsilon_2 = \pm 1 } \frac{1}{(1-\epsilon_1 \epsilon_2 \mu_1 t^2)(1-  \mu_2 t^2)(1-\epsilon_2   \mu_3 t^2)(1-\epsilon_1   \mu_4 t^2)}
\end{equation}
which evaluates to $\mathrm{PE}[ \mu_2 t^2 + (\mu_1^2 + \mu_3^2 + \mu_4^2) t^4 + \mu_1 \mu_3 \mu_4 t^6 -  \mu_1^2 \mu_3^2 \mu_4^2 t^{12}]$. One can check that this is indeed the HWG for the closure of the $[3,2^2,1]$ orbit of $\mathfrak{so}(8)$. An alternative way of seeing the same computation relies on the fact that $\mathbb{C}^3/\mathbb{Z}_2^2$ is a weighted hypersurface in $\mathbb{C}^4$.

\subsection{Higgs branch of wreathed quivers}\label{sec:higgs-symm}

In this subsection, we turn to the Higgs branch of wreathed quivers. This is in contrast with the rest of the paper, which focuses on the Coulomb branch of the $3d$ $\mathcal{N}=4$ theories, but it serves several purposes. First, it demonstrates that wreathed quivers do indeed provide a well-defined hyper-K\"{a}hler quotient, which can be associated with a gauge theory whose gauge group is disconnected. Secondly, we explain how to compute the Hilbert series of such quivers, using an averaging procedure. Finally, it allows the study of the geometric action of wreathing on the Higgs branch and contrasts it with the parallel action on the Coulomb branch. 

We focus on a simple but rich example, the affine $D_4$ quiver, and compute the Higgs branch of all the wreathed quivers that appear in Figure \ref{figHasseS4}. 
Let $\Gamma$ be a subgroup of $S_4$. We consider the wreathed quiver defined by this group acting on the four $\mathrm{U}(1)$ nodes. This produces (when $\Gamma$ is non-trivial) a disconnected gauge group, as follows directly from the definition (\ref{definitionWreath}). Disconnected gauge groups have been considered in the context of the plethystic program in \cite{bourget_non-connected_2017}, where groups were extended by outer automorphisms, following a formula of Wendt \cite{wendt_weyls_2001}. Here the context is different but the techniques spelled out in \cite{bourget_non-connected_2017} apply. In fact, the case considered here is particularly easy to handle because the groups which are being wreathed are all $\mathrm{U}(1)$ groups, therefore the Haar measure is not modified. We pick fugacities $z_i$ ($i=1,2,3,4$) for the $\mathrm{U}(1)$ factors and fugacity $y$ for the $\mathrm{U}(2)$ factor (after ungauging a diagonal $\mathrm{U}(1)$). It follows that the Higgs branch Hilbert series is obtained via a Molien-Weyl integral which is written explicitly as 
\begin{equation}
\label{formulaHiggs}
    \HS^{\Higgs}_{\Gamma} (t) = \frac{1}{|\Gamma|} \sum\limits_{\gamma \in \Gamma} \int_{z_i,y} \mathrm{d} \mu (z_i,y) \mathcal{F}^\flat (z_i , y , t , \gamma) \, , 
\end{equation}
where the measure is 
\begin{equation}
     \mathrm{d} \mu (z_i,y) = \frac{\mathrm{d} z_1}{2 \pi i z_1 } \frac{\mathrm{d} z_2}{2 \pi i z_2 } \frac{\mathrm{d} z_3}{2 \pi i z_3 } \frac{\mathrm{d} z_4}{2 \pi i z_4 }  \frac{(1-y^2) \mathrm{d} y}{2 \pi i y }
\end{equation}
and 
\begin{equation}
    \mathcal{F}^\flat (z_i , y , t , \gamma) = \frac{\det \left( \mathbf{1}_4 - \gamma t^2 \right) (1-t^2)(1-t^2 y^2)(1-t^2 y^{-2})}{\det \left( \mathbf{1}_4 - \gamma t y D \right)\det \left( \mathbf{1}_4 - \gamma t y^{-1} D \right)\det \left( \mathbf{1}_4 - \gamma t y D^{-1} \right)\det \left( \mathbf{1}_4 - \gamma t y^{-1} D^{-1} \right)}
\end{equation}
with $D$ the diagonal matrix $\mathrm{Diag}(z_1,z_2,z_3,z_4)$. The integral over the $z_i$ and $y$ fugacities are performed over the contours $|z_i|=|y|=1$. Note that (\ref{formulaHiggs}) makes it manifest that $\gamma \in \Gamma$ can be considered as a discrete fugacity for the disconnected gauge group $\mathrm{U}(1) \wr \Gamma$. The integrals (\ref{formulaHiggs}) are readily evaluated for each of the 11 subgroups of $S_4$, and the resulting Hilbert series are presented in Figure \ref{figHasseS4Higgs}. 

We make a few comments on the results. First, the Hilbert series coincide with those of Du Val singularities $\mathbb{C}^2/J$, with $J$ a finite subgroup of $\mathrm{SU}(2)$, of ADE type. Specifically, four instances occur, namely $J=D_4,D_6,E_6,E_7$, that can be identified using the degrees of invariants of the corresponding groups. In particular, this shows that the quaternionic dimension of the Higgs branches of all these quivers is 1. 

\begin{landscape}
\begin{figure}
    \centering
\vspace*{-1.5cm}\hspace*{-3cm}\begin{tikzpicture}
			\node[draw] (S4) at (0,-2) {\begin{tabular}{c}
			     24 \hfill $S_4$ \hfill $E_7$ \\$\qquad \mathrm{PE}[t^8+t^{12}+t^{18}-t^{36}]\qquad $	\end{tabular} }; 
			\node[draw] (A4) at (-1,1) {\begin{tabular}{c}
			     12 \hfill $A_4$  \hfill $E_6$ \\$\qquad \mathrm{PE}[t^6+t^8+t^{12}-t^{24}]\qquad $	\end{tabular} }; 
			\node[draw] (D4) at (-9,1) {\begin{tabular}{c}
			     8 \hfill $\mathrm{Dih}_4$ \hfill $D_6$ \\$\qquad \mathrm{PE}[t^4+t^8+t^{10}-t^{20}]\qquad $	\end{tabular} }; 
			\node[draw] (S3) at (6,2.5) {\begin{tabular}{c}
			     6 \hfill $S_3$ \hfill $E_7$ \\$\qquad \mathrm{PE}[t^8+t^{12}+t^{18}-t^{36}]\qquad $	\end{tabular} }; 
			\node[draw] (Z4) at (-15,7) {\begin{tabular}{c}
			    4 \hfill  $\mathbb{Z}_4$ \hfill $D_6$ \\$\qquad \mathrm{PE}[t^4+t^8+t^{10}-t^{20}]\qquad $	\end{tabular} }; 
			\node[draw] (NK) at (-9,5) {\begin{tabular}{c}
			     4 \hfill Normal Klein  \hfill $D_4$  \\$\qquad \mathrm{PE}[2 t^4+t^6-t^{12}]\qquad $	\end{tabular} }; 
			\node[draw] (NNK) at (-3,7) {\begin{tabular}{c}
			    4 \hfill  Non-normal Klein \hfill $D_6$  \\$\qquad \mathrm{PE}[t^4+t^8+t^{10}-t^{20}] \qquad$	\end{tabular} }; 
			\node[draw] (Z3) at (6,8.5) {\begin{tabular}{c}
			    3 \hfill $\mathbb{Z}_3$ \hfill $E_6$ \\$\qquad \mathrm{PE}[t^6+t^8+t^{12}-t^{24}]\qquad $	\end{tabular} }; 
			\node[draw] (DT) at (-9,10) {\begin{tabular}{c}
			     2 \hfill Double Transposition \hfill $D_4$ \\  $\qquad \mathrm{PE}[2 t^4+t^6-t^{12}]\qquad $
			\end{tabular} }; 
			\node[draw] (S2) at (-1,10) {\begin{tabular}{c}
			    2 \hfill $S_2$ \hfill $D_6$ \\$\qquad \mathrm{PE}[t^4+t^8+t^{10}-t^{20}]\qquad $ \end{tabular} }; 
			\node[draw] (S1) at (0,13) {\begin{tabular}{c}
			    1 \hfill Trivial \hfill $D_4$ \\$\qquad \mathrm{PE}[2 t^4+t^6-t^{12}]\qquad $ \end{tabular} }; 
			\draw (S4)--(D4)--(Z4)--(DT)--(NK)--(D4)--(NNK)--(DT)--(S1)--(S2)--(S3)--(S4);
			\draw (NK)--(A4)--(S4);
			\draw (A4)--(Z3)--(S1);
			\draw (S2)--(NNK);
			\draw (S3)--(Z3);
\end{tikzpicture}
    \caption[Higgs branches]{Hasse diagram of the 11 conjugacy classes of subgroups of $S_4$ with the Hilbert series for the Higgs branch of the corresponding wreathed affine $D_4$ quiver. In each box, the number on the left is the order of the group $\Gamma$, and the group on the right denotes the ADE type of the finite subgroup $J$ of $\mathrm{SU}(2)$ such that the Higgs branch is $\mathbb{C}^2/J$.   }
    \label{figHasseS4Higgs}
\end{figure}
\end{landscape}

\paragraph{Gauge invariant operators}
As a check of the computations presented in Figure \ref{figHasseS4Higgs}, we briefly show how the same results can be obtained from a counting of invariant operators. We call $A_i$ and $B_i$ the scalars in the chiral multiplets transforming as bifundamentals of $\mathrm{U}(2)$ and $\mathrm{U}(1)_i$, for $i=1,2,3,4$, $A_i$ being a column vector and $B_i$ being a row vector. For simplicity, we ungauge one of the $\mathrm{U}(1)$ groups, say $\mathrm{U}(1)_4$, and study the action of the wreath product by a subgroup $\Gamma$ of $S_3$ permuting the three remaining $\mathrm{U}(1)$ gauge groups. 

The F-term equations on $\mathrm{U}(1)_i$ are 
\begin{equation}
\label{fterms1}
  \textrm{For } i=1,2,3 , \qquad   B_i A_i = 0 \, . 
\end{equation}
The F-term equations on the $\mathrm{U}(2)$ group are 
\begin{equation}
\label{fterms2}
    \sum\limits_{i=1}^4 A_i B_i = 0 \, . 
\end{equation}
Taking the trace of (\ref{fterms2}) and combining with (\ref{fterms1}) we obtain 
\begin{equation}
\label{fterms3}
B_4 A_4 = 0 \, . 
\end{equation}
Gauge invariants are paths in the quiver of the form $B_4 \alpha_{i_1} \cdots \alpha_{i_r} A_4$ subject to the relations above, using the shorthand notation $\alpha_i = A_i B_i$. An irreducible gauge invariant is one that can not be written as a product of other non-trivial gauge invariants, so it can be written $B_4 \alpha_{i_1} \cdots \alpha_{i_r} A_4$ where the indices can not take the value $4$. The F-term relations imply that 
\begin{equation}
    \alpha_i \alpha_i = 0 \textrm{  and  } \sum\limits_{i=1}^4 \alpha_i = 0 \, . 
\end{equation}
In particular an irreducible gauge invariant can not contain three $\alpha_i$'s or more.\footnote{Consider for instance $B_4 \alpha_i \alpha_j \alpha_k A_4$ with $i\neq j$, $j \neq k$ and $i,j,k \neq 4$. If $i \neq k$ then one finds $B_4 \alpha_i \alpha_j \alpha_k A_4 = - B_4 \alpha_i (\alpha_i + \alpha_k + \alpha_4) \alpha_k A_4 = B_4 \alpha_i \alpha_4 \alpha_k A_4 = (B_4 \alpha_i A_4)(B_4 \alpha_k A_4)$. If $i=k$ then $B_4 \alpha_i \alpha_j \alpha_i A_4 = - B_4 \alpha_i \alpha_j \alpha_l A_4$ with $l \neq i,j,4$ and we're back in the previous case. } So generators of the Higgs branch coordinate ring contain either one or two $\alpha_i$'s. The generators containing one $\alpha_i$ are $X_i = B_4 \alpha_i A_4$ ($i=1,2,3$) subjected to $X_1+X_2+X_3 = 0$, and transform in the irreducible two-dimensional representations of $S_3$. The generators with two $\alpha_i$'s are built from $Y_{ij} = B_4 \alpha_i \alpha_j A_4$. Note that $Y_{ij}= -Y_{ji}$ and that $Y_{12}=Y_{23}=Y_{31}$, which shows that the $Y_{ij}$ transform in the $\varepsilon$ representation of $S_3$. Finally, there is a relation between the two families, for instance in the form
\begin{equation}
    X_1 X_2 X_3 = B_4 \alpha_1 \alpha_4 \alpha_2 \alpha_4 \alpha_3 A_4 = B_4 \alpha_1 \alpha_3 \alpha_2 \alpha_1 \alpha_3 A_4 = -Y_{12}^2 \, . 
\end{equation}
Putting all this together, we obtain the Hilbert series $\mathrm{PE}[2t^4 + t^6 - t^{12}]$ for the affine $D_4$ quiver (the $X_i$ have weight 4 while the $Y_{ij}$ have weight 6). 
To deal with the wreathed quivers, we have to impose the additional gauge invariance under the discrete factor $\Gamma$. The spectrum of operators on the Higgs branch is a subset of the one determined above for trivial $\Gamma$. The results are gathered in Figure \ref{tableHiggsBranchesEq}.

\begin{figure}
\begin{equation*}
    \begin{array}{|c|c|c|c|} \hline 
   \textrm{Group } \Gamma & \textrm{Generators} & \textrm{Relation} & \textrm{Geometry} \\ \hline 
        S_1 & \begin{array}{ccc}
           t^4 : & \qquad & x =  X_1  \\
           t^4 : & \qquad & y = X_2   \\
           t^6 : & \qquad & z = Y_{12} 
        \end{array}  & xy(x+y) = z^2 &  \mathbb{C}^2/D_4\\ \hline 
        S_2 & \begin{array}{ccc}
           t^4 : & \qquad & x =  X_1+X_2  \\
           t^8 : & \qquad & y =  X_1 X_2    \\
           t^{10} : & \qquad & z = Y_{12}(X_1-X_2) 
        \end{array}  & xy(x^2-4y) = z^2  &  \mathbb{C}^2/D_6  \\ \hline 
        \mathbb{Z}_3 & \begin{array}{ccc}
           t^6 : & \qquad & x =   Y_{12}  \\
           t^8 : & \qquad & y =  X_1^2 + X_1 X_2 + X_2^2    \\
           t^{12} : & \qquad & z = (X_1 - X_2) (2 X_1 + X_2) (X_1 + 2 X_2)
        \end{array}   &  27x^4-4y^3 + z^2 = 0  &  \mathbb{C}^2/E_6 \\ \hline 
        S_3 &\begin{array}{ccc}
           t^8 : & \qquad & x =  X_1^2 + X_1 X_2 + X_2^2     \\
           t^{12} : & \qquad & y =  Y_{12}^2    \\
           t^{18} : & \qquad & z = Y_{12}(X_1 - X_2) (2 X_1 + X_2) (X_1 + 2 X_2)
        \end{array}   &  -4 x^3 y + 27 y^3 + z^2 = 0   &  \mathbb{C}^2/E_7 \\ \hline 
    \end{array}
\end{equation*}
    \caption{Generators and relations for operators on the Higgs branch of the affine $D_4$ quiver wreathed by subgroups of $S_3$. }
    \label{tableHiggsBranchesEq}
\end{figure}

\paragraph{Comparison with adjoint matter}

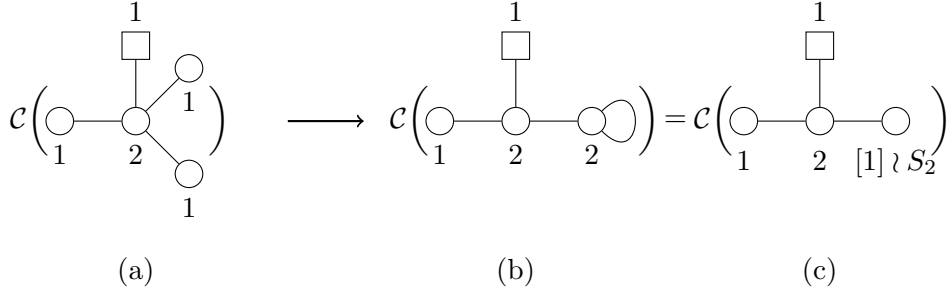
\begin{figure}[t]
    \centering
    \begin{tikzpicture}
		\node (coul1a) at (-0.4,0) {$\Coul \bigg($};
		\node (d4g1) at (0,0) [gauge, label=below:{1}]{}; 
		\node (d4g2) at (1,0) [gauge, label=below:{2}]{}; 
		\node (d4g3) at (1.7,0.7) [gauge, label=below:{1}]{}; 
		\node (d4g4) at (1.7,-0.7) [gauge, label=below:{1}]{}; 
		\node (d4f2) at (1,1) [flavor, label=above:{1}]{}; 
		\node (coul1b) at (2.1,0) {$\bigg)$};
		\draw (d4g1)--(d4g2)--(d4f2); 
		\draw (d4g3)--(d4g2)--(d4g4);

		\draw[->,thick] (3,0) -- (4,0);
		
		\node (coul2a) at (4.6,0) {$\Coul \bigg($};
		\node (loopg1) at (5,0) [gauge, label=below:{1}]{}; 
		\node (loopg2) at (6,0) [gauge, label=below:{2}]{}; 
		\node (loopg3) at (7,0) [gauge, label=below:{2}]{};
		\node (loopf2) at (6,1) [flavor, label=above:{1}]{}; 
		\node (coul2b) at (7.75,0) {$\bigg)$};
		\draw (loopg1)--(loopg2)--(loopf2); 
		\draw (loopg3)--(loopg2);
		\draw (loopg3) to [out=45,in=315,looseness=8] (loopg3);
		
		\node (or) at (8.1,0) {$=$};
		
		\node (coul3a) at (8.6,0) {$\Coul \bigg($};
		\node (symg1) at (9,0) [gauge, label=below:{\symmlabel{1}}]{}; 
		\node (symg2) at (10,0) [gauge, label=below:{\symmlabel{2}}]{}; 
		\node (symg3) at (11,0) [gauge, label=below:{\symmlabel{[1]\wr S_2}}]{};
		\node (symf2) at (10,1) [flavor, label=above:{1}]{}; 
		\node (coul3b) at (11.6,0) {$\bigg)$};
		\draw (symg1)--(symg2)--(symf2); 
		\draw (symg3)--(symg2);

		\node at (1,-2) []{(a)}; 
		\node at (6,-2) []{(b)}; 
		\node at (10,-2) []{(c)}; 
    \end{tikzpicture}
    \caption[$S_2$ orbifold of the affine $D_4$ quiver's Coulomb branch as the Coulomb branch of two distinct quivers]{The Coulomb branch of the $D_4$ quiver (left) is orbifolded by an $S_2$ action into a Coulomb branch shared by two distinct quivers.}
    \label{fig:d4-dg}
\end{figure}

Consider the case depicted in Fig. \ref{fig:d4-dg}. In \cite{hanany_discrete_2018,hanany_discrete_2018-1} it was pointed out that the Coulomb branch of the quiver (b) is an orbifold of the Coulomb branch of the quiver (a). We have argued that the Coulomb branch of the wreathed quiver (c) is also that very same orbifold. Let's look at the Higgs branch of quiver (b). 

The (quaternionic) dimension of the Higgs branch, when there is complete Higgsing, which is the case here, is equal to the number of matter multiplets minus the number of gauge multiplets. The $D_4$ quiver therefore has $\dim\ \Higgs_{D_4} = (4 \cdot 2 \cdot 1) - (3\cdot 1 + 2^2)=1$, as do all the wreathed quivers. The quiver (b) has a Higgs branch of quaternionic dimension $\dim\ \Higgs^{U(2)}_{\mathrm{loop}}=(2\cdot 2 \cdot 1 + 2 \cdot 2 + 2^2)-(1+2 \cdot 2^2)=3$, of which one dimension is a free factor $\mathbb{H}$ from the trivial factor in the adjoint loop. We can be more precise and compute the Hilbert series using the hyper-K\"{a}hler quotient, finding 
\begin{equation}
    \mathrm{PE}[2t] \mathrm{PE}[3 t^2 + 2 t^5 - t^{12}] \, . 
\end{equation}
The first term comes from a free contribution $\mathbb{H}$ which can be discarded. The second term can be identified as the Hilbert series for an intersection of a Slodowy slice and the nilpotent cone in the $C_3$ algebra, namely the transverse slice between the maximal orbit (of dimension 9) and the $\mathcal{O}_{[4,1^2]}$ orbit of dimension 7, see Table 12 in \cite{hanany_quiver_2020} (labelled  $[210]$ therein). The global symmetry on this space is $Sp(1)$ under which the generators of the chiral ring transform in the [2] and the [1] representations, respectively. This space makes a rare occurrence of a symplectic singularity which is also a hypersurface in $\mathbb{C}^5$. In fact it has been suggested that all hypersurface symplectic singularities of dimension 2 are intersections of Slodowy slices of the nilpotent orbit $\mathcal{O}_{[2n-2,1^2]}$ and the nilpotent cone in $C_n$ \cite{yamagishi_four-dimensional_2019}. This family appears in the context of trivertex theories where the rank of $C_n$ is interpreted as the genus of a Riemann surface ($A_1$ class S theory on a Riemann surface of genus $n$ and one puncture). See section 7.2 of \cite{hanany_tri-vertices_2011} and equation (7.12) for the hypersurface equation. The same family also appears as a Coulomb branch of the mirror quiver in the work of \cite{goncharov_coulomb_2019} where the identification as a transverse slice is made, as well as an explicit form of the hypersurface equation. The Hasse diagram is  
\begin{equation}
\raisebox{-.5\height}{\begin{tikzpicture}
		\tikzstyle{hasse} = [circle, fill,inner sep=2pt];
		\node [hasse] (1) [label=right:\footnotesize{$2$}] {};
		\node [hasse] (2) [below of=1, label=right:\footnotesize{$1$}] {};
		\node [hasse] (3) [below of=2, label=right:\footnotesize{$0$}] {};
		\draw (1) edge [] node[label=left:\footnotesize{$D_{n+1}$}] {} (2);
		\draw (2) edge [] node[label=left:\footnotesize{$A_1$}] {} (3);
	\end{tikzpicture} }
\end{equation}
  
In summary, the two quivers on the right of Figure \ref{fig:d4-dg} share the same Coulomb branch, but only the wreathed quiver's Higgs branch shares the original quiver's Higgs branch dimension, as one would expect from discrete gauging.

\subsection{Mirror symmetry and discrete gauging}

\begin{figure}[t]
    \centering
    \begin{tabular}{|c|c||c|c|} \hline 
        Coulomb Quiver & Discretely Gauged & Higgs Quiver & Discretely Gauged \\ \hline \hline
        \begin{tikzpicture}
			\node (g1) [gauge, label=below:{1}]{}; 
			\node (gn-2) [gauge, right of=g1, label=below:{2}]{}; 
			\node (gn-1) [gauge, above right of=gn-2, label=below:{1}]{}; 
			\node (gn) [gauge, below right of=gn-2, label=below:{1}]{}; 
			\node (f2) [flavor, above of=gn-2,label=above:{1}]{}; 
			\draw (g1)--(gn-2); 
			\draw (gn-2)--(gn-1); 
			\draw (gn-2)--(gn); 
			\draw (gn-2)--(f2); 
		\end{tikzpicture} &  \begin{tikzpicture}
			\node (g0) at (1,1) [flavor, label=left:{1}]{};
			\node (g1) at (0,0) [gauge, label=below:{\symmlabel{1}}]{}; 
			\node (g2) at (1,0) [gauge, label=below:{\symmlabel{2}}]{};
			\node (g3) at (2,0) [gauge, label=below:{\symmlabel{[1] \wr S_2}}]{};
			\draw (g1)--(g2)--(g3);
			\draw (g0)--(g2);
		\end{tikzpicture} & \begin{tikzpicture}
			\node (g0) at (0,1) [flavor, label=left:{$O_8$}]{};
			\node (g1) at (0,0) [gauge, label=below:{$C_1$}]{}; 
			\draw (g0)--(g1);
		\end{tikzpicture} & \begin{tikzpicture}
			\node (g0) at (0,1) [flavor, label=left:{$O_7$}]{};
			\node (g1) at (0,0) [gauge, label=below:{$C_1$}]{}; 
			\node (g2) at (1,0) [gauge, label=below:{$O_1$}]{}; 
			\draw (g0)--(g1)--(g2);
		\end{tikzpicture} \\  
		 $\Coul=\overline{\mathrm{min}\ D_4}$ & $\Coul=\overline{\mathrm{n.min}\ B_3} = \frac{\overline{\mathrm{min}\ D_4}}{S_2}$ &
		$\Higgs=\overline{\mathrm{min}\ D_4}$ & $\Higgs=\overline{\mathrm{n.min}\ B_3} = \frac{\overline{\mathrm{min}\ D_4}}{S_2}$ \\[5pt]
	\hline 
    \end{tabular}
    \caption[Discrete ungauging on electric and magnetic quivers]{Illustration of the relation between i) discrete gauging's effects on the Coulomb branch and ii) discrete gauging's effects on the Higgs branch of a corresponding electric quiver.}
    \label{table:higgs}
\end{figure}

Many $3d\ \mathcal{N}=4$ quiver theories admit a dual description as a theory whose Higgs branch is the original's Coulomb branch and vice versa; this property is known as \emph{$3d$ mirror symmetry} \cite{intriligator_mirror_1996,feng_mirror_2000,gaiotto_s-duality_2009} and is a consequence of S-duality for theories with brane realisations. One should therefore expect to be able to find the mirror dual of discrete gauging. As it turns out, it is already known.

Let us consider the paradigmatic case of quivers in Figure \ref{table:higgs}. The Coulomb branch of the quiver in the first column is the minimal nilpotent orbit of $D_4$. Its dual is depicted in the third column of the same figure; the symmetry of its Higgs branch is the same as the symmetry on the flavor node. Each matter hypermultiplet is coupled to a mass which can be viewed as a background vector multiplet. This vector can in turn be gauged, turning the quiver into the one depicted in the fourth column; such an operation was first reported as ``the case $O(1)$" in \cite{kobak_classical_1996}. The new gauge node $O(1) \cong \mathbb{Z}_2 \cong S_2$ represents the discrete symmetry of the gauged vector. In this case the gauge group is \emph{enlarged}. We claim this is the mirror dual of the process covered in the previous section. Somewhat confusingly, both procedures are called discrete gauging\footnote{We are not aware of a physics reference for discrete gauging on the Higgs branch but believe it to be fairly well known among physicists interested in Higgs branches of quiver gauge theories.} but they act differently. On the left quiver an automorphism is gauged; on the right we gauge a background vector.

If the enhancement of the mirror is discrete, so must be the original's. Moreover, since discrete gauging of background vectors is a genuine action on quiver theories, so is its mirror dual.

\section{Quiver folding}
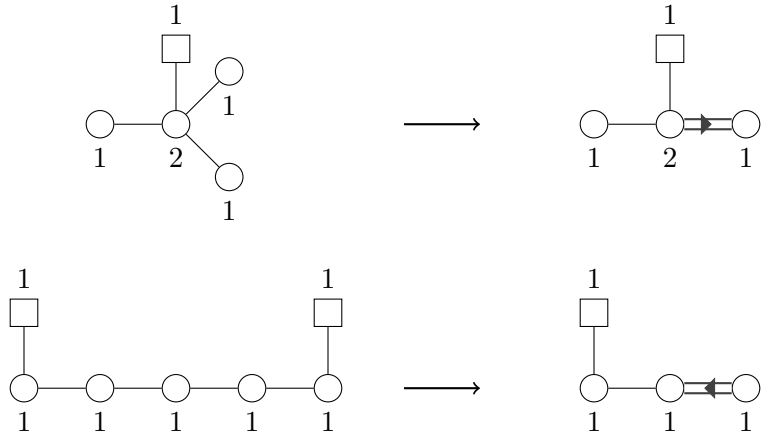
\begin{figure}
	[t] 
\centering 
	\begin{tikzpicture}
		\node (dg1) at (-1,0) [gauge, label=below:{1}]{}; 
		\node (dg2) at (0,0) [gauge, label=below:{2}]{}; 
		\node (dg3) at (.7,.7) [gauge, label=below:{1}]{};
		\node (dg4) at (.7,-.7) [gauge, label=below:{1}]{}; 
		\node (df2) at (0,1) [flavor, label=above:{1}]{}; 
		\draw (dg1)--(dg2); 
		\draw (dg2)--(dg3); 
		\draw (dg2)--(dg4); 
		\draw (dg2)--(df2); 
		
		\node (bg1) at (5.5,0) [gauge, label=below:{1}]{}; 
		\node (bg2) at (6.5,0) [gauge, label=below:{2}]{}; 
		\node (bg3) at (7.5,0) [gauge, label=below:{1}]{}; 
		\node (bf2) at (6.5,1) [flavor, label=above:{1}]{}; 
		\draw (bg1)--(bg2); 
		\draw[doublearrow] (bg2)-- node {\midarrow} (bg3); 
		\draw (bg2)--(bf2); 
		
		\draw[->,thick] (3,0) -- (4,0);
		
		\node (ag1) at (-2,-3.5) [gauge, label=below:{1}]{}; 
		\node (ag2) at (-1,-3.5) [gauge, label=below:{1}]{}; 
		\node (ag3) at (0,-3.5) [gauge, label=below:{1}]{}; 
		\node (ag4) at (1,-3.5) [gauge, label=below:{1}]{}; 
		\node (ag5) at (2,-3.5)  [gauge, label=below:{1}]{}; 
		\node (af1) at (-2,-2.5) [flavor, label=above:{1}]{}; 
		\node (af5) at (2,-2.5) [flavor, label=above:{1}]{}; 
		\draw (ag1)--(ag2)--(ag3)--(ag4)--(ag5); 
		\draw (ag1)--(af1); 
		\draw (ag5)--(af5); 
		
		\node (cg1) at (5.5,-3.5) [gauge, label=below:{1}]{}; 
		\node (cg2) at (6.5,-3.5) [gauge, label=below:{1}]{}; 
		\node (cg3) at (7.5,-3.5) [gauge, label=below:{1}]{}; 
		\node (cf1) at (5.5,-2.5) [flavor, label=above:{1}]{}; 
		\draw (cg1)--(cg2); 
		\draw (cg1)--(cf1); 
		\draw[doublearrow] (cg3)-- node {\midarrowrev} (cg2); 
		
		\draw[->, thick] (3,-3.5) -- (4,-3.5);
	\end{tikzpicture}
	\caption[Folding examples]{Quivers on the left fold into quivers on the right.} \label{fig:folding-examples}
\end{figure}

The next discrete operation allows for a natural interpretation of non–simply laced quivers, which were identified in \cite{cremonesi_coulomb_2014} through the use of the monopole formula. It was already well established \cite{cremonesi_monopole_2014} that many $ADE$ nilpotent orbits could be recovered as Coulomb branches of unitary quiver theories and that there is a robust connection between choice of quiver and the resulting nilpotent orbit. In particular, the quiver should be balanced and shaped like the desired symmetry algebra's Dynkin diagram. Consequently one might assume that quivers whose Coulomb branches reproduce $BCFG$ nilpotent orbits would resemble the non-simply laced $BCFG$ Dynkin diagrams. \cite{cremonesi_coulomb_2014} conjectured a minimal modification to the monopole formula which reflected the enigmatic multiple link, checking against earlier tentative results of \cite{hanany_hilbert_2013} on $F_4$ and $G_2$ spaces. Although the conjecture was highly successful in its goal, giving support to the existence of non-simply laced quivers and allowing further study \cite{dey_three-dimensional_2017}, precise details of multiple links remained elusive\footnote{According to \cite{cremonesi_coulomb_2014}, Jan Troost suggested that quivers of this type might be understood as folded simply laced quivers, an idea that ultimately finds validation in \cite{nakajima_coulomb_2019} and our results.}. They have made an appearance in the study of little string theory \cite{haouzi_abcdefg_2017} or gauge-vortex duality \cite{haouzi_supersymmetric_2019} and $W$-algebras associated to them were studied in \cite{kimura_fractional_2018}. A mathematical treatment of folding and Coulomb branches of non-simply laced quivers was recently provided in \cite{nakajima_coulomb_2019}. Some of the phenomena in \cite{zafrir_compactifications_2017,ohmori_compactifications_2019} can be reinterpreted as folding the five-dimensional theories' magnetic quivers \cite{cabrera_magnetic_2019}. 

In this section we show (using an alternative approach to \cite{nakajima_coulomb_2019}) that the multiple link can be interpreted as the result of \emph{quiver folding}; see Fig. \ref{fig:folding-examples} for examples. We first utilise abelianisation to show that Coulomb branches of $A_{2n-1}$ ($D_{n+1}$) quivers fold into spaces with $C_n$ ($B_n$) symmetry and derive the effects of folding on the monopole formula. We then reinterpret folding as an action on the quiver itself, showing that it produces non-simply laced quivers; in particular, our analysis of the monopole formula on examples reproduces the form in \cite{cremonesi_coulomb_2014,nakajima_coulomb_2019}.

Note that the examples below focus on nilpotent orbit quivers only because they are most easily studied using tools we have developed. We expect folding to be a completely general operation. For example, the quiver of Section 4.1.2 in \cite{hanany_hwg_2017} folds into the quiver in (7.1) of \cite{hanany_ungauging_2020} as can be guessed by mapping $\mu_{2N-i}\mapsto\mu_i$ for $i < N$ in the former's HWG and comparing to the HWG of the latter quiver.

\subsection{Action on the Coulomb branch}

Although one can fold a quiver directly, the operation can also be performed on a discretely gauged Coulomb branch. The prerequisites for folding and discrete gauging are identical: a quiver with an automorphism. We start yet again with the example of a $D_4$ quiver in the bottom left of Fig. \ref{fig:folding-examples}. Recall that in (\ref{eq:B3-2}) the discretely gauged quiver's operator $\te{0,0,1}$ is defined as $\e{0,0,1,0} + \e{0,0,0,1} = \up{3}+\up{4}$. A space is folded by restricting it to the subspace fixed under the action of the symmetry, which in this case generates the constraints $\up{3}=\up{4}$ as well as $\p{3}=\p{4}$ and so on; we denote this space $\hatCoul$ and in general use hats to denote variables on the folded space. Note that mass parameters must be set to identical values across folded legs; sometimes this removes all independent mass parameters but one and, as a result, even though the original space is mass-deformable, the folded space is not.

As long as we stay on $\hatCoul$ there is no more need to track each individual wreathed variable. To reduce to a minimal necessary set we introduce the \emph{folding map} 
\begin{align}
	F(\x{i,a}) &= \frac{\hatx{I,a}}{\#_i} \\
	F(x+y) &= F(x) + F(y) \\
	F(c x^m y^n) &= c F(x)^m F(y)^n
\end{align}
where the \emph{multiplicity} $\#_i$ denotes the number of nodes that fold onto the same node as node $i$, $x$ and $y$ are arbitrary operators, $c$ is a complex number and $I = \min_{j} \lbrace j : \pi(j) = \pi(i) \rbrace$. In particular, $F(\up{3}) = F(\up{4}) = \frac{\hatup{3}}{2}$. As a result, $F(\te{0,0,1}) = \hatup{3} = \hate{0,0,1}$. 

The folding map has a simple interpretation. Abelianised variables of the initial, unfolded quiver, partition into orbits of the automorphism. The folding map merely sets every single variable in that orbit to the same value; for convenience, basic abelianised variables are normalised by node multiplicity. In other words, the folded Coulomb branch is a restricted subspace of the discretely gauged quiver's Coulomb branch. 

While abelianised variables fold in a completely trivial manner, composite operators are more interesting. For example, let's fold the operator in (\ref{eq:B3-2}):
\be\begin{split}
	\hate{0,1,2} &= F(\te{0,1,2})  \\
	&= F(\e{0,1,1,1})  \\
	&= \sum_{a=1,2}\frac{F(\up{2,a}\up{3}\up{4})}{F(\p{2,a}-\p{3})F(\p{2,a}-\p{4})}  \\
	&=\sum_{a=1,2} \frac{\hatup{2,a} \hatup{3}{}^2/4}{(\p{2,a} - \hatp{3}/2)^2}  \\
	&=\sum_{a=1,2} \frac{\hatup{2,a} \hatup{3}{}^2}{(2\p{2,a}-\hatp{3})^2} 
\end{split}\ee

If the folded space is to retain the original's hyper-K\"{a}hler property, the symplectic property in particular must be preserved and the Poisson brackets on the folded space must close. In other words for any $\hat{f}, \hat{g} \in \co \lbrack \hatCoul \rbrack$ we require $\pois{\hat{f},\hat{g}} \in \co \lbrack \hatCoul \rbrack$, ie. $\pois{\hat{f},\hat{g}} = \pi\left( \pois{\hat{f},\hat{g}} \right)$. It is enough to show that generators $\hatx{i, a}$ of the Poisson algebra satisfy this property:
\be
\pois{\x{i,a},\x{j,b}} = f(\x{k,c}) = f(\x{\pi(k),c}) = \pi \left(f(\x{k,c}) \right) = \pi\left( \pois{\x{i,a},\x{j,b}} \right) \ee
where we restrict to the folding locus
\be
\x{i,a}=\x{\pi(i),a}, \forall \pi \in \Gamma \subset \mathrm{Aut}\ Q.
\ee
where $\Gamma$ is the subgroup by whose action we fold. 

So we have in our hands two pieces: a ``folded" subspace (with its coordinate ring) and a Poisson bracket on this space. If we assume that the complex structures also properly restrict to the subspace, we have a new hyper-K\"{a}hler space to study. What is it? What is its symmetry?

Now we re-establish contact with discrete gauging. For $\Tilde{\mathcal{O}}_i \in \co \lbrack \Tilde{\Coul} \rbrack_2$ and $\hat{\mathcal{O}}_i \in \co \lbrack \hat{\Coul} \rbrack_2$, we have
\be
\pois{\Tilde{\mathcal{O}}_i, \Tilde{\mathcal{O}}_j} = \sum_k \tensor{c}{_{i j}^k} \Tilde{\mathcal{O}}_k
\ee
and therefore the relations in particular hold on the automorphism's fixed point, which is the folded subspace:
\be
\pois{\hat{\mathcal{O}}_i, \hat{\mathcal{O}}_j} = \sum_k \tensor{c}{_{i j}^k} \hat{\mathcal{O}}_k.
\ee
Therefore, unless some folded $\hat{\mathcal{O}}_k$ identically vanish, the two algebras have identical structure constants and are in fact isomorphic as Lie algebras. A simple proof in appendix \ref{appendix:folded_lie_alg} shows that $\hat{\mathcal{O}}_k$ is not 0 everywhere on the folded space so we conclude that folded spaces have the same continuous symmetries as their discretely gauged counterparts.

In particular, a $A_{2n-1}$ ($D_{n+1}$) quiver's Coulomb branch folds into a $C_n$ ($B_n$)-symmetric space of strictly lower dimension and the minimal nilpotent orbit of $D_4$ folds into the minimal nilpotent orbit of $B_3$. Of course this space is just the Coulomb branch of a non-simply laced quiver, and we claim this is no coincidence: although we have so far only explored folding as an action on the Coulomb branch, we conjecture it is in fact merely one facet of an action on the \emph{quiver theory} and that all non-simply laced quivers can be understood as folded simply laced quivers.

As was hinted in Section \ref{sec:algebra}, in some special cases a $B_3$ non-simply laced quiver, eg. the bottom right quiver in Fig. \ref{fig:folding-examples}, can fold into $G_2$ despite the lack of an obvious symmetry. There is one major difference however: multiplicities are assigned in a more involved manner. As a prerequisite, the ``short root" (i.e. third) gauge node must have the same rank and number of flavors as the ``vector root" (i.e. first node). We can unfold the $B_3$ quiver into a $D_4$ shape by simply reversing the folding procedure. Let us denote the variables of that quiver's Coulomb branch e.g. $\p{i}^{D_4}$, with $\p{i}^{B_3}$ and $\p{i}^{G_2}$ the partially and fully folded counterparts. Then at the $D_4\rightarrow G_2$ folding locus the following holds:
\be
    \p{1}^{D_4} = \p{3}^{D_4} = \p{4}^{D_4} = \p{1}^{B_3} = \frac{\p{3}^{B_3}}{2} = \frac{\p{1}^{G_2}}{3}
\ee
So the $B_3$ quiver can fold to $G_2$ as if $\mu_1=3$ and $\mu_3=\frac{3}{2}$.

\subsection{Monopole formula: examples}

To show that folded quivers become non-simply laced, we compute two explicit examples and conjecture that the pattern generalises.

\subsubsection{\texorpdfstring{$\mathrm{min}\ A_3\rightarrow\mathrm{min}\ C_2$}{min A3 -> min C2}}

The first check will be done on quivers in Figure \ref{tab:a3min-c2} by folding two $U(1)$ nodes.

Let $\HS^A$ and $\HS^C$ be the Hilbert series of the initial and folded quivers, respectively: 
\begin{equation}\label{eq:hs-A}
	\HS^A (t,x,y,z) = \frac{1}{(1-t^2)^3} \sum\limits_{q_1,q_2,q_3 \in \mathbb{Z}} t^{|q_1|+|q_1-q_2|+|q_2-q_3|+|q_3|} (xy)^{q_1} \left( \frac{x}{y}\right)^{q_3} z^{q_2} 
\end{equation}
\begin{equation}
	\HS^C (t,x,z) = \frac{1}{(1-t^2)^2} \sum\limits_{r_1,r_2 \in \mathbb{Z}} t^{|r_1|+|r_1-2r_2|} x^{r_1} z^{r_2} \,. 
\end{equation}
The unrefined Hilbert series are:
\begin{equation}
	\HS^A (t,1,1,1) = \frac{\left(1+t^2\right) \left(1+8t^2+ t^4\right)}{\left(1-t^2\right)^6} 
\end{equation}
\begin{equation}
	\HS^C (t,1,1) = \frac{1+6t^2+ t^4}{\left(1-t^2\right)^4} \,. 
\end{equation}
Note the unusual fugacity $y$ in $\HS^A$ which is crucial in the following calculations. By comparison with known Hilbert series, we find that the two Coulomb branches are the (closures of the) minimal nilpotent orbits of $A_3$ and $C_2$.

We will now derive the action $\HS^A\rightarrow \HS^C$ in two steps. 

At the level of bare monopole operators, many become duplicate. For example, $(\up{1})^2$, $\up{1}\up{3}$ and $(\up{3})^2$ all fold to $\frac{(\hatup{1})^2}{4}$. More generally, a bare monopole monomial in the $A_3$ theory can be expressed (not necessarily uniquely) as a product of generators
\be
\mathcal{O}_{q_1,q_2,q_3} = \prod_i \e{q^i_1,q^i_2,q^i_3}
\ee
where $q_j = \sum_i q_j^i$. Note that $|q^i_1-q^i_3|\in \lbrace 0, 1 \rbrace$. The $S_2$ symmetry exchanges $\e{q^i_1,q^i_2,q^i_3} \leftrightarrow \e{q^i_3,q^i_2,q^i_1}$ and acting with it on \emph{any number} of operators in the product produces a monopole in the $A_3$ theory which folds to the exact same monopole in the $C_2$ theory. ``Flipping" a single operator in this way leaves $q_1$ and $q_3$ unchanged or changes both by $\pm 1$ with opposite signs so that $q_1+q_3$ is preserved. \emph{Sequential} action on all the monopoles in the product produces $\mathcal{O}_{q_3,q_2,q_1}$. It follows that in this chain of flips there is an operator $\mathcal{O}_{\frac{q_1+q_3}{2},q_2,\frac{q_1+q_3}{2}}$ or $\mathcal{O}_{\frac{q_1+q_3 +1}{2},q_2,\frac{q_1+q_3-1}{2}}$, depending on the parity of $q_1+q_3$. Since all operators in the chain fold to the same operator, the $C_2$ monopole formula better count \emph{precisely one of them}. We will pick the one with $q_1$ closest to $q_3 \leq q_1$.

To accomplish this we must extract only the terms constant and linear in $y$, as can be seen from (\ref{eq:hs-A}): terms constant in $y$ come from the charge sublattice $q_1=q_3$ while linear terms all satisfy $q_1=q_3+1$. To set up later generalisation we further slightly modify the prescription to an equivalent form: we will extract every operator at order $y^0$ and \emph{average over} operators at order $y$ and $y^{-1}$.

The second step corrects for scalar dressing: one extraneous scalar field must be removed since $\p{1}=\p{3}=\frac{\hatp{1}}{2}$. We need only multiply the entire expression with $1-t^2$ to remove the newly duplicate $U(1)$ dressing factor $\frac{1}{1-t^2}$.

We conjecture that these two modifications are sufficient to represent the action of folding on the Hilbert series.

To implement them, we multiply the (unsummed) monopole formula by the kernel $\frac{1}{2\pi i y} \left(1+\frac{1}{2}\left(y+\frac{1}{y}\right) \right)$ and and integrate around $y=0$, picking up the desired contributions by the residue theorem. Finally we multiply by the scalar factor $(1-t^2)$:
\begin{equation}
	 \HS^C (t,x,z) = (1-t^2) \oint \frac{\mathrm{d} y}{2 \pi i y} \,  \left(1+\frac{1}{2}\left(y+\frac{1}{y}\right) \right) \HS^A (t,x,y,z) 
\end{equation}

And indeed: 
\begin{eqnarray*}
	RHS &=& \frac{1}{2} (1-t^2)^{-2} \oint \frac{\mathrm{d} y}{2 \pi i y} \, \left(y+\frac{1}{y} + 2 \right) \sum\limits_{q_1,q_2,q_3 \in \mathbb{Z}} t^{|q_1|+|q_1-q_2|+|q_2-q_3|+|q_3|} (xy)^{q_1} \left( \frac{x}{y}\right)^{q_3} z^{q_2} \\
	&=& \frac{1}{2} (1-t^2)^{-2} \oint \frac{\mathrm{d} y}{2 \pi i y} \, \sum\limits_{q_1,q_2,q_3 \in \mathbb{Z}} t^{|q_1|+|q_1-q_2|+|q_2-q_3|+|q_3|} x^{q_1+q_3} y^{q_1-q_3-1} z^{q_2} + \\
	& & \frac{1}{2} (1-t^2)^{-2} \oint \frac{\mathrm{d} y}{2 \pi i y} \, \sum\limits_{q_1,q_2,q_3 \in \mathbb{Z}} t^{|q_1|+|q_1-q_2|+|q_2-q_3|+|q_3|} x^{q_1+q_3} y^{q_1-q_3+1} z^{q_2} + \\
	& & (1-t^2)^{-2} \oint \frac{\mathrm{d} y}{2 \pi i y} \, \sum\limits_{q_1,q_2,q_3 \in \mathbb{Z}} t^{|q_1|+|q_1-q_2|+|q_2-q_3|+|q_3|} x^{q_1+q_3} y^{q_1-q_3} z^{q_2} \\
	&=& \frac{1}{2} (1-t^2)^{-2} \sum\limits_{r_2 \in \mathbb{Z}, r_1 \in( 2 \mathbb{Z}+1)} t^{|(r_1+1)/2|+|(r_1+1)/2-r_2|+|r_2-(r_1-1)/2|+|(r_1-1)/2|} x^{r_1} z^{r_2} + \\
	& & \frac{1}{2} (1-t^2)^{-2} \sum\limits_{r_2 \in \mathbb{Z}, r_1 \in( 2 \mathbb{Z}+1)} t^{|(r_1-1)/2|+|(r_1-1)/2-r_2|+|r_2-(r_1+1)/2|+|(r_1+1)/2|} x^{r_1} z^{r_2} + \\
	& & (1-t^2)^{-2} \sum\limits_{r_2 \in \mathbb{Z}, r_1 \in( 2 \mathbb{Z})} t^{|r_1/2|+|r_1/2-r_2|+|r_2-r_1/2|+|r_1/2|} x^{r_1} z^{r_2} \\
	&=& (1-t^2)^{-2} \sum\limits_{r_2 \in \mathbb{Z}, r_1 \in( 2 \mathbb{Z}+1)} t^{|(r_1+1)/2|+|(r_1+1)/2-r_2|+|r_2-(r_1-1)/2|+|(r_1-1)/2|} x^{r_1} z^{r_2} + \\
	& & (1-t^2)^{-2} \sum\limits_{r_2 \in \mathbb{Z}, r_1 \in( 2 \mathbb{Z})} t^{|r_1|+|r_1-2r_2|} x^{r_1} z^{r_2} \\
	&=& (1-t^2)^{-2} \sum\limits_{r_1,r_2 \in \mathbb{Z}} t^{|r_1|+|r_1-2r_2|} x^{r_1} z^{r_2} \\
\end{eqnarray*}

In particular note the appearance of 2 in $|r_1-2r_2|$, the novel feature in non-simply laced quivers' monopole formulas.

\subsubsection{\texorpdfstring{$\min D_4 \rightarrow \min G_2$}{min D4 -> min G2}}

We now look at the folding of three $U(1)$ gauge nodes of the $D_4$ minimal nilpotent orbit quiver. We again assign fugacities to the nodes: call $z$ the fugacity for the $U(2)$ node, and $x y_1$, $x \frac{y_2}{y_1}$, $x \frac{1}{y_2}$ the fugacities for the three $U(1)$ nodes. This parametrisation is chosen so that folding corresponds to an integration over the $y_i$, which have an $A_2$ symmetry. Note that this prescription generalises the previous example, where the ``folding fugacity" appeared as $y$ and $y^{-1}$, which are related by an $A_1$ symmetry.

The folding equation becomes 
\begin{equation*}
	\HS^{G_2} (t,x,z) = (1-t^2)^2 \oint \frac{\mathrm{d} y_1}{2 \pi i y_1} \frac{\mathrm{d} y_2}{2 \pi i y_2} \, f(y_1 , y_2) H^{D_4} (t,x,y_1,y_2,z) 
\end{equation*}
with 
\begin{equation*}
	f(y_1 , y_2) = 1 + \frac{1}{3} \left( y_1 + \frac{1}{y_1} + y_2 + \frac{1}{y_2} + \frac{y_2}{y_1} + \frac{y_1}{y_2} \right) \,. 
\end{equation*}

Note that this kernel is a natural generalization of the previous case $f(y) = 1 + \frac{1}{2}(y + y^{-1})$. We conjecture that the monopole formula of a quiver with $n$ $U(1)$ legs folds by integration over the kernel
\be
f(y_1,\dots y_{n-1})= 1+\frac{1}{n}\chi^{A_{n-1}}_f(y_1,\dots,y_{n-1})
\ee
where $\chi^{A_{n-1}}_f$ is the character of the $A_{n-1}$
 fundamental representation.
 
 The steps outlined above can be generalised to longer legs, larger gauge groups and, presumably, to completely arbitrary legs. However, rather than undertaking this task ourselves, we refer to \cite{nakajima_coulomb_2019} for a systematic look at the link between folding and the modified monopole formula of \cite{cremonesi_coulomb_2014}.
 
\subsection{Non-simply laced quivers}

It is possible to generalise abelianisation, including the Poisson structure, directly to non-simply laced framed quivers; the generalisation of the monopole formula was already achieved in \cite{cremonesi_coulomb_2014}. The input data are a list of gauge nodes with optional fundamental matter and a connectivity matrix $\kappa$ defined precisely like the Cartan matrix of a Dynkin diagram. One can always unfold the quiver $\hat{Q}$ into a simply laced quiver $Q$. Keeping with the term's use in previous sections, the number of nodes of $Q$ which fold onto the $i$-th node of $\hat{Q}$ is called the \emph{multiplicity $\#_i$ of node $i$}.

Each node still contributes three abelianised variables $\hatupm{i,a}$ and $\hatp{i,a}$ but the relations are slightly modified. They can be derived by demanding consistency with folding; recall that $x_{i,a}=\hat{x}_{i,a}/\#_i$ on the subspace preserved by discrete action. For simplicity we present them in the case of quivers with one multiple edge:
\be \hatup{i,a} \hatum{i,a} = -\#_i^2 \frac{\prod_{w\in \mathcal{R}}\langle{w,\vec{\hat\varphi}/\vec\#}\rangle^{g_i(w)\left| w_{i,a} \right|}}{\prod_{\alpha\in\Phi}\langle\alpha,\vec{\hat\varphi}/\vec\#\rangle^{\left| \alpha_{i,a} \right|}}\label{eq:nonsimplylacedabelrel} \ee
where $\mathcal{R}$ is defined as if the quiver were simply laced (ie. the multiple link were replaced with one simple link), $\vec{\hat\varphi}/\vec\#$ denotes a vector of $\hatp{i,a}/\#_i$ and $g_i(w)$ is an auxiliary function defined as
\be
g_i(w) = \begin{cases}|\kappa_{ji}| &\text{ if $w$ connects the node $i$ to node $j$,}\\ 1 &\text{ otherwise}\end{cases}
\ee
and $\kappa$ is the Cartan matrix of the non-simply laced quiver.

\begin{figure}[t]
	\centering
	\begin{tikzpicture}
		\node (hiddenL1) []{}; 
		\node (hiddenL2) [right = 3.3cm of hiddenL1]{};
		\node (g1) [gauge, above = 0.5cm of hiddenL1, label=above:{$\upm{1},\p{1}$}]{}; 
		\node (g2) [gauge, right = 1.5cm of g1, label=above:{$\upm{2},\p{2}$}]{}; 
		\node (g3) [gauge, below = 0.5cm of hiddenL1, label=below:{$\upm{3},\p{3}$}]{}; 
		\node (g4) [gauge, right = 1.5cm of g3, label=below:{$\upm{4},\p{4}$}]{};
		\draw (g1)--(g2); 
		\draw (g3)--(g4); 
		
		\node (hiddenR) [right = 2cm of hiddenL2]{};	
		\node (g1f) [gauge, right = 1.3cm of hiddenR, label=above:{$\hatupm{1},\hatp{1}$}]{}; 
		\node (g2f) [gauge, right = 1.5cm of g1f, label=above:{$\hatupm{2},\hatp{2}$}]{}; 
		\draw (g1f)--(g2f); 		
		
		\draw[thick, ->] (hiddenL2) -- (hiddenR);
	\end{tikzpicture}
	\caption[Folding two parallel links]{Example of folding two ``parallel" links which do not originate from the same node. Note that folding does not introduce a multiple link in this case.} 
\label{fig:toyquiver} \end{figure}
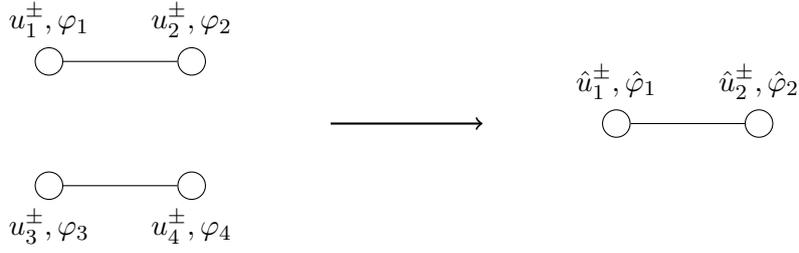

The derivation of Poisson brackets is slightly more subtle. As a concrete example, consider a quiver with nodes 1 to 4 (plus possibly others) such that 1 and 2 , resp. 3 and 4 are connected, and 3 and 4 fold onto 1 and 2, respectively (see Fig \ref{fig:toyquiver}). Then
\begin{multline}
\pois{\hatp{1},\hatup{1}}=\restr{\pois{\p{1}+\p{3},\up{1}+\up{3}}}{x_1=x_3} = \restr{(\pois{\p{1},\up{1}}+\pois{\p{3},\up{3}})}{x_1=x_3} =\\= \restr{\p{1}+\p{3}}{\substack{x_1=x_3\\x_2=x_4}} = \hatp{1}
\end{multline}
Similarly, and keeping to the same quiver for this example,
\be
\pois{\hatup{1},\hatup{2}} = \restr{\pois{\up{1}+\up{3},\up{2}+\up{4}}}{\substack{x_1=x_3\\x_2=x_4}} = 2\pois{\up{1},\up{2}} = 2\kappa_{12} \frac{\up{1}\up{2}}{\p{1}-\p{2}} = 2\kappa_{12} \frac{\hatup{1}/2\  \hatup{2}/2}{\hatp{1}/2 - \hatp{2}/2}
\ee
Note that the factor of 2 comes from the two links which fold onto each other.

Now that the procedure is clear it readily generalises:
\begin{align}
	\pois{\hatp{i,a},\hatupm{i,a}} &= \pm\hatupm{i,a} \\
	\pois{\hatup{i,a},\hatum{i,a}} &= \#_i^2 \frac{\partial}{\partial \hatp{i,a}} \frac{\prod_{w\in \mathcal{R}}\langle{w,\vec{\hat{\varphi}}/\vec{\#}}\rangle^{g_i(w)\left| w_{i,a} \right|}}{\prod_{\alpha\in\Phi}\langle\alpha,\vec{\hat{\varphi}}/\vec{\#}\rangle^{\left| \alpha_{i,a} \right|}} \\
	\pois{\hatupm{i,a},\hatupm{j,b}} &= \pm \kappa^S_{ij}\frac{\#_{ij}}{\#_i \#_j}\frac{\hatupm{i,a}\hatupm{j,b}}{\hatp{i,a}/\#_i-\hatp{j,b}/\#_j}  
\end{align}
where $\kappa^S$ is a ``simply laced" Cartan matrix defined as $\kappa^S_{ij} = \max (\kappa_{ij},\kappa_{ji})$ (essentially throwing away information about multiplicity of edges) and $\#_{ij}$ is the \emph{link multiplicity} of the edge between nodes $i$ and $j$ defined as the number of its pre-images in the unfolded quiver. Remember that just as in the case of abelianised relations this form is appropriate for quivers with one multiple edge.

\section{Examples}

In this section we study several cases of nilpotent orbit quivers, ie. quiver theories whose Coulomb branches are nilpotent orbits. Their chiral rings are generated by moment maps, which we explicitly construct; recall that such moment maps transform in the coadjoint representation $\mathrm{coadj}(\mathfrak{g})\simeq \mathrm{adj}(\mathfrak{g})$ of the Coulomb branch symmetry. The chiral ring data is completed by providing a set of relations which also form representations of the Coulomb branch symmetry. We discretely gauge and fold such quivers and examine the resulting Coulomb branches in turn.

Most spaces encountered in this section are nilpotent orbits; their coordinate rings are generated by a single coadjoint representation. But there are a few cases which are not nilpotent orbits: their Coulomb branches are generated not only by $\mathrm{coadj}(\mathfrak{g})$ but also by chiral ring elements in other representations of $\mathfrak{g}$. If the quiver is balanced, for the examples studied in this paper, we find that the remaining generators \emph{also} assemble into coadjoint (or sometimes trivial) representations and the bulk of our techniques still applies. One such case appears in Sec. \ref{sec:g2-z3}. The resulting spaces are not as comprehensively tabulated as nilpotent orbits and we generally have to turn to more varied sources to find their Hilbert series or highest weight generating functions.

\subsection{\texorpdfstring{$\mathrm{min}\ A_3 \rightarrow \mathrm{(n.)min}\ C_2$}{min A3 -> n.min C2}}
\label{section:a3-to-c2}

\begin{figure}[t]
    \centering
    \begin{tabular}
    {| >{\centering\arraybackslash} m{3.5cm} | >{\centering\arraybackslash} m{3.5cm} |  >{\centering\arraybackslash} m{3.5cm} |}
    \hline 
        Initial & Discretely Gauged & Folded \\ \hline \hline
			\vspace{0.3cm}\begin{tikzpicture}
			\node (f1) at (0,1) [flavor, label=left:{1}]{};
			\node (g1) at (0,0) [gauge, label=below:{1}]{}; 
			\node (g2) at (1,0) [gauge, label=below:{1}]{}; 
			\node (g3) at (2,0) [gauge, label=below:{1}]{};
			\node (f3) at (2,1) [flavor, label=right:{1}]{};
			\draw (f1)--(g1)--(g2)--(g3)--(f3);
		\end{tikzpicture} & \vspace{1.3cm}\begin{tikzpicture}
			\node (f1) at (-1,0) [flavor, label=below:{\symmlabel{[ 1}}]{};
			\node (g1) at (0,0) [gauge, label=below:{\symmlabel{\qquad 1] \wr S_2}}]{}; 
			\node (g2) at (1.2,0) [gauge, label=below:{\symmlabel{1}}]{}; 
			\draw (f1)--(g1)--(g2);
		\end{tikzpicture} & \vspace{0.3cm}\begin{tikzpicture}
			\node (f1) at (0,1) [flavor, label=left:{1}]{};
			\node (g1) at (0,0) [gauge, label=below:{1}]{}; 
			\node (g2) at (1,0) [gauge, label=below:{1}]{}; 
			\draw (f1)--(g1);
			\draw[doublearrow] (g1)-- node {\midarrowrev} (g2);
		\end{tikzpicture} \\ 
		\begin{tikzpicture}
		\tikzstyle{hasse} = [circle, fill,inner sep=2pt];
		\node [hasse] (2) [label=right:\footnotesize{$3$}] {};
		\node [hasse] (3) [below of=2,label=right:\footnotesize{$0$}] {};
		\draw (2) edge [] node[label=left:\footnotesize{$a_3$}] {} (3);
	\end{tikzpicture}
	& \begin{tikzpicture}
		\tikzstyle{hasse} = [circle, fill,inner sep=2pt];
		\node [hasse] (1) [label=right:\footnotesize{$3$}] {};
		\node [hasse] (2) [below of=1, label=right:\footnotesize{$2$}] {};
		\node [hasse] (3) [below of=2, label=right:\footnotesize{$0$}] {};
		\draw (1) edge [] node[label=left:\footnotesize{$c_1$}] {} (2);
		\draw (2) edge [] node[label=left:\footnotesize{$c_2$}] {} (3);
	\end{tikzpicture} & 
	 	\begin{tikzpicture}
		\tikzstyle{hasse} = [circle, fill,inner sep=2pt];
		\node [hasse] (2) [label=right:\footnotesize{$2$}] {};
		\node [hasse] (3) [below of=2,label=right:\footnotesize{$0$}] {};
		\draw (2) edge [] node[label=left:\footnotesize{$c_2$}] {} (3);
	\end{tikzpicture} \\ 
	 \begin{tabular}{cc}
	     $\mu_1 \mu_3 t^2$ & ($A_3$) \\
	     $(\mu_1^2 + \mu_2)t^2$ & ($C_2$)
	 \end{tabular}
	  &  $\mu_1^2 t^2 + \mu_2^2 t^4$ & $\mu_1^2 t^2$  \\ 
	\hline 
    \end{tabular}
    \caption{$A_3$ minimal nilpotent orbit and its discrete reductions.  }
    \label{tab:a3min-c2}
\end{figure}

$A$-type quivers tend to have very simple moment maps which can be presented in reasonably compact form, allowing us to present the action of discrete gauging and folding.

The quivers we choose, as exhibited in Figure \ref{tab:a3min-c2}, also exhibit an interesting pattern of complex mass deformation. As a general rule, all $\p{i,a}$ abelian moduli and $\m{i,a}$ parameters only appear in the abelian algebra as differences and as a result the moduli space is invariant under reparametrisations $\p{i,a}\rightarrow\p{i,a}+c$, $\m{i,a}\rightarrow \m{i,a}+c$. Since there are precisely two mass parameters, the moduli space relations can be modified by terms proportional to $\m{1}-\m{3}$, ie. a complex mass deformation. However both discrete gauging and folding remove one half of mass parameters by forcing $\m{1}=\m{3}$, which can in turn be set to 0 by the reparametrisation above. As a result only the original space can be deformed by one triplet of mass parameters.

\subsubsection{Initial quiver}

To remind the reader we reproduce abelianised relations restricting $\upm{i}, \p{i}$ for $i=1,2,3$:
\begin{align}
    \up{1}\um{1} &= - (\p{1}-\m{1})(\p{1}-\p{2}) \\
    \up{2}\um{2} &= - (\p{2}-\p{1})(\p{2}-\p{3}) \\
    \up{3}\um{3} &= - (\p{3}-\p{2})(\p{3}-\m{3})
\end{align}

The Coulomb branch is generated by 
\be \mommap{A_3}= \left( 
\begin{matrix}
	\p{1} - \frac{3\m{1} +\m{3}}{4} & \um{1} & -\frac{\um{1}\um{2}}{\p{1}-\p{2}} & \frac{\um{1}\um{2}\um{3}}{(\p{1}-\p{2})(\p{2}-\p{3})}\\
	\up{1} & -\p{1} + \p{2} + \frac{\m{1}-\m{3}}{4} & \um{2} & -\frac{\um{2}\um{3}}{(\p{2}-\p{3})}\\
	-\frac{\up{1}\up{2}}{\p{1}-\p{2}} & \up{2} & -\p{2} + \p{3} + \frac{\m{1}-\m{3}}{4} & \um{3}\\
	\frac{\up{1}\up{2}\up{3}}{(\p{1}-\p{2})(\p{2}-\p{3})} & -\frac{\up{2}\up{3}}{(\p{2}-\p{3})} & \up{3} & -\p{3} + \frac{\m{1} + 3 \m{3}}{4}
\end{matrix}
\right) \ee
and one can read its relations either from the HWG \cite{hanany_quiver_2016}
\be
\HWG(t,\mu_i) = \frac{1}{1-\mu_1 \mu_3 t^2}
\ee
or simply from the Joseph relations, which are obeyed by any minimal nilpotent orbit:
\begin{align}
    t^4\left(\lbrack 101\rbrack+\lbrack 000\rbrack\right)&: N^2 = -\frac{1}{2}(\m{1}-\m{3})N + \frac{3}{16}(\m{1}-\m{3})^2 \mathbf{1}\\
    t^4\lbrack 020 \rbrack&: \sum_{a',b'}\varepsilon_{a'b'\lbrack cd} N^{a'}_{a} N^{b'}_{b\rbrack}=-\frac{1}{16}(\m{1}-\m{3})^2 \varepsilon_{abcd}
\end{align}

\subsubsection{Discrete Gauging}

The $A_3$ moment map discretely gauges to the following expression:
\be \mommap{C_2}= \left( 
\begin{matrix}
	\frac{1}{2}(\tp{1}+\tp{3}) & \frac{1}{2}(\tum{1}+\tum{3}) & \frac{1}{2}(-\frac{\tum{1}\tum{2}}{\tp{1}-\tp{2}}+\frac{\tum{2}\tum{3}}{\tp{2}-\tp{3}}) & \frac{\tum{1}\tum{2}\tum{3}}{(\tp{1}-\tp{2})(\tp{2}-\tp{3})}\\
	\frac{1}{2}(\tup{1}+\tup{2}) & -\frac{1}{2}(\tp{1}+\tp{3}) + \tp{2} & \tum{2} & \frac{1}{2}(\frac{\tum{1}\tum{2}}{\tp{1}-\tp{2}}-\frac{\tum{2}\tum{3}}{\tp{2}-\tp{3}}) \\
	\frac{1}{2}(-\frac{\tup{1}\tup{2}}{\tp{1}-\tp{2}}+\frac{\tup{2}\tup{3}}{\tp{2}-\tp{3}}) & \tup{2} & -\tp{2} + \frac{1}{2} (\tp{1}+\tp{3}) & \frac{1}{2}(\tum{1}+\tum{3})\\
	\frac{\tup{1}\tup{2}\tup{3}}{(\tp{1}-\tp{2})(\tp{2}-\tp{3})} & \frac{1}{2}(\frac{\tup{1}\tup{2}}{\tp{1}-\tp{2}}-\frac{\tup{2}\tup{3}}{\tp{2}-\tp{3}}) & \frac{1}{2}(\tup{1}+\tup{2}) & \frac{1}{2}(-\tp{1}-\tp{3})
\end{matrix}
\right) \ee
and the resulting space is expected to have quaternionic dimension 3 and exhibit $C_2$ symmetry. The next-to-minimal nilpotent orbit of $C_2$ is a suitable candidate. Its HWG reads \cite{hanany_quiver_2016}
\be
\HWG(t,\mu_i)=\frac{1}{(1 - \mu_1^2 t^2) (1 - \mu_2^2 t^4)}
\ee
suggesting several relations:
\begin{align}
    t^4 \left(\lbrack 00 \rbrack + \lbrack 01 \rbrack \right)&: N^2 = 0 \\
    t^6 \lbrack 20 \rbrack&: \rank(N) \leq 2
\end{align}
(Note that in our convention we multiply $C_n$ matrices as ordinary matrices, ie. without insertion of an $\epsilon$ tensor.) The second of these relations can be written equivalently as $$\sum_{b,c,d,b',c',d'} \varepsilon_{a'b'c'd'} \varepsilon^{abcd} N^{b'}_b N^{c'}_c N^{d'}_d = 0.$$ In other words, an explicit algebraic description of the Coulomb branch of the  discretely gauged quiver is 
\begin{equation}
  \{ N \in \mathfrak{gl}(4,\mathbb{C}) | N^2=0 , \quad \rank(N) \leq 2 , \quad  N^T J - J N = 0 \} \, . 
\end{equation}

\subsubsection{Folding}
The folded moment map is similar:
\be \mommap{C_2}= \left( 
\begin{matrix}
	\frac{1}{2}\hatp{1} & \frac{1}{2}\hatum{1} & -\frac{\hatum{1}\hatum{2}}{\hatp{1}-2\hatp{2}} & -\frac{(\hatum{1})^2\hatum{2}}{(\hatp{1}-2\hatp{2})^2}\\
	\frac{1}{2}\hatup{1} & -\frac{1}{2}\hatp{1} + \hatp{2} & \hatum{2} & \frac{\hatum{2}\hatum{1}}{(\hatp{1}-2\hatp{2})}\\
	-\frac{\hatup{1}\hatup{2}}{\hatp{1}-2\hatp{2}} & \hatup{2} & -\hatp{2} + \frac{1}{2}\hatp{1} & \frac{1}{2}\hatum{1}\\
	-\frac{(\hatup{1})^2\hatup{2}}{(\hatp{1}-2\hatp{2})^2} & \frac{\hatup{2}\hatup{1}}{(\hatp{1}-2\hatp{2})} & \frac{1}{2}\hatup{1} & -\frac{1}{2}\hatp{1} 
\end{matrix}
\right) \ee
The Coulomb branch has dimension 2 and $C_2$ symmetry, which agrees with the minimal nilpotent orbit with HWG \cite{hanany_quiver_2016}
\be
\HWG(t,\mu_i) = \frac{1}{1-\mu_1^2 t^2}
\ee
This space satisfies slightly more stringent (Joseph) relations:
\begin{align}
    t^4 \left(\lbrack 00 \rbrack + \lbrack 01 \rbrack \right)&: N^2 = 0 \\
    t^4 \lbrack 02 \rbrack&: \rank(N) \leq 1
\end{align}
The second of these relations can be written equivalently as $N^{a'}_{\lbrack a} N^{b'}_{b \rbrack} = 0$. In other words, an explicit algebraic description of the Coulomb branch of the folded quiver is 
\begin{equation}
   \{ N \in \mathfrak{gl}(4,\mathbb{C}) | N^2=0 , \quad \rank(N) \leq 1 , \quad  N^T J - J N = 0 \} \, . 
\end{equation}

\subsection{\texorpdfstring{$\mathrm{min}\ D_4 \rightarrow G_2$}{min D4 -> G2}} \label{sec:g2}

$G_2$ is small yet non-trivial enough to serve as an excellent illustration of the techniques studied in this paper. Since it is only fourteen-dimensional, we provide the complete folding prescription from both $D_4$ and $B_3$:
$$
    G_2 = \mathrm{span}_\co\left(E^{G_2}_{\pm 1^3 2^2},E^{G_2}_{\pm 1^3 2}, E^{G_2}_{\pm 1^2 2},E^{G_2}_{\pm 1 2},E^{G_2}_{\pm 1},E^{G_2}_{\pm 2},H^{G_2}_1, H^{G_2}_2 \right)
$$

\begin{align*}
    E^{G_2}_{\pm 1^3 2^2} &= E^{D_4}_{\pm 1 2^2 3 4} = E^{B_3}_{\pm 1 2^2 3^2} \\
    E^{G_2}_{\pm 1^3 2} &= E^{D_4}_{\pm 1 2 3 4} = E^{B_3}_{\pm 1 2 3^2} \\
    E^{G_2}_{\pm 1^2 2} &= - E^{D_4}_{\pm1 2 3} - E^{D_4}_{\pm 1 2 4} + E^{D_4}_{\pm 2 3 4} =  - E^{B_3}_{\pm1 2 3} + E^{B_3}_{\pm 2 3^2} \\
    E^{G_2}_{\pm 1 2} &= E^{D_4}_{\pm 1 2} - E^{D_4}_{\pm 2 3} - E^{D_4}_{\pm 2 4} = E^{B_3}_{\pm 1 2} - E^{B_3}_{\pm 2 3} \\
    E^{G_2}_{\pm 1} &= E^{D_4}_{\pm 1} + E^{D_4}_{\pm 3} + E^{D_4}_{\pm 4} =  E^{B_3}_{\pm 1} + E^{B_3}_{\pm 3} \\
    E^{G_2}_{\pm 2} &= E^{D_4}_{\pm 2} =  E^{B_3}_{\pm 2} \\
    H^{G_2}_{\pm 1} &= H^{D_4}_{\pm 1} + H^{D_4}_{\pm 3} + H^{D_4}_{\pm 4} =  H^{B_3}_{\pm 1} + H^{B_3}_{\pm 3} \\
    H^{G_2}_{\pm 2} &= H^{D_4}_{\pm 2} = H^{B_3}_{\pm 2} \\
\end{align*}

Recall that $G_2$ is characterised as the subalgebra of $B_3$ which preserves a particular rank 3 antisymmetric tensor $\phi$; for more details see Section \ref{sec:algebra}.

The goal of this subsection is to identify quivers whose Coulomb bran\-ches are generated by operators in one $G_2$ coadjoint representation $\lbrack 01 \rbrack$; such spaces are necessarily nilpotent orbits. We also study one related space whose coordinate ring is generated by coadjoint generators but is not a nilpotent orbit. The following sections should be read alongside Figures \ref{table:g2-npo} and \ref{table:g2-reductions}.

Note that because the quiver has only flavor node of rank 1, the $G_2$ spaces studied below cannot be deformed by a complex mass.

We provide the first few symmetric products of the (co)adjoint representation for reference:
\begin{align}
    \mathrm{Sym}^2 \lbrack 01 \rbrack &= \lbrack 20 \rbrack + \lbrack 00 \rbrack + \lbrack 02 \rbrack \label{eq:s2g2adj} \\ 
    \mathrm{Sym}^3 \lbrack 01 \rbrack &= \lbrack 30 \rbrack + \lbrack 21 \rbrack + \lbrack 01 \rbrack + \lbrack 10 \rbrack + \lbrack 03 \rbrack \label{eq:s3g2adj} \\ 
    \mathrm{Sym}^4 \lbrack 01 \rbrack &= \lbrack 40 \rbrack + \lbrack 31 \rbrack + \lbrack 22 \rbrack + \lbrack 11 \rbrack + 2 \lbrack 20 \rbrack + \lbrack 00 \rbrack + 2 \lbrack 02 \rbrack + \lbrack 04 \rbrack \label{eq:s4g2adj} 
\end{align}

\begin{landscape}
    \begin{figure}[p]
    \centering
    \begin{tabular}
    {| >{\centering\arraybackslash} m{0.8cm} | >{\centering\arraybackslash} m{1.7cm} |  >{\centering\arraybackslash} m{4cm} | >{\centering\arraybackslash} m{6cm} |  >{\centering\arraybackslash} m{4cm} | >{\centering\arraybackslash} m{2.2cm} |}
    \hline 
        Label & Dimension & Relations & PL(HWG) & Hilbert Series & Quiver \\\hline \hline
        [00] & 0 & $N=0$  & 0 & 1 &  \\ \hline 
        [01] & 3 & $N^2=0$ & $\mu_2 t^2$ & $\frac{(1+t^2) (1+7 t^2+t^4)}{(1-t^2)^6}$ & 
        \begin{tikzpicture}
			\node (f2) at (1,1) [flavor, label=above:{1}]{};
			\node (g1) at (0,0) [gauge, label=below:{1}]{}; 
			\node (g2) at (1,0) [gauge, label=below:{2}]{}; 
			\draw (g1)--(g2)--(f2);		
			\draw[doublearrow] (g2)-- node {\midarrowrev} (g1);
			\draw (g2)-- node {\midarrowrev} (g1);
		\end{tikzpicture}\\ \hline 
        [10] & 4 & \begin{tabular}{cc}
            $\tr(N^2) = 0$ \\
            $N \wedge N \wedge N = 0$  \\
            $\rank(N^2) \leq 1$
        \end{tabular} & $\mu_2 t^2 + \mu_1^2 t^4 + \mu_1^3 t^6 - \mu_1^6 t^{12}$ & $\frac{1+6t^2+20t^4+43t^6-7t^8-7t^{10}}{(1-t^2)^8}$& 
        \begin{tikzpicture}
			\node (f2) at (1,1) [flavor, label=above:{1}]{};
			\node (g1) at (0,0.5) [gauge, label=left:{1}]{}; 
			\node (g2) at (1,0) [gauge, label=below:{2}]{}; 
			\node (g3) at (0,-0.5) [gauge, label=left:{1}]{};
			\draw (g1)--(g2)--(f2);
			\draw[doublearrow] (g2)-- node {\midarrowdiag} (g3);
			\draw[<->] (0,0.25) -- (0,-0.25);
		\end{tikzpicture}\\ \hline 
        [02] & 5 & \begin{tabular}{cc}
            $\tr(N^2) =0$ \\
            $N \wedge N \wedge N = 0$
        \end{tabular} & $\mu_2 t^2 + \mu_1^2 t^4 + \mu_1^3 t^6 + \mu_2^2 t^8 - \mu_1^3 \mu_2 t^{10}$ & $\frac{(1+t^2)(1+3t^2+6t^4+3t^6+t^8)}{(1-t^2)^{10}}$& 
        \hspace{-.3cm}\begin{tikzpicture}
			\node (f2) at (1,1) [flavor, label=above:{\symmlabel{1}}]{};
			\node (g1) at (0,0) [gauge, label=below:{\symmlabel{[1]\wr S_3}}]{}; 
			\node (g2) at (1,0) [gauge, label=below:{\symmlabel{2}}]{}; 
			\draw (g1)--(g2)--(f2);
		\end{tikzpicture}\\ \hline 
        [22] & 6 & $\tr(N^2) = \tr(N^6)=0$ & too long to display & $\frac{(1+t^2)^2 (1+t^4+t^8)}{(1-t^2)^{12}}$& \\ \hline 
    \end{tabular}
    \caption[$G_2$ nilpotent orbits]{$G_2$ nilpotent orbits. We use the convention that the root $\alpha_1$ is short and the root $\alpha_2$ is long, so that $\mu_1$ is the fundamental and $\mu_2$ is the adjoint. $N$ is a matrix in the fundamental 7-dimensional representation of the $\mathfrak{g}_2$, $N \in \mathfrak{so}(7,\mathbb{C})$. The relation $N \wedge N \wedge N = 0$ stands for $N_{ab}N_{cd}N_{ef} \varepsilon^{abcdefgh} = 0$. The Hilbert series in the last column are computed using Macaulay2. They agree with the computation of \cite{hanany_quiver_2017}. The Hasse diagram for nilpotent orbits of $G_2$ was computed in \cite{fu_generic_2017}. The [10] quiver is constructed with an unusual combination of folding and discrete gauging as describe in Sec. \ref{sec:g2-4dim}.}
    \label{table:g2-npo}
\end{figure}

\begin{figure}
    \centering
    \begin{tabular}
    {| >{\centering\arraybackslash} m{5cm} | >{\centering\arraybackslash} m{9cm} |  >{\centering\arraybackslash} m{5cm} |}
    \hline 
        Initial & Discretely Gauged & Folded \\ \hline \hline
        \multirow{3}{*}{
          \begin{tikzpicture}
			\node (g0) at (-1,0) [flavor, label=below:{1}]{};
			\node (g1) at (0,0) [gauge, label=below:{2}]{}; 
			\node (g2) at (1,1) [gauge, label=right:{1}]{}; 
			\node (g3) at (1,0) [gauge, label=right:{1}]{}; 
			\node (g4) at (1,-1) [gauge, label=right:{1}]{};
			\draw (g0)--(g1)--(g2);
			\draw (g3)--(g1)--(g4);
		\tikzstyle{hasse} = [circle, fill,inner sep=2pt];
		\node [hasse] (2) at (0,-1.5) [label=right:\footnotesize{$5$}] {};
		\node [hasse] (3) at (0,-3) [label=right:\footnotesize{$0$}] {};
		\draw (2) edge [] node[label=left:\footnotesize{$d_4$}] {} (3);
		\node at (-.5,-4) {$\mu_2 t^2$};\node at (2,-4) {$D_4$};
		\node at (-.5,-4.5) {$(\mu_1 + \mu_2) t^2$};\node at (2,-4.5) {$B_3$};
		\node at (-.5,-5) {$(2\mu_1 + \mu_2) t^2 + \mu_2 t^4$};\node at (2,-5) {$G_2$};
		\end{tikzpicture}} 
	 & 
	\begin{tikzpicture}
			\node (f2) at (0,1) [flavor, label=above:{1}]{};
			\node (g1) at (-1,0) [gauge, label=below:{\symmlabel{1}}]{}; 
			\node (g2) at (0,0) [gauge, label=below:{\symmlabel{2}}]{}; 
			\node (g3) at (1,0) [gauge, label=below:{\symmlabel{[1]\wr S_2}}]{};
			\draw (g1)--(g2)--(f2);
			\draw (g2)--(g3);
		\tikzstyle{hasse} = [circle, fill,inner sep=2pt];
		\node [hasse] (1) at (3,1) [label=right:\footnotesize{$5$}] {};
		\node [hasse] (2) at (3,0.5) [label=right:\footnotesize{$4$}] {};
		\node [hasse] (3) at (3,-0.5) [label=right:\footnotesize{$0$}] {};
		\draw (1) edge [] node[label=left:\footnotesize{$a_1$}] {} (2);
		\draw (2) edge [] node[label=left:\footnotesize{$b_3$}] {} (3);
		\node at (0,-1.5) {$\mu_2 t^2 + \mu_1^2 t^4$};\node at (4.5,-1.5) {$B_3$};
		\node at (0,-2) {$(\mu_1+\mu_2) t^2 + \mu_1^2 t^4 + \mu_1\mu_2 t^6 + \mu_2^2 t^8 - \mu_1^2 \mu_2^2 t^{12}$};\node at (4.5,-2) {$G_2$};
		\end{tikzpicture}
		&
	\begin{tikzpicture}
			\node (f2) at (1,1) [flavor, label=above:{1}]{};
			\node (g1) at (0,0) [gauge, label=below:{1}]{}; 
			\node (g2) at (1,0) [gauge, label=below:{2}]{}; 
			\node (g3) at (2,0) [gauge, label=below:{1}]{};
			\draw (g1)--(g2)--(f2);
			\draw (g2)--(g3);
			\draw[doublearrow] (g2)-- node {\midarrow} (g3);
		\tikzstyle{hasse} = [circle, fill,inner sep=2pt];
		\node [hasse] (1) at (3.5,1) [label=right:\footnotesize{$4$}] {};
		\node [hasse] (2) at (3.5,0) [label=right:\footnotesize{$0$}] {};
		\draw (1) edge [] node[label=left:\footnotesize{$b_3$}] {} (2);
		\node at (1,-1.5) {$\mu_2 t^2 $};\node at (3.5,-1.5) {$B_3$};
		\node at (1,-2) {$(\mu_1+\mu_2) t^2$};\node at (3.5,-2) {$G_2$};
		\end{tikzpicture}
		\\     \cline{2-3}
	 &
	\begin{tikzpicture}
			\node at (0,2) {};
			\node (f2) at (0,1) [flavor, label=above:{1}]{};
			\node (g1) at (-1,0) [gauge, label=below:{\symmlabel{[1]\wr \mathbb{Z}_3}}]{}; 
			\node (g2) at (0,0) [gauge, label=below:{\symmlabel{2}}]{}; 
			\draw (g1)--(g2)--(f2);
		\node at (0,-1.5) {$\mu_2 t^2 + (\mu_1^2 + \mu_2) t^4 + 2 \mu_1^3 t^6 - \mu_1^6 t^{12}$};\node at (4.5,-1.5) {$G_2$};
		\tikzstyle{hasse} = [circle, fill,inner sep=2pt];
		\node [hasse] (1) at (2,1.5) [label=right:\footnotesize{$5$}] {};
		\node [hasse] (3) at (2,0.5) [label=right:\footnotesize{$3$}] {};
		\node [hasse] (4) at (2,-.5) [label=right:\footnotesize{$0$}] {};
		\draw (1) edge [] node[label=left:\footnotesize{$cg_2$}] {} (3);
		\draw (3) edge [] node[label=left:\footnotesize{$g_2$}] {} (4);
		\end{tikzpicture}
		&          
		\multirow{2}{*}[-0.6cm]{	\begin{tikzpicture}
			\node (f2) at (1,1) [flavor, label=above:{1}]{};
			\node (g1) at (0,0) [gauge, label=below:{1}]{}; 
			\node (g2) at (1,0) [gauge, label=below:{2}]{}; 
			\draw (g1)--(g2)--(f2);		
			\draw[doublearrow] (g2)-- node {\midarrowrev} (g1);
			\draw (g2)-- node {\midarrowrev} (g1);
		\tikzstyle{hasse} = [circle, fill,inner sep=2pt];
		\node [hasse] (1) at (2.5,1) [label=right:\footnotesize{$3$}] {};
		\node [hasse] (2) at (2.5,0) [label=right:\footnotesize{$0$}] {};
		\draw (1) edge [] node[label=left:\footnotesize{$g_2$}] {} (2);
		\node at (.5,-1) {$\mu_2 t^2$};\node at (2,-1) {$G_2$};
		\end{tikzpicture} }  
		\\    
		\cline{2-2}
	 & 
	\begin{tikzpicture}
			\node (f2) at (0,1) [flavor, label=above:{1}]{};
			\node (g1) at (-1,0) [gauge, label=below:{\symmlabel{[1]\wr S_3}}]{}; 
			\node (g2) at (0,0) [gauge, label=below:{\symmlabel{2}}]{}; 
			\draw (g1)--(g2)--(f2);
		\node at (0,-2) {$\mu_2 t^2 + \mu_1^2 t^4 + \mu_1^3 t^6 + \mu_2^2 t^8 + \mu_1^3 \mu_2 t^{10} - \mu_1^6 \mu_2^2 t^{20}$};\node at (4.5,-2) {$G_2$};
		\tikzstyle{hasse} = [circle, fill,inner sep=2pt];
		\node [hasse] (1) at (2,1) [label=right:\footnotesize{$5$}] {};
		\node [hasse] (2) at (2,0.5) [label=right:\footnotesize{$4$}] {};
		\node [hasse] (3) at (2,0) [label=right:\footnotesize{$3$}] {};
		\node [hasse] (4) at (2,-1) [label=right:\footnotesize{$0$}] {};
		\draw (1) edge [] node[label=left:\footnotesize{$a_1$}] {} (2);
		\draw (2) edge [] node[label=left:\footnotesize{$m$}] {} (3);
		\draw (3) edge [] node[label=left:\footnotesize{$g_2$}] {} (4);
		\end{tikzpicture}
		&   \\ 
	\hline 
    \end{tabular}
    \caption[Discrete actions on the $D_4$ minimal nilpotent orbit quiver by $S_2$, $S_3$, $\Z_3$]{Overview of discrete actions on the $D_4$ minimal nilpotent orbit quiver by groups $S_2$, $S_3$ and $\mathbb{Z}_3$, previously studied in \cite{hanany_ungauging_2020}.  The non-normal transverse slice $m$ was introduced in \cite{fu_generic_2017} and the slice $cg_2$ was described in \cite{malkin_minimal_2003}. Its Hilbert series is expressed in characters of $SU(2)$: $\sum_{n\not=1}[n]t^n$. Highest weight generating functions of the corresponding moduli space are written in terms of fugacities for the algebras $D_4, B_3, G_2$, as specified next to them.} \label{table:g2-reductions}
\end{figure}
\end{landscape}

\subsubsection{Initial quiver}
The next few examples share the quiver on the left of Figure \ref{table:g2-reductions} as the common starting point. Its Coulomb branch is the minimal nilpotent orbit of $D_4$ which is parametrised by a coadjoint (antisymmetric) matrix $M$ subject to the Joseph relations
\begin{align}
    \left(\lbrack 2000 \rbrack + \lbrack 0000 \rbrack \right) t^4:&\ N^2 = 0 \\
    \left(\lbrack 0020 \rbrack + \lbrack 0002 \rbrack \right) t^4:&\ N \wedge N = 0 
\end{align}
We refer the reader to its treatment in \cite{hanany_nilpotent_2019} for more details.

\subsubsection{Folding}
The minimal nilpotent orbit of $D_4$ folds into the minimal nilpotent orbit of $G_2$ whose quiver is depicted in Figure \ref{table:g2-npo} under the label $[01]$. To verify this claim we can look at the highest weight generating function of the minimal nilpotent orbit of $G_2$ \cite{hanany_quiver_2016}
\be
\HWG(t) = \frac{1}{1-\mu_2 t^2} = 1 + \mu_2 t^2 + \mu_2^2 t^4 + \dots
\ee
or recall that the Joseph relations tell us that the coadjoint generator is constrained by the quadratic relation
\be
    \left(\lbrack 20 \rbrack + \lbrack 00 \rbrack\right)t^4:\ N^2 = 0.
\ee

Direct computation shows that the relation is satisfied by $N$ defined either by folding the moment map of the $D_4$ minimal nilpotent orbit quiver or directly using the non-simply laced prescription.

\subsubsection{\texorpdfstring{$S_3$}{S3} discrete gauging}
The five-dimensional subregular orbit of $G_2$ is known to be the $S_3$ quotient of the minimal nilpotent orbit of $D_4$ \cite{hanany_discrete_2018} so it should be the Coulomb branch of the appropriate $D_4$ quiver after discrete gauging, see row $[02]$ of Figure \ref{table:g2-npo}. One can either symmetrise the $D_4$ moment map using the projector defined in (\ref{eq:projector}) or, given the $G_2$ Chevalley Serre basis $\lbrace X_i \rbrace$, form the $G_2$ moment map $\mommap{G_2}$ from its $D_4$ counterpart $\mommap{D_4}$ as
$$
\mommap{G_2} = \sum_i X^*_i\  \tr\left(\mommap{D_4} X_i\right).
$$

The highest weight generating function is
\begin{multline}
\HWG(t) = \frac{1 + \mu_1^3 \mu_2 t^{10}}{(1 - \mu_2 t^2) (1 - \mu_1^2 t^4) (1 - \mu_1^3 t^6) (1 - \mu_2^2 t^8)} =\\ 1+\mu _2 t^2+\left(\mu _1^2+\mu _2^2\right) t^4+\left(\mu _1^3+\mu _2 \mu _1^2+\mu _2^3\right) t^6+\left(\mu _1^4+\mu _2 \mu _1^3+\mu _2^2 \mu _1^2+\mu _2^4+\mu _2^2\right) t^8+ \dots
\end{multline}

Two relations are needed this time:
\begin{align}
    \lbrack 00 \rbrack t^4:&\ \tr N^2 = 0 \\
    \lbrack 10 \rbrack t^6:&\ N \wedge N \wedge N = 0 
\end{align}
and both are satisfied by the coadjoint $\mommap{G_2}$.

\subsubsection{Mixed folding and \texorpdfstring{$S_2$}{S2} gauging} \label{sec:g2-4dim}
Midway between the two previous examples lies a nilpotent orbit of dimension 4. It is known \cite{hanany_quiver_2017} to be non-normal and hence not expected to be the Coulomb branch of any quiver since both simply and non-simply laced quivers are necessarily normal \cite{nakajima_coulomb_2019,braverman_towards_2019}. However we conjecture that it \emph{can} be recovered by using a specific and non-generic discrete operation on the minimal nilpotent orbit quiver of $B_3$, which is itself four-dimensional. This would make our construction the first non-normal Coulomb branch in the literature.

We first construct the moment map $\mommap{B_3}$ of the underlying $B_3$ quiver. The quiver has no obvious automorphism so rather than using the projector form in (\ref{eq:projector}) we define the Chevalley-Serre basis $\lbrace X_i \rbrace$ of $G_2$ and project using the trick from the previous quiver calculation:
\be
\mommap{G_2} = \sum_i X^*_i\  \tr\left(\mommap{B_3} X_i\right)
\ee
We depict the conjectured quiver theory in Figure \ref{table:g2-npo} on row $10$.

The HWG of this orbit is given by\cite{hanany_quiver_2016}\footnote{We can compare this expression with the HWG for the minimal $B_3$ orbit, written in terms of $G_2$ fugacities, which reads  \cite{hanany_quiver_2016}
\begin{multline}
    \HWG(t)=\frac{1}{\left(1-\mu _1 t^2\right)\left(1-\mu _2 t^2\right)}    =1+ \left( \mu_1 + \mu _2 \right) t^2+\left(\mu _1^2 + \mu _1\mu _2 +\mu _2^2\right) t^4 \\ +\left(\mu _1^3+ \mu _1^2 \mu _2 + \mu _1 \mu _2^2 +\mu _2^3\right) t^6  +\left(\mu _1^4+ \mu _1^3\mu _2+ \mu _1^2\mu _2^2+ \mu _1\mu _2^3+\mu _2^4\right) t^8+\dots
\end{multline}
The difference between the two expressions is 
\begin{equation}
\frac{\mu_1 t^2}{1-\mu _2 t^2}    = \mu_1 t^2 + \mu_1 \mu _2 t^4 + \mu_1 \mu _2^2 t^6  + \mu_1 \mu _2^3 t^8 + ...
\end{equation} 
}
\begin{multline}
\HWG(t)=\frac{1-\mu _1^6 t^{12}}{\left(1-\mu _2 t^2\right) \left(1-\mu _1^2 t^4\right) \left(1-\mu _1^3 t^6\right)}   \\=1+\mu _2 t^2+\left(\mu _1^2+\mu _2^2\right) t^4+\left(\mu _1^3+ \mu _1^2 \mu _2+\mu _2^3\right) t^6+\left(\mu _1^4+ \mu _1^3\mu _2+ \mu _1^2\mu _2^2+\mu _2^4\right) t^8+\dots
\end{multline}

Compared to the subregular nilpotent orbit we find an extra relation at $t^8$ in the $[02]$ representation. The condition that $N^2$ is of rank at most 1 is of this type.

In total the moment map is expected to satisfy three relations:
\begin{align}
    \lbrack 00 \rbrack t^4:&\ \tr (N^2) = 0 \\
    \lbrack 10 \rbrack t^6:&\ N \wedge N \wedge N = 0 \\
    \lbrack 02 \rbrack t^8:&\ \rank(N^2) \leq 1 \label{rank3}
\end{align}
and indeed all are met by our coadjoint $\mommap{G_2}$. The last relation (\ref{rank3}) can be written as $\sum_{m,n}\left( \mommap{am}\mommap{mb}\mommap{cn}\mommap{nd}- \mommap{am}\mommap{md}\mommap{cn}\mommap{nb} \right) = 0 $. We have checked analytically that the three relations above form a complete set of relations.   

\subsubsection{\texorpdfstring{$\mathbb{Z}_3$}{Z3} discrete gauging}\label{sec:g2-z3}
Although elsewhere in the paper we discretely gauge or fold $S_m$ quiver automorphisms, discrete gauging by a subset of $S_m$ is perfectly well defined. Here we consider the $\mathbb{Z}_3$ discrete gauging of the $D_4$ quiver studied in this section. Its Coulomb branch was previously investigated in \cite{hanany_ungauging_2020} under the name $\Coul_{D_4^{(3)}}$. The plethystic logarithm of its highest weight generating function was reported as\footnote{Paper \cite{hanany_ungauging_2020} also follows the opposite root convention to the present paper.}
\be
PL(t)= \lbrack 01\rbrack t^2 +(\lbrack 01 \rbrack - \lbrack 00 \rbrack ) t^4 - (\lbrack 01 \rbrack + \lbrack 10 \rbrack + \lbrack 20 \rbrack + \lbrack 00 \rbrack) t^6 - (\lbrack 01 \rbrack + \lbrack 10 \rbrack - \lbrack 02 \rbrack)t^8 + O(t^{10}).
\ee
This space is not a nilpotent orbit. It is generated by two coadjoint matrices at quadratic and quartic order in $t$ respectively. The lower coadjoint matrix $N$ is also the moment map and looks precisely like the one obtained by $S_3$ symmetric gauging. Since $\mathbb{Z}_3 \subset S_3$, there are operators in this theory which are removed if the remaining $S_2\subset S_3$ symmetry is imposed. One of the simplest operators is
\be
\tilde{e}^4_{\langle 10 \rangle} = \up{1}(\p{4}-\p{3}) + \up{3}(\p{1}-\p{4}) + \up{4}(\p{3}-\p{1}).
\ee
As its label suggests, $\tilde{e}^4_{\langle 10 \rangle}$ is a $t^4$ operator which acts as the first simple root under action of the moment map's components. And just as one can ``rotate" a simple root into any other root by repeated action of the Lie bracket, it is possible to repeatedly act with the Poisson bracket on $\tilde{e}^4_{\langle 10 \rangle}$ to generate an entire $t^4$ adjoint representation's worth of operators which can be bundled together to form the second coadjoint matrix $R$. For example:
\be
\tilde{e}^4_{\langle 01 \rangle} = -\pois{\tilde{e}_{\langle -10 \rangle},\pois{\tilde{e}_{\langle 01 \rangle},\tilde{e}^4_{\langle 10 \rangle}}}
\ee

The plethystic logarithm suggests several relations between $N$ and $R$ but we find it is not too helpful in this case. For example, its syzygies obscure several relations at order $t^8$. Accordingly, we opt for a different approach to identify the relations. \cite{hanany_ungauging_2020} identifies a non-simply laced quiver with the same Coulomb branch, which is itself a folded version of the quiver in Figure 8 of \cite{cheng_coulomb_2017}; the latter paper reports matrix relations. In general folded relations follow the form of the original quiver's; indeed they must as they are merely the original relations restricted to the folded subspace. Accounting for several coincidences in $G_2$ (eg $N^3 \propto (\tr N^2)N, \lbrace N,R \rbrace \propto N\wedge R$) and a different numerical factor in the last relation, we are left with the following relations:
\begin{align}
    \lbrack 00 \rbrack t^4:&\ \tr N^2 = 0 \\
    \lbrack 10 \rbrack t^6:&\ N \wedge N \wedge N = 0 \\
    \lbrack 01 \rbrack t^6:&\ \lbrack N, R \rbrack = 0 \\
    (\lbrack 20 \rbrack + \lbrack 00 \rbrack) t^6:&\ \lbrace N, R \rbrace = 0 \\
    (\lbrack 20 \rbrack + [00])  t^8:&\ R^2 = 0 \\
    \lbrack 02 \rbrack t^8:&\ (N^2)_{\lbrack a}^{\ \lbrack b} (N^2)_{c\rbrack }^{\ d\rbrack } = \frac{1}{54} R_a^{\ b} R_c^{\ d}
\end{align}
We are able to verify all of them symbolically, but cannot guarantee that they form a minimal set of relations as our current techniques run against a computational limit.

\subsection{\texorpdfstring{$D_5 \rightarrow B_4$}{D5 -> B4}} \label{section:d5-to-b4}
We close off by studying discrete gauging and folding on a family of quivers. Figures \ref{tab:d5min-b4}, \ref{tab:d5-nmin-b4} and \ref{tab:d5nnmin-b4} present results of discrete gauging and folding on three $D_5$ nilpotent orbit quivers. The Hilbert series, HWGs and quivers were originally reported in \cite{hanany_quiver_2016,cabrera_nilpotent_2017}. 

Figures \ref{tab:d5min-b4}-\ref{tab:d5nnmin-b4} follow the same pattern. The first line shows the unitary magnetic quivers. The second line shows the equivalent orthosymplectic magnetic quivers (ie. with the same Coulomb branch); our discrete gauging appears to be the unitary analogue of gauging an $O(1)$ group in these quivers as studied in \cite{cabrera_nilpotent_2017}. The third line shows an electric quiver, by which we mean a classical quiver theory whose Higgs branch is the Coulomb branch under study. Several quivers may share this property; in particular the ones chosen here need not be the $3d$ mirrors. Note in those electric quivers the appearance of an $O_1=\mathbb{Z}_2$ gauge group in the middle column. The last lines show the Hasse diagrams, HWG and relations. The HWG use $B_4$ fugacities except in the first column where $D_5$ fugacities are also used.

We draw the reader's attention to several interesting properties.

Firstly, a $D$–type moment map in the Chevalley-Serre basis is too long to print but both discrete gauging and folding have clear and discernible effects on it. The original, unfolded moment map transforms in the coadjoint (antisymmetric) matrix representation of $\so{10}$. Upon either discrete operation, all components along the last row and column vanish and the originally $10\times 10$ matrix effectively becomes a $9\times 9$ antisymmetric matrix padded by zeroes – and hence transforms in the coadjoint representation of $\so{9}$.

Secondly, in the case of the next-to-next-to-minimal nilpotent orbit we wreathe a $U(2)$ node rather than the simple and well understood case of $U(1)$, demonstrating that discrete gauging generalises to gauge ranks higher than 1. 
Finally, in the same example, each wreathed $U(2)$ node comes with one flavor so the triplet of spaces exhibits interesting complex mass deformation behaviour analogous to that of Section \ref{section:a3-to-c2}: only the initial space can be deformed by complex mass, and turning on two inequivalent mass parameters spoils the $S_2$ symmetry required for both discrete gauging and folding.

Note that notation of the form $N\wedge \cdots \wedge N$ denotes antisymmetrisation over all indices, or equivalently contraction with the appropriate Levi-Civita tensor.

In Figures \ref{tab:d5min-b4}-\ref{tab:d5nnmin-b4}, we have colored the terms of the HWG which are charged under the $\mathbb{Z}_2$ action in violet. 

\begin{landscape}
\begin{figure}
    \centering
    \begin{tabular}
    {| >{\centering\arraybackslash} m{7.2cm} | >{\centering\arraybackslash} m{7cm} |  >{\centering\arraybackslash} m{6.2cm} |}
    \hline 
        Initial & Discretely Gauged & Folded \\ \hline \hline
			\vspace{0.3cm}\begin{tikzpicture}
			\node (g0) at (1,1) [flavor, label=left:{1}]{};
			\node (g1) at (0,0) [gauge, label=below:{1}]{}; 
			\node (g2) at (1,0) [gauge, label=below:{2}]{}; 
			\node (g3) at (2,0) [gauge, label=below:{2}]{}; 
			\node (g4) at (3,.5) [gauge, label=below:{1}]{};
			\node (g6) at (3,-.5) [gauge, label=below:{1}]{}; 
			\draw (g1)--(g2)--(g3)--(g4);
			\draw (g3)--(g6);
			\draw (g0)--(g2);
		\end{tikzpicture} & \begin{tikzpicture}
			\node (g0) at (1,1) [flavor, label=left:{1}]{};
			\node (g1) at (0,0) [gauge, label=below:{\symmlabel{1}}]{}; 
			\node (g2) at (1,0) [gauge, label=below:{\symmlabel{2}}]{}; 
			\node (g3) at (2,0) [gauge, label=below:{\symmlabel{2}}]{}; 
			\node (g4) at (3,0) [gauge, label=below:{\symmlabel{[1]\wr S_2}}]{};
			\draw (g1)--(g2)--(g3)--(g4);
			\draw (g0)--(g2);
		\end{tikzpicture} & \begin{tikzpicture}
			\node (g0) at (1,1) [flavor, label=left:{1}]{};
			\node (g1) at (0,0) [gauge, label=below:{1}]{}; 
			\node (g2) at (1,0) [gauge, label=below:{2}]{}; 
			\node (g3) at (2,0) [gauge, label=below:{2}]{}; 
			\node (g4) at (3,0) [gauge, label=below:{$1$}]{};
			\draw (g1)--(g2)--(g3);
			\draw (g0)--(g2);
			\draw[doublearrow] (g3)-- node {\midarrow} (g4);
		\end{tikzpicture} \\ \hline
			\vspace{0.2cm}\begin{tikzpicture}
			\node (g1) at (0,0) [gauge, label=below:{$SO_2$}]{}; 
			\node (g2) at (1,0) [gauge, label=below:{$Sp_1$}]{}; 
			\node (g3) at (2,0) [gauge, label=below:{$SO_3$}]{}; 
			\node (g4) at (3,0) [gauge, label=below:{$Sp_1$}]{}; 
			\node (g5) at (4,0) [gauge, label=below:{$SO_3$}]{}; 
			\node (g6) at (5,0) [gauge, label=below:{$Sp_1$}]{}; 
			\node (g7) at (6,0) [gauge, label=below:{$SO_2$}]{}; 
			\node (f2) at (1,1) [flavor, label=right:{$SO_1$}]{};
			\node (f6) at (5,1) [flavor, label=right:{$SO_1$}]{};
			\draw(g1)--(g2)--(g3)--(g4)--(g5)--(g6)--(g7);
			\draw (g2)--(f2);
			\draw (g6)--(f6);
		\end{tikzpicture} 	
		&  	
		\vspace{0.2cm}\begin{tikzpicture}
			\node (g1) at (0,0) [gauge, label=below:{$SO_2$}]{}; 
			\node (g2) at (1,0) [gauge, label=below:{$Sp_1$}]{}; 
			\node (g3) at (2,0) [gauge, label=below:{$SO_3$}]{}; 
			\node (g4) at (3,0) [gauge, label=below:{$Sp_1$}]{}; 
			\node (g5) at (4,0) [gauge, label=below:{$SO_3$}]{}; 
			\node (g6) at (5,0) [gauge, label=below:{$Sp_1$}]{}; 
			\node (g7) at (6,0) [gauge, label=below:{$O_2$}]{}; 
			\node (f2) at (1,1) [flavor, label=right:{$SO_1$}]{};
			\node (f6) at (5,1) [flavor, label=right:{$SO_1$}]{};
			\draw(g1)--(g2)--(g3)--(g4)--(g5)--(g6)--(g7);
			\draw (g2)--(f2);
			\draw (g6)--(f6);
		\end{tikzpicture} 
		&  Non-special \\ \hline 
    	\vspace{0.2cm}\begin{tikzpicture}
			\node (g1) at (0,0) [gauge, label=below:{$Sp_1$}]{}; 
			\node (f1) at (0,1) [flavor, label=right:{$SO_{10}$}]{};
			\draw (f1)--(g1);
		\end{tikzpicture}	& 
		\vspace{0.2cm}\begin{tikzpicture}
			\node (g1) at (0,0) [gauge, label=below:{$Sp_1$}]{}; 
			\node (g2) at (1,0) [gauge, label=below:{$O_1$}]{}; 
			\node (f1) at (0,1) [flavor, label=right:{$SO_{9}$}]{};
			\draw (f1)--(g1)--(g2);
		\end{tikzpicture} &
		\vspace{0.2cm}\begin{tikzpicture}
			\node (g1) at (0,0) [gauge, label=below:{$Sp_1$}]{}; 
			\node (f1) at (0,1) [flavor, label=right:{$SO_{9}$}]{};
			\draw (f1)--(g1);
		\end{tikzpicture}
		\\ \hline 
		\begin{tikzpicture}
		\tikzstyle{hasse} = [circle, fill,inner sep=2pt];
		\node [hasse] (1) [label=right:\footnotesize{$7$}] {};
		\node  (2) [below of=1] {};
		\node [hasse] (3) [below of=2,label=right:\footnotesize{$0$}] {};
		\draw (1) edge [] node[label=left:\footnotesize{$d_5$}] {} (3);
	\end{tikzpicture} 
	\hspace{1.5cm}\raisebox{1cm}{$\bar{\mathcal{O}}^{\mathfrak{so}(10)}_{[2^2,1^6]}$}
	& 	\begin{tikzpicture}
		\tikzstyle{hasse} = [circle, fill,inner sep=2pt];
		\node [hasse] (1) [label=right:\footnotesize{$7$}] {};
		\node [hasse] (2) [below of=1,label=right:\footnotesize{$6$}] {};
		\node [hasse] (3) [below of=2,label=right:\footnotesize{$0$}] {};
		\draw (1) edge [] node[label=left:\footnotesize{$a_1$}] {} (2);
		\draw (2) edge [] node[label=left:\footnotesize{$b_4$}] {} (3);
	\end{tikzpicture}
	\hspace{1.5cm}\raisebox{1cm}{$\bar{\mathcal{O}}^{\mathfrak{so}(9)}_{[3,1^6]}$} & 
	 	\begin{tikzpicture}
		\tikzstyle{hasse} = [circle, fill,inner sep=2pt];
		\node [hasse] (2) [label=right:\footnotesize{$6$}] {};
		\node [hasse] (3) [below of=2,label=right:\footnotesize{$0$}] {};
		\draw (2) edge [] node[label=left:\footnotesize{$b_4$}] {} (3);
	\end{tikzpicture} 
	   \hspace{1.5cm}\raisebox{0.4cm}{$\bar{\mathcal{O}}^{\mathfrak{so}(9)}_{[2^2,1^5]}$} \\ \hline
	\begin{tabular}{cc}
        $\mu_2 t^2$ & $D_5$ \\
        $(\mu_2+\textcolor{violet}{\mu_1})t^2$ & $B_4$
    \end{tabular}
	  &  $\mu_2 t^2 + \textcolor{violet}{\mu_1^2 t^4} \qquad B_4$ & $\mu_2 t^2 \qquad B_4$  \\  \hline
 	 \begin{tabular}{ll}
    $t^4 \left(\lbrack 20000 \rbrack + \lbrack 00000 \rbrack\right):$& $N^2 = 0$ \\
    $t^4 \lbrack 00011 \rbrack:$& $N \wedge N = 0$
\end{tabular} 
&  
\begin{tabular}{ll}
    $t^4 \lbrack 0000 \rbrack:$& $\tr{N^2} = 0$ \\
    $t^4 \lbrack 0002 \rbrack:$& $N \wedge N = 0$\\
    $t^6 \lbrack 0100 \rbrack:$& $N^3 = 0$
\end{tabular}
& 
 	 \begin{tabular}{ll}
    $t^4 \left(\lbrack 2000 \rbrack + \lbrack 00000 \rbrack\right):$& $N^2 = 0$ \\
    $t^4 \lbrack 0002 \rbrack:$& $N \wedge N = 0$
\end{tabular}\\ 
	\hline 
    \end{tabular}
    \caption[$D_5$ minimal nilpotent orbit quiver and its discrete reductions]{$D_5$ minimal nilpotent orbit quiver and its discrete reductions. The folded space is non-special \cite{cabrera_nilpotent_2017} and therefore without an orthosymplectic realisation.}
    \label{tab:d5min-b4}
\end{figure}

\begin{figure}[p]
    \centering
    \hspace*{-1cm}
    \begin{tabular}
    {| >{\centering\arraybackslash} m{8.55cm} | >{\centering\arraybackslash} m{7cm} |  >{\centering\arraybackslash} m{7cm} |}
    \hline 
        Initial & Discretely Gauged & Folded \\ \hline \hline
			\vspace{0.3cm}\begin{tikzpicture}
			\node (g0) at (-1,0) [flavor, label=below:{2}]{};
			\node (g1) at (0,0) [gauge, label=below:{2}]{}; 
			\node (g2) at (1,0) [gauge, label=below:{2}]{}; 
			\node (g3) at (2,0) [gauge, label=below:{2}]{}; 
			\node (g4) at (3,.5) [gauge, label=below:{1}]{};
			\node (g6) at (3,-.5) [gauge, label=below:{1}]{}; 
			\draw (g0)--(g1)--(g2)--(g3)--(g4);
			\draw (g3)--(g6);
		\end{tikzpicture} & \vspace{0.35cm}\begin{tikzpicture}
			\node (g0) at (-1,0) [flavor, label=below:{\symmlabel{2}}]{};
			\node (g1) at (0,0) [gauge, label=below:{\symmlabel{2}}]{}; 
			\node (g2) at (1,0) [gauge, label=below:{\symmlabel{2}}]{}; 
			\node (g3) at (2,0) [gauge, label=below:{\symmlabel{2}}]{}; 
			\node (g4) at (3,0) [gauge, label=below:{\symmlabel{[1]\wr S_2}}]{};
			\draw (g0)--(g1)--(g2)--(g3)--(g4);
		\end{tikzpicture} & \vspace{0.35cm}\begin{tikzpicture}
			\node (g0) at (-1,0) [flavor, label=below:{2}]{};
			\node (g1) at (0,0) [gauge, label=below:{2}]{}; 
			\node (g2) at (1,0) [gauge, label=below:{2}]{}; 
			\node (g3) at (2,0) [gauge, label=below:{2}]{}; 
			\node (g4) at (3,0) [gauge, label=below:{$1$}]{};
			\draw (g0)--(g1)--(g2)--(g3);
			\draw[doublearrow] (g3)-- node {\midarrow} (g4);
		\end{tikzpicture} \\ \hline  
			\vspace{0.2cm}\begin{tikzpicture}
			\node (g1) at (0,0) [gauge, label=below:{$SO_2$}]{}; 
			\node (g2) at (1,0) [gauge, label=below:{$Sp_1$}]{}; 
			\node (g3) at (2,0) [gauge, label=below:{$SO_3$}]{}; 
			\node (g4) at (3,0) [gauge, label=below:{$Sp_1$}]{}; 
			\node (g5) at (4,0) [gauge, label=below:{$SO_3$}]{}; 
			\node (g6) at (5,0) [gauge, label=below:{$Sp_1$}]{}; 
			\node (g7) at (6,0) [gauge, label=below:{$SO_3$}]{}; 
			\node (g8) at (7,0) [gauge, label=below:{$Sp_1$}]{};
			\node (f2) at (1,1) [flavor, label=right:{$SO_1$}]{};
			\node (f8) at (7,1) [flavor, label=right:{$SO_3$}]{};
			\draw (g1)--(g2)--(g3)--(g4)--(g5)--(g6)--(g7)--(g8);
			\draw (f2)--(g2);
			\draw (f8)--(g8);
		\end{tikzpicture} 	&  	? (not a nilpotent orbit) & 
		\vspace{0.2cm}\begin{tikzpicture}
			\node (g1) at (0,0) [gauge, label=below:{$SO_2$}]{}; 
			\node (g2) at (1,0) [gauge, label=below:{$Sp_1$}]{}; 
			\node (g3) at (2,0) [gauge, label=below:{$SO_3$}]{}; 
			\node (g4) at (3,0) [gauge, label=below:{$Sp_1$}]{}; 
			\node (g5) at (4,0) [gauge, label=below:{$SO_3$}]{}; 
			\node (g6) at (5,0) [gauge, label=below:{$Sp_1$}]{}; 
			\node (g7) at (6,0) [gauge, label=below:{$O_2$}]{}; 
			\node (f2) at (1,1) [flavor, label=right:{$SO_2$}]{};
			\node (f6) at (5,1) [flavor, label=right:{$SO_2$}]{};
			\draw(g1)--(g2)--(g3)--(g4)--(g5)--(g6)--(g7);
			\draw (g2)--(f2);
			\draw (g6)--(f6);
		\end{tikzpicture}
		\\ \hline 
	    \vspace{0.2cm}\begin{tikzpicture}
			\node (g1) at (0,0) [gauge, label=below:{$Sp_1$}]{}; 
			\node (g2) at (1,0) [gauge, label=below:{$O_1$}]{}; 
			\node (f1) at (0,1) [flavor, label=right:{$SO_{10}$}]{};
			\draw (f1)--(g1)--(g2);
		\end{tikzpicture}	& 
		\vspace{0.2cm}\begin{tikzpicture}
			\node (g1) at (0,0) [gauge, label=below:{$Sp_1$}]{}; 
			\node (g2) at (1,0) [gauge, label=below:{$O_1$}]{}; 
			\node (g3) at (-1,0) [gauge, label=below:{$O_1$}]{};
			\node (f1) at (0,1) [flavor, label=right:{$SO_{9}$}]{};
			\draw (f1)--(g1)--(g2);
			\draw (g3)--(g1);
		\end{tikzpicture} &
		\vspace{0.2cm}\begin{tikzpicture}
			\node (g1) at (0,0) [gauge, label=below:{$Sp_1$}]{}; 
			\node (g2) at (1,0) [gauge, label=below:{$O_1$}]{}; 
			\node (f1) at (0,1) [flavor, label=right:{$SO_{9}$}]{};
			\draw (f1)--(g1)--(g2);
		\end{tikzpicture}\\   \hline 
		\begin{tikzpicture}
		\tikzstyle{hasse} = [circle, fill,inner sep=2pt];
		\node [hasse] (1) [label=right:\footnotesize{$8$}] {};
		\node [hasse] (2) [below of=1,label=right:\footnotesize{$7$}] {};
		\node [hasse] (3) [below of=2,label=right:\footnotesize{$0$}] {};
		\draw (1) edge [] node[label=left:\footnotesize{$a_1$}] {} (2);
		\draw (2) edge [] node[label=left:\footnotesize{$d_5$}] {} (3);
	\end{tikzpicture}
	\hspace{1.5cm}\raisebox{1cm}{$\bar{\mathcal{O}}^{\mathfrak{so}(10)}_{[3,1^7]}$} 
	& 
	\begin{tikzpicture}
		\tikzstyle{hasse} = [circle, fill,inner sep=2pt];
		\node [hasse] (1) [label=right:\footnotesize{$8$}] at (0,0) {};
		\node [hasse] (2) [label=left:\footnotesize{$7$}] at (0,-1) {};
		\node [hasse] (3) [label=right:\footnotesize{$7$}] at (1,-1) {};
		\node [hasse] (4) [label=right:\footnotesize{$6$}] at (0,-2) {};
		\node [hasse] (5) [label=right:\footnotesize{$0$}] at (0,-3) {};
		\draw (1) edge [] node[label= left:\footnotesize{$a_1$}] {} (2);
		\draw (1) edge [] node[label=above right:\footnotesize{$a_1$}] {} (3);
		\draw (2) edge [] node[label= left:\footnotesize{$a_1$}] {} (4);
		\draw (3) edge [] node[label=below right:\footnotesize{$a_1$}] {} (4);
		\draw (4) edge [] node[label=left:\footnotesize{$b_4$}] {} (5);
	\end{tikzpicture}
	& 
	 	\begin{tikzpicture}
		\tikzstyle{hasse} = [circle, fill,inner sep=2pt];
		\node [hasse] (1) [label=right:\footnotesize{$7$}] {};
		\node [hasse] (2) [below left of=1,label=right:\footnotesize{$6$}] {};
		\node [hasse] (3) [below of=2,label=right:\footnotesize{$0$}] {};
		\draw (1) edge [] node[label=left:\footnotesize{$a_1$}] {} (2);
		\draw (2) edge [] node[label=left:\footnotesize{$b_4$}] {} (3);
	\end{tikzpicture}
	\hspace{1.5cm}\raisebox{1cm}{$\bar{\mathcal{O}}^{\mathfrak{so}(9)}_{[3,1^6]}$} \\ \hline 
	 \begin{tabular}{cc}
	     $\mu_2 t^2 + \mu_1^2 t^4$ & $D_5$ \\
	     $(\textcolor{violet}{\mu_1} + \mu_2) t^2 + (1 + \mu_1^2 + \textcolor{violet}{\mu_1} ) t^4 - \mu_1^2 t^8$ & $B_4$ 
	 \end{tabular} & 
	 $\mu_2 t^2 + (1 +  \mu_1^2  +  \textcolor{violet}{\mu_1^2} )t^4 + \textcolor{violet}{\mu_1^2 t^6} - \textcolor{violet}{\mu_1^4 t^{12}} \qquad B_4$ & $\mu_2 t^2  + \mu_1^2 t^4 \qquad B_4$  \\ 
	\hline  
	\begin{tabular}{ll}
    $t^4 \lbrack 00000 \rbrack:$& $\tr{N^2} = \frac{1}{2}(\m{1,1}-\m{1,2})^2$ \\
    $t^4 \lbrack 00011 \rbrack:$& $N \wedge N = 0$\\
    $t^6 \lbrack 01000 \rbrack:$& $N^3 = \frac{1}{4}(\m{1,1}-\m{1,2})^2 N$
\end{tabular}
	&
	\begin{tabular}{ll}
    $t^4 \lbrack 0002 \rbrack:$& $N \wedge N = 0$\\
    ? & ?
    \end{tabular}
	& 
	\begin{tabular}{ll}
    $t^4 \lbrack 0000 \rbrack:$& $\tr{N^2} = \frac{1}{2} (\m{1,1}-\m{1,2})^2$ \\
    $t^4 \lbrack 0002 \rbrack:$& $N \wedge N = 0$\\
    $t^6 \lbrack 0100 \rbrack:$& $N^3 = \frac{1}{4}(\m{1,1}-\m{1,2})^2 N$
\end{tabular}\\ 
	\hline 
    \end{tabular}
    \caption[$D_5$ next-to-minimal nilpotent orbit quiver and its discrete reductions]{$D_5$ next-to-minimal nilpotent orbit quiver and its discrete reductions. $\m{1,1}$ and $\m{1,2}$ are mass parameters associated to the flavor node of the top quiver. The discretely gauged space is generated by a coadjoint matrix $N$ and another generator $R$ transforming as $\lbrack 2000 \rbrack$. The methods in the present paper need to be generalised to non-coadjoint representations to verify relations between $N$ and $R$ and such work is beyond the current scope. Its Hasse diagram should also be regarded as a conjecture.}
    \label{tab:d5-nmin-b4}
\end{figure}

\begin{figure}[p]
    \centering
    \vspace*{-2cm}\begin{tabular}
    {| >{\centering\arraybackslash} m{8.55cm} | >{\centering\arraybackslash} m{7.6cm} |  >{\centering\arraybackslash} m{4.95cm} |}
    \hline 
        Initial & Discretely Gauged & Folded \\ \hline \hline
		\vspace{0.3cm}\begin{tikzpicture}
			\node (g1) at (0,0) [gauge, label=below:{1}]{}; 
			\node (g2) at (1,0) [gauge, label=below:{2}]{}; 
			\node (g3) at (2,0) [gauge, label=below:{3}]{}; 
			\node (g4) at (3,.5) [gauge, label=below:{2}]{}; 
			\node (g5) at (4,.5) [flavor, label=below:{1}]{}; 
			\node (g6) at (3,-.5) [gauge, label=below:{2}]{}; 
			\node (g7) at (4,-.5) [flavor, label=below:{1}]{}; 
			\draw (g1)--(g2)--(g3)--(g4)--(g5);
			\draw (g3)--(g6)--(g7);
		\end{tikzpicture} & \vspace{0.32cm}\begin{tikzpicture}
			\node (g1) at (0,0) [gauge, label=below:{\symmlabel{1}}]{}; 
			\node (g2) at (1,0) [gauge, label=below:{\symmlabel{2}}]{}; 
			\node (g3) at (2,0) [gauge, label=below:{\symmlabel{3}}]{}; 
			\node (g4) at (3,0) [gauge, label=below:{\symmlabel{[2}}]{}; 
			\node (g5) at (4,0) [flavor, label=below:{\symmlabel{\qquad 1]\wr S_2}}]{}; 
			\draw (g1)--(g2)--(g3)--(g4)--(g5);
		\end{tikzpicture} & \vspace{0.3cm}\begin{tikzpicture}
			\node (g1) at (0,0) [gauge, label=below:{1}]{}; 
			\node (g2) at (1,0) [gauge, label=below:{2}]{}; 
			\node (g3) at (2,0) [gauge, label=below:{3}]{}; 
			\node (g4) at (3,0) [gauge, label=below:{2}]{}; 
			\node (g5) at (4,0) [flavor, label=below:{1}]{}; 
			\draw (g1)--(g2)--(g3);
			\draw (g4)--(g5);
			\draw[doublearrow] (g3)-- node {\midarrow} (g4); 
		\end{tikzpicture} \\   \hline 
		\vspace{0.2cm}\begin{tikzpicture}
			\node (g1) at (0,0) [gauge, label=below:{$SO_2$}]{}; 
			\node (g2) at (1,0) [gauge, label=below:{$Sp_1$}]{}; 
			\node (g3) at (2,0) [gauge, label=below:{$SO_4$}]{}; 
			\node (g4) at (3,0) [gauge, label=below:{$Sp_2$}]{}; 
			\node (g5) at (4,0) [gauge, label=below:{$SO_4$}]{}; 
			\node (g6) at (5,0) [gauge, label=below:{$Sp_1$}]{}; 
			\node (g7) at (6,0) [gauge, label=below:{$SO_2$}]{}; 
			\node (g8) at (3,1) [flavor, label=right:{$SO_2$}]{};
			\draw (g1)--(g2)--(g3)--(g4)--(g5)--(g6)--(g7);
			\draw (g4)--(g8);
		\end{tikzpicture} 	&  	\vspace{0.2cm}\begin{tikzpicture}
			\node (g1) at (0,0) [gauge, label=below:{$SO_2$}]{}; 
			\node (g2) at (1,0) [gauge, label=below:{$Sp_1$}]{}; 
			\node (g3) at (2,0) [gauge, label=below:{$SO_4$}]{}; 
			\node (g4) at (3,0) [gauge, label=below:{$Sp_2$}]{}; 
			\node (g5) at (4,0) [gauge, label=below:{$[SO_4$}]{}; 
			\node (g6) at (5,0) [gauge, label=below:{$Sp_1$}]{}; 
			\node (g7) at (6,0) [gauge, label=below:{$SO_2]$}]{}; 
			\node (g8) at (3,1) [flavor, label=right:{$SO_2$}]{};
			\draw (g1)--(g2)--(g3)--(g4)--(g5)--(g6)--(g7);
			\draw (g4)--(g8);
		\end{tikzpicture} &   Non-special
		\\ \hline 
		\vspace{0.2cm}\begin{tikzpicture}
			\node (g4) at (3,0) [gauge, label=below:{$Sp_2$}]{}; 
			\node (g8) at (3,1) [flavor, label=right:{$SO_{10}$}]{};
			\draw (g4)--(g8);
		\end{tikzpicture} 	&  	\vspace{0.2cm}\begin{tikzpicture}
			\node (g4) at (3,0) [gauge, label=below:{$Sp_2$}]{}; 
			\node (g5) at (4,0) [gauge, label=below:{$O_1$}]{}; 
			\node (g8) at (3,1) [flavor, label=right:{$SO_{9}$}]{};
			\draw (g5)--(g4)--(g8);
		\end{tikzpicture}  &   	\vspace{0.2cm}\begin{tikzpicture}
			\node (g4) at (3,0) [gauge, label=below:{$Sp_2$}]{}; 
			\node (g8) at (3,1) [flavor, label=right:{$SO_{9}$}]{};
			\draw (g4)--(g8);
		\end{tikzpicture} 
		\\ \hline 
		\begin{tikzpicture}
		\tikzstyle{hasse} = [circle, fill,inner sep=2pt];
		\node [hasse] (1) [label=right:\footnotesize{$10$}] at (0,3) {};
		\node [hasse] (2) [label=right:\footnotesize{$7$}] at (0,2) {};
		\node [hasse] (3) [label=right:\footnotesize{$0$}] at (0,0) {};
		\draw (1) edge [] node[label=left:\footnotesize{$d_3$}] {} (2);
		\draw (2) edge [] node[label=left:\footnotesize{$d_5$}] {} (3);
	\end{tikzpicture} 
	\hspace{1.5cm}\raisebox{1.3cm}{$\bar{\mathcal{O}}^{\mathfrak{so}(10)}_{[2^4,1^2]}$}  & \begin{tikzpicture}
		\tikzstyle{hasse} = [circle, fill,inner sep=2pt];
		\node [hasse] (1) [label=left:\footnotesize{$10$}] at (0,3) {};
		\node [hasse] (2) [label=left:\footnotesize{$7$}] at (0,2) {};
		\node [hasse] (3) [label=left:\footnotesize{$0$}] at (0,0) {};
		\node [hasse] (4) [label=left:\footnotesize{$6$}] at (0,1) {};
		\node [hasse] (5) [label=right:\footnotesize{$8$}] at (1,2.4) {};
		\draw (1) edge [] node[label=left:\footnotesize{$d_3$}] {} (2);
		\draw (2) edge [] node[label=left:\footnotesize{$a_1$}] {} (4);
		\draw (4) edge [] node[label=left:\footnotesize{$b_4$}] {} (3);
		\draw (5) edge [] node[label=right:\footnotesize{$b_2$}] {} (4);
		\draw (1) edge [] node[label=above right:\footnotesize{$c_2$}] {} (5);
	\end{tikzpicture} 
	\hspace{1cm}\raisebox{1.3cm}{$\bar{\mathcal{O}}^{\mathfrak{so}(9)}_{[3,2^2,1^2]}$} & 
	\begin{tikzpicture}
		\tikzstyle{hasse} = [circle, fill,inner sep=2pt];
		\node [hasse] (3) [label=left:\footnotesize{$0$}] at (0,0) {};
		\node [hasse] (4) [label=left:\footnotesize{$6$}] at (0,1) {};
		\node [hasse] (5) [label=right:\footnotesize{$8$}] at (1,2.4) {};
		\draw (4) edge [] node[label=left:\footnotesize{$b_4$}] {} (3);
		\draw (5) edge [] node[label=right:\footnotesize{$b_2$}] {} (4);
	\end{tikzpicture}
	\hspace{0.8cm}\raisebox{1.3cm}{$\bar{\mathcal{O}}^{\mathfrak{so}(9)}_{[2^4,1]}$} 
	\\ 	\hline 
\begin{tabular}{cc}
$\mu_2 t^2 + \mu_4 \mu_5 t^4$ & $D_5$ \\
$(\mu_2+\textcolor{violet}{\mu_1}) t^2 + (\mu_4^2+\textcolor{violet}{\mu_3}) t^4$ & $B_4$
\end{tabular}
& 
$\mu_2 t^2 + (\textcolor{violet}{\mu_1^2} + \mu_4^2) t^4  + \textcolor{violet}{\mu_1 \mu_3} t^6 + \textcolor{violet}{\mu_3^2} t^8 - \textcolor{violet}{\mu_1^2 \mu_3^2} t^{12}$ 
& $\mu_2 t^2 + \mu_4^2 t^4$ \\ \hline
\begin{tabular}{ll}
    $t^4 \lbrack 20000 \rbrack:$& $N^2 = \frac{1}{16} (\m{4}-\m{5})^2 \mathbf{1}$ \\
    $t^6 \lbrack 00011 \rbrack:$& $N \wedge N \wedge N = \frac{3 \im (\m{4}-\m{5})}{2 \cdot 6!}  \star \left( N \wedge N \right)$
\end{tabular}
& 
\begin{tabular}{ll}
    $t^4 \lbrack 0000 \rbrack:$& $\tr(N^2) = 0$ \\
    $t^6 \lbrack 0100 \rbrack:$& $N^3 = 0$ \\
    $t^6 \lbrack 0010 \rbrack:$& $N \wedge N \wedge N = 0$\\
    $t^8 \lbrack 0200 \rbrack:$& $\rank(N^2)\leq 1$
\end{tabular}
& 
\begin{tabular}{ll}
    $t^4 \lbrack 0100 \rbrack:$& $N^2 = 0$ \\
    $t^6 \lbrack 0010 \rbrack:$& $N \wedge N \wedge N = 0$
\end{tabular}
\\ \hline
    \end{tabular}
    \caption[$D_5$ next-to-next-to-minimal nilpotent orbit quiver and its discrete reductions]{$D_5$ next-to-next-to-minimal nilpotent orbit quiver and its discrete reductions. $\m{4}$ and $\m{5}$ are mass parameters associated to the flavor nodes of the top quiver. The folded space is non-special \cite{cabrera_nilpotent_2017} and therefore without an orthosymplectic realisation.}
    \label{tab:d5nnmin-b4}
\end{figure}
\end{landscape}

\section*{Acknowledgements}
The authors are grateful to Julius Grimminger and Zhenghao Zhong for clearing up an important point about discrete gauging and useful comments on the draft. D.M. is supported by STFC DTP research studentship grant ST/N504336/1 while A.B. and A.H. are supported by STFC grant ST/P000762/1. We are grateful to organisers of the 11th Joburg Workshop on String Theory for their hospitality, and the MIT-Imperial College London Seed Fund for support.

\appendix

\section{Folded Lie algebras are the same as discretely gauged Lie algebras}\label{appendix:folded_lie_alg}

As mentioned in the main text, the Lie algebra of the discretely gauged space is given by
\be
\pois{\Tilde{\mathcal{O}}_i, \Tilde{\mathcal{O}}_j} = \sum_k \tensor{c}{_{i j}^k} \Tilde{\mathcal{O}}_k.
\ee
for $\Tilde{\mathcal{O}}_i$ which form a basis of $\co \lbrack \Tilde{\Coul} \rbrack_2$. In particular these operators vary across the moduli space. Restricting to the folded subspace, we find
\be
\pois{\hat{\mathcal{O}}_i, \hat{\mathcal{O}}_j} = \sum_k \tensor{c}{_{i j}^k} \hat{\mathcal{O}}_k.
\ee
This does not necessarily mean that the two Lie algebras are isomorphic as some of the RHS terms could vanish if $\hat{\mathcal{O}}_k$ vanishes identically. We will now prove that this does not happen.

$\Tilde{\mathcal{O}}_k$ is a non-constant symmetric function in variables attached to wreathed legs; call them $\vec{x}_i$ where $i$ labels the leg. So we can rewrite the operator as $f(\vec{x}_1, \dots \vec{x}_n)$ for some $n$. At the fixed point $\vec{x}_i = \vec{x}$, so the operator becomes $f(\vec{x}, \dots, \vec{x})$. Assume it vanishes everywhere. Then

\be
\nabla_{\vec{x}} f(\vec{x}, \dots, \vec{x}) = \restr{\sum_i \nabla_{\vec{x}_i} f(\vec{x}_1,\dots \vec{x}_n)}{\vec{x}_j = x}.
\ee

However all the summands are identical under the restriction:
\be \begin{split}
\left(\nabla_{\vec{x}_i} f(\vec{x}_1,\dots \vec{x}_n)\right)_j &= \lim_{\varepsilon \rightarrow 0}\frac{f(\vec{x}_1,\dots, \vec{x}_i + \varepsilon e_i, \dots, \vec{x}_n) - f(\vec{x}_1,\dots, \vec{x}_i, \dots, \vec{x}_n)}{\varepsilon} \\
&= \lim_{\varepsilon \rightarrow 0}\frac{f(\vec{x}_i + \varepsilon e_i, \dots, \vec{x}_1, \dots, \vec{x}_n) - f(\vec{x}_i,\dots, \vec{x}_1, \dots \vec{x}_n)}{\varepsilon} \\
&= \restr{\left(\nabla_{\vec{x}'_1} f(\vec{x}'_1,\dots \vec{x}'_n)\right)_j}{\begin{split}\vec{x}'_1 &= \vec{x}_i \\ \vec{x}'_i &= \vec{x}_1 \\ \vec{x}'_{j\neq 1,i} &= \vec{x}_j \end{split}}
\end{split} \ee
so
\be
\restr{\left(\nabla_{\vec{x}_i} f(\vec{x}_1,\dots \vec{x}_n)\right)_j}{\vec{x}_j = \vec{x}} = \restr{\left(\nabla_{\vec{x}_1} f(\vec{x}_1,\dots \vec{x}_n)\right)_j}{\vec{x}_j = \vec{x}}.
\ee
Then
\be
\nabla_{\vec{x}} f(\vec{x}, \dots, \vec{x}) = \restr{n \nabla_{\vec{x}_1} f(\vec{x}_1,\dots \vec{x}_n)}{\vec{x}_j = x} \neq 0
\ee
unless $\restr{\nabla_{\vec{x}_i} f(\vec{x}_1,\dots \vec{x}_n)}{\vec{x}_j = \vec{x}}$ vanishes, ie. $f(\vec{x}_1,\dots \vec{x}_n)$ is a constant, which contradicts the assumption that $\Tilde{\mathcal{O}}_k$ is non-constant. It follows that both the discretely gauged and folded spaces have isomorphic Lie algebras and hence share the same continuous symmetry.

\newpage 

\section{\texorpdfstring{Computation of Hilbert series with $S_4$ wreathing}{Computation of Hilbert series with S4 wreathing}}
\label{appendixS4}

The computation of the exact Hilbert series presented in Figure \ref{figHasseS4} can be done in principle using (\ref{monopoleSym}). However it is often useful to massage this formula until a more manageable form can be used in practice. In this appendix, we give the result of such manipulations in the case of the quiver at hand. Derivations use simple algebra and are not detailed here. 

Using the notations of (\ref{notationD4}), but using the gauge group (\ref{gaugegroupS4}), one can set $m_{a,2}=0$ and $m_{a,1} \equiv m_a$ and the conformal dimension can be expressed in terms of \be m=(m_a,m_b,m_c,m_d,m_e)\ee as 
\begin{equation}
    2 \Delta (m) =  |m_a - m_b| + |m_a - m_c| + |m_a - m_d| + |m_a - m_e|  + |m_b| + |m_c| + |m_d| + | m_e| - 2 |m_a | \,  . 
\end{equation}
One then computes the auxiliary sums 
\begin{equation}
\label{sumsS4}
    \Sigma_i = \sum\limits_{m_a = 0}^{\infty}  \sum\limits_{(m_b,m_c,m_d,m_e) \in \mathrm{Range}_i} t^{ 2 \Delta (m) }
\end{equation}
where $\mathrm{Range}_i$ is defined in Figure \ref{tableRanges}. The exact value of the sums $\Sigma_i$ is straightforward to compute (note the absence of Casimir factors!) and is given in Figure \ref{tableRanges} as well. 

Let's now pick a subgroup $\Gamma$ of $S_4$. For $\mu \in \mathbb{Z}^4$ we call $O^{S_4}(\mu)$ the orbit of $\mu$ under the action of $S_4$. This orbit can be written as a disjoint union of $n(\mu)$ orbits under $\Gamma$, 
\begin{equation}
    O^{S_4}(\mu) = \coprod\limits_{j=1}^{n(\mu)} O^{\Gamma}(\mu_j)
\end{equation}
where the $\mu_j \in \mathbb{Z}^4$ are representatives of these orbits, (not uniquely!) determined by the above equation. Using the notation (\ref{casimirGamma}), that we recall here, 
\begin{equation}
    P_{\Gamma} (t^2 ; \mu) = \frac{1}{|\Gamma|} \sum\limits_{\gamma \in \Gamma (\mu)} \frac{1}{\det \left( 1-t^2 \gamma \right)} \, , 
\end{equation}
one can define the modified Casimir factor 
\begin{equation}
    \tilde{P}_{\Gamma} (t^2 ; \mu) = \sum\limits_{j=1}^{n(\mu)}  P_{\Gamma} (t^2 ; \mu_j) \, . 
\end{equation}
The rationale behind this definition is that we have evaluated the sums (\ref{sumsS4}) which are adapted to the full group $S_4$, and the Casimir factors for the group $\Gamma$ have to be collected accordingly. This being done, the Hilbert series for the Coulomb branch of the wreathed quivers are simply 
\begin{equation}
    \HS_{\Gamma}(t) = \sum\limits_{i=1}^6 \tilde{P}_{\Gamma} (t^2 ; \mu_i) \Sigma_i \, ,  
\end{equation}
where $\mu_i \in \mathbb{Z}^4$ is any element satisfying the condition $\textrm{Range}_i$. Using this formula, all the Hilbert series of Figure \ref{figHasseS4} are evaluated in a fraction of a second on a standard computer. 

\begin{figure}[t]
    \centering
    \begin{tabular}{|c|c|c|}\hline 
       $i$ & $\textrm{Range}_i$ & $\Sigma_i$ \\ \hline 
        1 & $m_b<m_c<m_d<m_e$ &   $\frac{t^6 \left(4+16 t^2+18 t^4+13 t^6+4 t^8+t^{10}\right)}{\left(1-t^2\right)^6 \left(1+t^2\right)^3} $  \\ \hline 
        2 & \begin{tabular}{c}
            $m_b=m_c<m_d<m_e$  \\
            $m_b<m_c<m_d=m_e$
        \end{tabular} &    $ \frac{t^4 \left(2+9 t^2+10 t^4+9 t^6+3 t^8+t^{10}\right)}{\left(1-t^2\right)^5 \left(1+t^2\right)^3}$  \\ \hline 
        3 & $m_b<m_c=m_d<m_e$ &    $\frac{t^4 \left(4+7 t^2+7 t^4+3 t^6+t^8\right)}{\left(1-t^2\right)^5 \left(1+t^2\right)^2}$  \\ \hline 
        4 & \begin{tabular}{c}
            $m_b=m_c=m_d<m_e$  \\
            $m_b<m_c=m_d=m_e$
        \end{tabular} &    $\frac{t^2 \left(2+3 t^2+4 t^4+2 t^6+t^8\right)}{\left(1-t^2\right)^4 \left(1+t^2\right)^2}$  \\ \hline 
        5 & $m_b=m_c<m_d=m_e$ &    $\frac{t^2+5 t^4+5 t^6+6 t^8+2 t^{10}+t^{12}}{\left(1-t^2\right)^4 \left(1+t^2\right)^3}$  \\ \hline 
        6 & $m_b=m_c=m_d=m_e$ &    $\frac{\left(1-t+t^2\right) \left(1+t+t^2\right) \left(1+t^4\right)}{\left(1-t^2\right)^3 \left(1+t^2\right)^2}$  \\ \hline 
    \end{tabular}
    \caption{Definitions of the ranges involved in the sums (\ref{sumsS4}), and exact values of these sums. When there are two possible ranges, this means that the two choices lead to the same sums. }
    \label{tableRanges}
\end{figure}

\newpage 

\bibliographystyle{JHEP} 
\bibliography{bibli.bib}

\end{document}